\documentclass[12pt, article, leqno]{article}%
\usepackage[font=small]{caption}
\usepackage[utf8]{inputenc}
\usepackage{amssymb}
\usepackage{natbib}
\usepackage[titletoc]{appendix}
\usepackage{color,bm}
\usepackage{pdflscape}
\usepackage{booktabs}
\usepackage{graphicx,rotating}
\usepackage{amsthm,amsmath,amsfonts,amssymb, xr}
\usepackage{graphics}
\usepackage{epsfig}
\usepackage{booktabs}
\usepackage{array,multirow}
\usepackage{verbatim}
\usepackage{bm}
\usepackage{lscape}
\usepackage{sanitize-umlaut}
\usepackage{diagbox}
\usepackage{latexsym}
\usepackage[bottom]{footmisc}
\usepackage{hyperref}
\usepackage{pdfpages}
\usepackage{lscape}
\usepackage{adjustbox}
\usepackage{amsmath}
\usepackage{amssymb}
\usepackage{enumerate,enumitem}
\usepackage{float}
\usepackage{setspace}
\usepackage{longtable}
\usepackage{graphics}
\usepackage{epsfig}
\usepackage{amsfonts}
\usepackage{graphicx}
\usepackage{pdflscape}
\usepackage{textcomp}
\usepackage{titling}
\usepackage[resetlabels, labeled]{multibib}
\usepackage[mathscr]{eucal}
\usepackage{geometry}
\usepackage[compact]{titlesec}
\usepackage{titlesec}
\usepackage{setspace}
\usepackage{setspace}
\usepackage{protecteddef}
\usepackage{authblk}%
\hypersetup{
	colorlinks=true,
	linkcolor=blue,
	filecolor=blue,
	urlcolor=blue,
	citecolor=blue,
}
\providecommand{\U}[1]{\protect\rule{.1in}{.1in}}
\providecommand{\U}[1]{\protect
	\rule{.1in}{.1in}}

\def \1{\hbox{\bf 1}}

\def \Var{\var \,}

\def \beqn{\begin{displaymath}}
\def \eeqn{\end{displaymath}}

\newcommand{\mockalph}[1]{}

\newtheorem{theorem}{\sc Theorem}
\newtheorem{lemma}{\sc Lemma}

\newtheorem{assumption}{\sc Assumption}[section]
\newtheorem{definition}{\sc Definition}
\newtheorem{@example}{\sc Example}[section]

\newtheorem{@remark}{\sc Remark}
\newenvironment{remark}{\begin{@remark}\rm}{\end{@remark}}
{\catcode `\@=11 \global \let \AddToReset=\@addtoreset}

\newcommand{\var}{\mathrm{Var}}

\renewcommand{\theequation}{\arabic{equation}}

\renewcommand{\thelemma}{\arabic{lemma}}

\def \1{{\bf 1}}

\newcites{OA}{References}

\usepackage{geometry}
\allowdisplaybreaks

\AtBeginDocument{\addtolength\abovedisplayskip{-0.2\baselineskip}\addtolength\belowdisplayskip{-0.2\baselineskip}\addtolength\abovedisplayshortskip{-0.5\baselineskip}\addtolength\belowdisplayshortskip{-0.5\baselineskip}}

\begin{document}
	
	\title{Heterogeneous Grouping Structures in Panel Data}
	\author[1]{Katerina Chrysikou 
		\thanks{ {katerina.chrysikou@kcl.ac.uk}} } 
	\renewcommand\Authands{ and }
	\author[1]{George Kapetanios 
		\thanks{{george.kapetanios@kcl.ac.uk}.  }}
	\affil[1]{	King's Business School , 	King's College London}
	\date{{ \today}}
	\maketitle
	
	\begin{abstract}
		In this paper we examine the existence of heterogeneity within a group, in
		panels with latent grouping structure. The assumption of within group
		homogeneity is prevalent in this literature, implying that the formation of
		groups alleviates cross-sectional heterogeneity, regardless of the prior
		knowledge of groups. While the latter hypothesis makes inference powerful, it
		can be often restrictive. We allow for models with richer heterogeneity that
		can be found both in the cross-section and within a group, without imposing
		the simple assumption that all groups must be heterogeneous. We further
		contribute to the method proposed by \cite{su2016identifying}, by showing that
		the model parameters can be consistently estimated and the groups, while
		unknown, can be identifiable in the presence of different types of
		heterogeneity. Within the same framework we consider the validity of assuming
		both cross-sectional and within group homogeneity,
		using testing procedures. Simulations demonstrate good finite-sample
		performance of the approach in both classification and estimation, while
		empirical applications across several datasets provide evidence of multiple
		clusters, as well as reject the hypothesis of within group homogeneity.
		
		
	\end{abstract}
	
	\affil{King's Business School , King's College London}
	
	\renewcommand\Authands{ and }
	
	\section{Introduction}
	
	\label{introduction}
	
	The use and analysis of panel datasets and associated models is a vital
	research topic in both econometric theory and applications; these models are
	frequently applied in many disciplines across economics and other social
	sciences. This type of dataset allows the consideration of both
	cross-sectional and time dimensions, and provides a rich source of empirical
	information. In its majority, panel data analysis relies heavily on the
	assumption of homogeneity across units making inference more powerful.
	
	Nevertheless, such an assumption is often restrictive and frequently rejected;
	see, among others, \cite{chudik2015common}, \cite{phillips2007transition},
	\cite{pesaran2006estimation}, and \cite{browning2014dynamic}. Ignoring
	heterogeneity, across units, can result in biased estimation and misleading
	inference; see, among others, Chapter 1 in \cite{baltagi2008econometric},
	\cite{moulton1986random}, and \cite{moulton1987diagnostics}. On the other
	hand, allowing complete heterogeneity leads to potentially inefficient panel
	data estimation and inference. To address this issue, researchers often
	consider intermediate ways of exploring panel data such as panel structure models.
	
	Panel structure models rely on the hypothesis that all, or a subset of,
	parameters are heterogeneous across groups but homogeneous within a group
	where neither the number of groups nor the individuals' group-membership is
	known. Examples of these models can be found in \cite{phillips2007transition}
	who accommodate the hypothesis of convergence of clubs; in this case different
	countries belong to different groups and behave differently with the group
	structures being latent rather than assumed to be observable.
	
	The identification of latent structures is not an easy task. It is a
	computationally expensive procedure and it can be even infeasible to try all
	the possible permutations of units across groups. A part of the literature
	considers that units in the dataset are grouped using an external observable
	classification; see, e.g., \cite{BESTER2016197}. This approach may suffer, as
	suitable external variables can be difficult to find in many empirical
	settings. Furthermore, a wrong choice for the external classifier may again
	lead to misleading inference. For this reason, it can be argued that it is
	more useful for the researcher to use data-driven classification procedures
	such as \emph{k-means}, see, e.g., \cite{sarafidis2015partially},
	\cite{bonhomme2015grouped}, \cite{lin2012estimation}, and \cite{ando2016panel}%
	; these studies consider \emph{k-means} in linear panel data models, with an unknown
	grouping structure. Although unsupervised classification is an appealing
	approach and has been theoretically proven to be asymptotically consistent,
	see, e.g., \cite{pollard1981strong}, alternative approaches based on the
	\emph{Classifier Lasso} (\textsc{c-lasso}), proposed by
	\cite{su2016identifying} (SSP) and extended by \cite{su2018identifying} and
	\cite{su2019sieve}, have been recently introduced in the literature. In
	particular, SSP propose a novel estimation technique which involves
	penalisation and simultaneously classifies the data and estimates the
	underlying parameters consistently. \cite{meh2022} provides a framework for
	joint estimation and identification of latent grouping structures in panel data
	models using a pairwise fusion penalized approach. \cite{mammen2022estimation}
	presents a new approach to grouping fixed effects in a linear panel model to
	reduce their dimensionality and ensure identifiability, by using unsupervised
	non-parametric density based clustering, cluster patterns including their
	location and number are not restricted.
	
	A different line of work is that of \cite{ren2022matrix} who propose a group
	matrix network autoregression (GMNAR) model, which assumes that the subjects
	in the same group share the same set of model parameters. In a similar vein,
	\cite{zhu2022simultaneous} study dynamic behaviours of heterogeneous
	individuals observed in a network, where the dynamic patterns are
	characterised by a network vector autoregression model with a latent grouping
	structure, where group-wise network effects and time-invariant fixed-effects
	can be incorporated. \cite{wu2022} develop methods to recover sparsity
	patterns and grouping structures in large panels with both individual and time effects.
	
	In a low dimensional setting, where $T>>N,p$, \cite{wanga2022panel} generalise
	their framework and consider panel model with interactive fixed effects such
	that individual heterogeneity is captured by latent grouping structure and time
	heterogeneity is captured by an unknown structural break. \cite{liu2022panel}
	propose a methodology for identifying and estimating explosive bubbles in
	mixed-root panel autoregressions with a latent grouping structure.
	
	Full homogeneity placed within groups, is a very common assumption in the literature, but this can be a potentially restrictive hypothesis.  It is conceivable and, as we will argue, often the
	case, in practice, that each group contains units with heterogeneous
	parameters whose mean is group specific and differs across groups.
	
	To fix ideas let the parameter of interest across $N$ panel units be
	$\boldsymbol{\beta}_{i} = (\boldsymbol{\beta}_{1,i}, \boldsymbol{\beta}_{2,i},
	\ldots, \boldsymbol{\beta}_{p, i})^{\prime}$, where $p$ is the number of
	covariates within each unit $i=1,\ldots,N$. Traditionally, under full homogeneity, $\boldsymbol{\beta
	}_{i}=\boldsymbol{\beta}$. This is full homogeneity. A heterogeneous extension
	is $\boldsymbol{\beta}_{i}=\boldsymbol{\beta}+\boldsymbol{\eta}_{i}$, where
	$\boldsymbol{\eta}_{i}$ is an i.i.d. process and focus is placed on estimation
	of and inference for $\boldsymbol{\beta}$. Grouping structures specify that
	$\boldsymbol{\beta}_{i}=\boldsymbol{\alpha}_{k}$, whereby unit $i$ belongs to
	group $k$, where $k=1,\ldots,K$ and $K$ is usually assumed finite. Focus here is placed
	on $\boldsymbol{\alpha}_{k}$. Our setting extends this by specifying that
	$\boldsymbol{\beta}_{i}=\boldsymbol{\alpha}_{k}+\boldsymbol{\eta}_{i}.$
	Clearly $\boldsymbol{\eta}_{i}=\boldsymbol{0}_{p\times1}$ and
	$\boldsymbol{\alpha}_{k}=\boldsymbol{\alpha}$, are both assumptions that are
	empirically verifiable and potentially invalid. The case where neither holds
	has not been explored in the literature and is the main focus, and  contribution, of this paper.
	
	Following the work of SSP, we extend the Classifier-Lasso framework to the
	above case and further propose an estimation technique based on the
	\emph{k-means} clustering method allowing for within-group heterogeneity.
	Our focus is mainly to showcase consistency  of the estimation of
	$\boldsymbol{\beta}_{i}$ and $\boldsymbol{\alpha}_{k}$ providing both  theoretical and  small sample evidence, rather than the identification of an exact
	grouping structure, which under our form of heterogeneity is obviously not
	consistently possible. We suggest the use of a simple unsupervised classifier,
	like the \emph{k-means}, to determine the grouping structure and, once
	determined, we estimate the unit-specific parameters with penalised methods.
	
	Our work extends SSP and contributes to the literature as follows: first, we
	allow for a \emph{less} restrictive model, second we allow for richer
	cross-sectional heterogeneity, and finally, we provide a data-driven approach
	to classify each cross-section. While the latter is a gain of our modelling
	approach, heterogeneity is still restrictive and, in certain cases,
	contradicts the basic principles of \emph{k-means} clustering. In
	particular, when the level of heterogeneity is high, causing the data to be
	more noisy cross-sectionally, the clustering process tends to be misleading
	since it violates one of the three basic assumptions of \emph{k-means}; that
	is, the variance between individuals must be constant and small. Because
	\emph{k-means} minimises the within cluster deviation (in terms of squared
	Euclidean distances), the optimisation method converges to local minima when
	the variance between individuals (cross-sections) is large.
	
	The latter is inconvenient when the empirical data is noisy or even
	correlated, but there are other unsupervised classifiers which deal with these
	types of cases; see, e.g., \cite{leisch1999bagged} who re-estimates the data
	in different partitions using bootstrapping. In this paper, we do not focus
	on such a task since our approach does not cover cases where data  exhibit
	autocorrelation or endogeneity. Obviously, an alternative to our approach
	would be the construction of factors instead of groups (clusters). This would
	lower the dimension of the data by choosing $K$ factors which explain
	considerably the level of heterogeneity in the cross-sections; see, e.g.,
	\cite{su2018identifying} for more details.
	
	A further and important contribution of our work is a set of  testing procedures, exploring the validity of $\boldsymbol\eta_{i}=\boldsymbol0$
	and $\boldsymbol\alpha_{k}=\boldsymbol\alpha$, using testing procedures in the former case. We use
	our methods in a variety of empirical settings to explore the validity of the
	restrictions and find that in most cases both are not valid, further
	showcasing the utility of our framework.
	
	The rest of the paper is organised as follows. Sections \ref{sec1} and
	\ref{FeasibleEstimation} outline our theoretical considerations.  In Section \ref{MC} we
	present extensive Monte Carlo simulations and discussion, Section
	\ref{Empirics} presents several empirical examples in support of our method.
	We conclude in Section \ref{Discussion}. Technical proofs,  of the main (auxiliary) results,  and descriptions of the empirical datasets are relegated to  the Appendix.
	
	\subsubsection*{Notation}
	\noindent\label{notation} For any vector $\boldsymbol{x}\in\mathbb{R}^{n}, $ we denote
	the $\ell_{p}$-, and $\ell_{\infty}$ as $\left\|  \boldsymbol{x} \right\|
	_{p}=\left(  \sum_{i=1}^{n} | x_{i} |^{p}\right)  ^{1/p}$, $\lVert
	\boldsymbol{x} \rVert_{\infty} = \max_{i=1,\ldots,n}|x_{i}|$. Further, $\boldsymbol{1}\{\cdot\}$ denotes an indicator function. We use "$\to_{P}$" to denote
	convergence in probability. For two deterministic sequences $a_{n}$ and
	$b_{n}$ we define asymptotic proportionality, "$\asymp$", by writing $a_{n}
	\asymp b_{n}$ if there exist constants $0 < a_{1} \leq a_{2}$ such that
	$a_{1}b_{n} \leq a_{n} \leq a_{2}b_{n}$ for all $n \geq1$. For any set $A$,
	$|A|$ denotes its cardinality, while $A^{c}$ denotes its complement. Define $\boldsymbol{A}$, a $n\times m $ matrix, we  denote $\Lambda_{\min}(\boldsymbol{A})$ as the smallest eigenvalue of $\boldsymbol{	{A}}$, and $\Lambda_{\max}(\boldsymbol{A})$ as the largest eigenvalue of $\boldsymbol{	{A}}$. 
	
	\section{Theoretical considerations}
	
	\label{sec1}
	
	We consider the following balanced panel model:
	\begin{align}
	y_{it}^{\ast}  &  =\mu_{i}^{0}+\boldsymbol{\beta}_{i}^{0\prime}\boldsymbol{x}%
	_{it}^{\ast}+\varepsilon_{it}^{\ast},\\
	\boldsymbol{\beta}_{i}^{0}  &  =\sum_{k=1}^{K}\boldsymbol{\alpha}_{k}%
	^{0}\boldsymbol{1}\{i\in\boldsymbol{G}_{k}^{0}\}+\boldsymbol{\eta}_{i},\quad
	i=1,\ldots,N,\quad t=1,\ldots T \label{model}%
	\end{align}
	where $\mu_{i}^{0}$ is an individual fixed effect, $\boldsymbol{x}_{it}^{\ast
	}=\left(  {x}_{1,it}^{\ast},\ldots,{x}_{p,it}^{\ast}\right)  ^{\prime}$is a
	$(p\times1)$ vector of covariates, with $E(x_{l,it}^{\ast}x_{j,it}^{\ast})=0$,
	$\varepsilon_{it}^{\ast}=(\varepsilon_{1t},\ldots,\varepsilon_{Nt}^{\ast})$ is
	the idiosyncratic error, $\boldsymbol{\alpha}^{0}=(\boldsymbol{\alpha}_{1}%
	^{0},\ldots,\boldsymbol{\alpha}_{K}^{0})^{\prime}$ a $K\times p$ matrix of
	group specific centres for each covariate $j$, where $K$ denotes the number of
	groups considered. $\boldsymbol{G}_{k}^{0}$, $k=1,\ldots,K$, are sets of indices,
	denoting groups of units. Note that in our analysis, both $N$ and $T$ can be
	large and $N$ can possibly grow faster than $T$. Further, $\boldsymbol{\alpha
	}_{j}^{0}\neq\boldsymbol{\alpha}_{k}^{0}$ for any $j\neq k,$ $\boldsymbol{G}%
	_{k}^{0}\subset\{1,2,\ldots,N\}$, $\cup_{k=1}^{K}\boldsymbol{G}_{k}%
	^{0}=\{1,2,\ldots,N\}$, and $\boldsymbol{G}_{k}^{0}\cap\boldsymbol{G}_{j}%
	^{0}=\varnothing$ for any $j\neq k$, $j=1,\ldots,K$. We denote the cardinality
	of the set $\boldsymbol{G}_{k}$ as $N_{k}=|\boldsymbol{G}_{k}^{0}|$. The slope
	heterogeneity, $\boldsymbol{\eta}_{i}=(\eta_{i,1},\ldots,\eta_{i,p})^{\prime}$
	is considered to be unknown and allowed to be group specific, along with the
	cluster membership. Intuitively, $K$ corresponds to the number of groups and
	units (indexed by $i$) within the same group share a common slope
	$\boldsymbol{\alpha}_{k}$ but are allowed to differ by $\boldsymbol{\eta}%
	_{i},i\in\boldsymbol{G}_{k},k=1,\ldots,K$. Further the grouping membership is
	common across different covariates. 
	
	Often in the panel data literature, a common characteristic can be proxied by
	factors (or interactive effects), e.g., as in \cite{bai2009panel}, because of
	their ability to effectively summarise information in large data sets. While
	existing work suggests that the grouping of different units in a panel model can
	be effective under cross-sectional dependence and heteroscedasticity where
	$N<T$, see, e.g., \cite{bai2020standard}, and can even accommodate time
	variation, see e.g. \cite{su2019sieve}. In this paper, we allow for a degree
	of within-group heterogeneity, implying that grouping does not account for the
	entirety of cross-sectional heterogeneity. While we are agnostic about the
	cross-sectional and within-group idiosyncratic effect, we do assume an
	\emph{a-priori} knowledge of the existence of grouping.
	
	For the linear model in \eqref{model} with ${E}\left(  \varepsilon
	_{it}^{\ast}\mid\boldsymbol{x}_{it}^{\ast},\mu_{i}^{0}\right)  =0$, we have
	$\widehat{\mu}_{i}=\bar{y}_{i.}^{\ast}-\boldsymbol{\beta}_{i}^{\prime}%
	\bar{\boldsymbol{x}}_{i}^{\ast}$, where $\bar{y}_{i.}^{\ast}=\frac{1}{T}%
	\sum_{t=1}^{T}y_{it}^{\ast},{y}_{it}=y_{it}^{\ast}-\bar{y}_{i}^{\ast}$, and
	${\bar{\boldsymbol{x}}}_{i}$, $\boldsymbol{x}_{it}$, $\bar{\varepsilon}_{it}$,
	and $\varepsilon_{it}$ are analogously defined. Then \eqref{model} becomes
	\begin{align}
	y_{it}=\boldsymbol{\beta}_{i}^{0}\boldsymbol{x}_{it}+\varepsilon_{it},\quad
	i=1,\ldots,N,\quad t=1,\ldots T\label{mod}%
	\end{align} 
	Following \cite{su2016identifying} and motivated by the literature on fused
	Lasso e.g. \cite{tibshirani2005sparsity}, we minimise the following penalised
	profile likelihood (PPL) to estimate the parameters of interest.
	\begin{equation}
	Q_{iNT,\lambda}\left(  \boldsymbol{{\beta}},\boldsymbol{{\alpha}}\right)
	=\frac{1}{NT}\sum_{i=1}^{N}\sum_{i=1}^{T}\frac{1}{2}\left(  y_{it}%
	-\boldsymbol{\beta}_{i}^{\prime}\boldsymbol{x}_{it}\right)  ^{2}+\frac
	{\lambda}{N}\sum_{i}\prod_{k=1}^{K}\left\Vert \boldsymbol{\beta}%
	_{i}-\boldsymbol{\alpha}_{k}\right\Vert _{2}, \label{minimisation}%
	\end{equation}
	where $\lambda>0$ is the regularisation parameter and $\boldsymbol{{\beta
			=(\boldsymbol{\beta}}}_{1},\ldots,\boldsymbol{{\beta}}_{N}\boldsymbol{{)}%
	}^{\prime}$. Then, \eqref{minimisation} is minimised for some value of
	$(\lambda,K)$ and produces $(\boldsymbol{\widehat{\beta}}%
	,\boldsymbol{\widehat{\alpha}})^{\prime}$. We generalise
	\cite{su2016identifying} (\textsc{c-lasso} hereafter) by allowing for heterogeneity within groups.
	
	We show that both the group and the unit-specific parameter can be
	consistently estimated even when the underlying generating process is
	\eqref{mod}. We obtain estimates of the (unit)group-specific parameters by
	using either the \textsc{c-lasso} iterative procedure or \emph{k-means}
	estimation, see e.g. \cite{macqueen1967}, \cite{pollard1981strong}.
	
	The penalty term in \eqref{minimisation} takes a mixed additive-multiplicative
	form, introduced in \cite{su2016identifying}. In the traditional Lasso
	literature, e.g. \cite{tibshirani1996regression}, the penalty term is
	additive, and imposes a degree of sparsity in favour of a parsimonious model
	at the cost of a degree of bias, whereas in this setting, parsimony results
	from grouping. The multiplicative nature of the penalty term in
	\eqref{minimisation} is essential as it produces $N$ additive terms for each
	of the $K$ separate penalties, such that $\{i=1\ldots,N,\;k=1,\ldots
	,K:\;i\in\boldsymbol{G}_{k},\boldsymbol{\beta}_{i}^{0}\in
	\boldsymbol{\mathcal{B}}_{k}\subseteq\boldsymbol{\mathcal{B}}\}$, where, for
	each group $k$, $\boldsymbol{\mathcal{B}}_{k}\subseteq\boldsymbol{\mathcal{B}%
	}$ and $\boldsymbol{{\beta}}^{0}\in\boldsymbol{\mathcal{B}}\subseteq
	\mathbb{R}^{N\times p}$. The parameters are assumed to exhibit a certain
	grouped pattern, the number of unknown slope parameters in
	$\{\boldsymbol{{\beta}}_{i}\}$ is of order $O(K)$, instead of $O(N)$, where,
	$K(<N)$ is typically a fixed constant in empirical applications.
	
	Contrary to \textsc{c-lasso}, we do not impose homogeneity within any of the
	groups $\boldsymbol{\mathcal{B}}_{k}$, hence the unit-specific
	$\boldsymbol{{\beta}}_{i}^{0}$ are not penalised down to a specific
	group-centre $\boldsymbol{{\alpha}}^{0}_{k}$, simply because
	$\boldsymbol{{\beta}}_{i}^{0}\neq\boldsymbol{{\alpha}}^{0}_{k}$, which is more
	indicative of a {fused-Lasso} penalty. The component of summation is essential
	in order to extract information from the $N$ cross-sectional units, resulting
	in the identification of both $\{ \boldsymbol{{\beta}}_{i}^{0} \}$ and $\{
	\boldsymbol{{\alpha}}^{0}_{k} \}$. The grouping, while fixed, is not known a-priori.
	
	We  present the necessary assumptions that ensure consistency of the proposed estimator:
	
	\begin{assumption}
		\label{AssonX}
		
		\begin{enumerate}[nolistsep]
			
			\item $\left\{  \boldsymbol{x}_{it}\right\}  $ is a $p$-dimensional ergodic
			sequence of r.v.'s such that $E\left(  \boldsymbol{x}_{it}|\mathcal{F}%
			_{t-1}\right)  =0$ a.s., $\forall\;i=1,\ldots,N$, $\sup_{i,t}E\left(
			{x}_{j,it}^{2}|\mathcal{F}_{t-1}\right)  =\sigma^{2}>0$ $\,$
			$\sup_{i}\sup_{t}E\left(  |{x}_{j,it}|^{2+\nu}\right)  <\infty$, for some $\nu>0$ and $j=1,\ldots, p$, where 
			$\mathcal{F}_{t-1}=\left\{	{x}_{j, it-1}, \ldots, {x}_{j, i1}\right\}$ is the information set at time $t-1$. \label{Xt}
			
			\item $\boldsymbol{\Sigma}_{\boldsymbol{x}}=\mathbb{E}\left[\boldsymbol{x}_{it} \boldsymbol{x}_{it}^{\prime} \mid \mathcal{F}_{t-1}\right] $, we assume that $ \boldsymbol{\Sigma}_{\boldsymbol{x}}=\operatorname{plim}_{T,N \rightarrow \infty} \frac{1}{NT} \sum_{i=1}^N \sum_{t=1}^T\boldsymbol{x}_{it}\boldsymbol{x}_{it}'$  exists and is positive definite, such that  $\Lambda_{\min}(\boldsymbol{\Sigma}_{\boldsymbol{x}})>m, \; m>0 $, where $\widehat{\boldsymbol{\Sigma}}_{\boldsymbol{x}}=\frac{1}{NT} \sum_{i=1}^N \sum_{t=1}^T\boldsymbol{x}_{it}\boldsymbol{x}_{it}'$ is the sample covariance matrix, and $\Lambda_{\min}(\boldsymbol{\Sigma}_{\boldsymbol{x}})$ the smallest eigenvalue of $\boldsymbol{\Sigma}_{\boldsymbol{x}}$.\label{posdef}
			
			\item $\left\{  \varepsilon_{it}\right\}  $ is an ergodic sequence of r.v.'s
			such that 
			$E\left(\varepsilon_{i t}\right)=0, E\left(\varepsilon_{i t}^2\right)=\sigma_{\varepsilon_i}^2$ and $E\left(\varepsilon_{i t}^{2+\nu}\right)<\infty$ for $i=1, \ldots, N ; t=1, \ldots, T$ and for some $\nu>0$. Let $\mathcal{F}_{-\varepsilon_{i t}}$ be the $\sigma$-field of all stochastic elements in the panel data model, apart from $\varepsilon_{i t}$. Then, $E\left(\varepsilon_{i t} \mid \mathcal{F}_{-\varepsilon_{i t}}\right)=0$.
			
			\label{et}
			
			\item $\{\varepsilon_{it}\}$ is uncorrelated with past, as well as future
			realisations of $\{\boldsymbol{\boldsymbol{x}_{it}}\}$, such that
			$E({\varepsilon}_{it}|\boldsymbol{x}_{i1},\ldots,\boldsymbol{x}_{iT})=0$,
			$\forall\;i=1,\ldots,N,\;t=1,\ldots,T$ \label{exog}
		\end{enumerate}
	\end{assumption}
	
	\begin{remark}
		In Assumption \ref{AssonX}.\ref{Xt} and \ref{AssonX}.\ref{et} we assume that $\{\boldsymbol{x}_{it} \}$ and $\{ \varepsilon_{it}\}$ have  finite $2+\nu$ moments for some $\nu>0.$   In Assumption \ref{AssonX}.\ref{posdef} we ensure positive definiteness of the covariance matrix, guaranteeing $\left\|\frac{1}{NT} \sum_{i=1}^N\sum_{t=1}^T\left[\boldsymbol{x}_{it} \boldsymbol{x}_{it} ^{\prime}-\boldsymbol\Sigma_{\boldsymbol{x}}\right]\right\|_2=o_P(1) .$ 
		Note that  in the current framework we do not accommodate panels with cross-sectional
		dependence.
		However, the latter can be relaxed 
		at the cost of imposing stronger boundedness conditions on the largest
		eigenvalue of the variance covariance matrix of $\{\varepsilon_{it}\}$.
	\end{remark}
	\begin{assumption}
		\label{Assonalpha}
		\begin{enumerate}[nolistsep]
			
			\item $E\left(\boldsymbol{\eta}_i\right)=0 \text { and } E\left(\boldsymbol{\eta}_i \boldsymbol{\eta}_i^{\prime}\right)=\boldsymbol{\Sigma}_{\eta \eta, i}$
			\item $E(\boldsymbol{\eta}_{i}\boldsymbol{\eta}_{j}')=0$ for ${i\neq j= 1, \ldots, N}$. \label{uu}
			
			\item $E\left( \boldsymbol{x}_{it} \boldsymbol{\eta
			}_{i}\right)=0  $.\label{etax}
		\end{enumerate}
	\end{assumption}
	\begin{remark}
		Substituting $\boldsymbol\beta_{i}^0$ in  \eqref{mod}, we have that $y_{it}={\boldsymbol{\alpha}_k^0}'\boldsymbol{	x}_{it }+\boldsymbol{	x}_{it }\boldsymbol{	\eta}'_{i }+\varepsilon_{it}$. Under Assumption \ref{Assonalpha}  we obtain a consistent estimation of ${\boldsymbol{\alpha}_k^0}$.  
	\end{remark}
	\begin{remark}
		In Assumption \ref{AssonX}.\ref{exog}, weak exogeneity can be considered in
		our framework by adding lagged dependent variables, for example considering
		dynamic panel models.
		Note that in the case of dynamic panel specification,   we may maintain the Assumption \ref{Assonalpha}.\ref{etax}, however we can no longer assume that $E\left(\boldsymbol{\eta}_i y_{i, t-1}\right)=\boldsymbol{0}$, where $y_{i, t-1}$ can be defined through continuous substitutions, such that 
		$
		y_{i, t-1}=\sum_{j=0}^{\infty}\left(\eta_{i 1}\right)^j \boldsymbol{x}_{i, t-j-1}^{\prime}\left(\boldsymbol{\alpha}_k^0+\boldsymbol\eta_{i 2}\right)+\sum_{j=0}^{\infty}\left(\eta_{i 1}\right)^j u_{i, t-j-1},
		$
		and  $\boldsymbol{\eta}_i=\left(\eta_{i 1}, \boldsymbol{\eta}_{i 2}^{\prime}\right)^{\prime}$. It follows that $E\left(\boldsymbol{\eta}_i y_{i, t-1}\right) \neq \boldsymbol{0}$.
		The violation of the independence between the covariates and the individual effects, $\boldsymbol{\eta}_i$, implies that  $y_{i t}|y_{i, t-1}$, and $\boldsymbol{x}_{i t}$ will result to inconsistent estimates of $\boldsymbol{\alpha}^0_k$, even for sufficiently large $T$ and $N$.
	\end{remark}
	
	\begin{assumption}
		\label{assKm1} \label{tuneparameters}
		
		\begin{enumerate}[nolistsep]

			\item The number of clusters, $K$, is fixed and $N_{k}/N\rightarrow\tau_{k}%
			\in(0,1)$ for each $k=1,\ldots,K$ as $N\rightarrow\infty$, where $N_{k}= |\boldsymbol{G}^0_k|$ in
			cluster $k$.\label{a21}
			
			\item $\lambda\asymp\log{T}^{-\alpha}$, for some $\alpha\in(0,1/2]$ as $(N, T)
			\rightarrow\infty$.\label{a22}
			
			\item $T \lambda^{2} /(\ln T)^{6+2 \nu} \rightarrow\infty$ and $\lambda(\ln
			T)^{\nu} \rightarrow0$ for some $\nu>0$ as $(N, T) \rightarrow\infty
			$.\label{Asskm11}
			
			\item $N^{- 1 / 2} T^{-1}(\ln T)^{9} \rightarrow0$ and $N^{2} T^{1-q / 2}
			\rightarrow c \in[0, \infty)$ as $(N, T) \rightarrow\infty$.\label{Asskm12}
		\end{enumerate}
	\end{assumption}
	
	\begin{remark}
		Assumption \ref{assKm1}.\ref{Asskm11} imposes conditions on $\lambda$, all of
		which hold if $\lambda\propto T^{-\delta} \quad$ for any $\delta\in(0,1 / 2)$.
		Assumption \ref{assKm1}.\ref{Asskm12} is needed to ensure some higher-order
		terms vanish asymptotically.
	\end{remark}
	
	We further introduce an additional assumption on the second moment of
	$\boldsymbol{\eta}_{i}$, which serves as a group-uniqueness condition towards
	cluster identifiability.\textsc{c-lasso}
	
	\begin{assumption}
		\label{clust}
		
		\begin{enumerate}[nolistsep]

			\item $\int\|\boldsymbol{\eta}_{i} \|_{2}^{2} P(d \boldsymbol{\eta}%
			_{i})<\infty$ \label{3cl} and the probability measure $P$ has a continuous
			density $f$ on $\mathbb{R}^{p}$.
			
			\item The group centres are pairwise different, such that $\boldsymbol{\alpha
			}^{0}_{k}\neq\boldsymbol{\alpha}^{0}_{j}, $ and groups are pairwise disjoint
			such that $\boldsymbol{G}_{j}^{0} \cap\boldsymbol{G}_{k}^{0} = \varnothing$,
			for $j\neq k , j,k=1,\ldots, K$.
			
		\end{enumerate}
	\end{assumption}
	
	\begin{remark}
		Assumption \ref{clust}.\ref{3cl} is an identifiability assumption and
		imposes finiteness conditions on $\boldsymbol{ {\eta}}_{i}$ for each
		$i=1,\ldots, N$.
	\end{remark} 
	With the following theorems, Theorem \ref{SSP_cons} and \ref{consistencyeta},
	we provide asymptotic guarantees for the proposed estimator, the
	\textsc{c-lasso} estimator and the feasible \emph{k-means} estimator. The
	following theorems establish consistency of the adjusted penalised profile
	likelihood estimates of the slope parameter ${\boldsymbol{\beta}}$ and the and
	group-specific parameter ${\boldsymbol{\alpha}}$.
	
	We consider two types of penalised estimators, the first is \textsc{c-lasso}. We
	obtain the set of estimated parameters $(\boldsymbol{\ddot{ \beta} },
	\boldsymbol{\ddot{ \alpha}})$, by minimising the objective in
	\eqref{minimisation}, $\arg\min_{\boldsymbol{\beta}\in\mathbb{R}^{N\times p},
		\boldsymbol{\alpha}\in\mathbb{R}^{p\times K}} Q_{iNT,\lambda}\left(
	\boldsymbol{{\beta}},\boldsymbol{{\alpha}}\right)  $ for a choice of tuning
	parameter, $\lambda$.
	In the following theorem  we establish asymptotic consistency of
	\textsc{c-lasso} under heterogeneous groups.
	
	\begin{theorem}
		\label{SSP_cons} Consider Assumptions \ref{AssonX} and \ref{Assonalpha}, the
		linear model in \eqref{mod} and the minimisation of \eqref{minimisation}.
		Then, for a tuning parameter $\lambda=o(1)$, we write
		\begin{align}
		\left\| {\boldsymbol{\ddot{\beta}}}_{i}- \boldsymbol{\beta}^{0}_{i}\right\|
		_{2} & =O_{P}(T^{-1/2}+\lambda),\label{s1}\\
		\Vert\boldsymbol{\ddot{\alpha}} _{k}-\boldsymbol{\alpha} _{k}^{0}%
		\Vert_{2} & =O_{P}\left( T^{-1/2} \right) , \label{s2}\\
		(\boldsymbol{\ddot{\alpha}}_{1}, \ldots, \boldsymbol{\ddot{\alpha}%
		}_{K}) - ( \boldsymbol{\alpha}_{1}^{0}, \ldots, \boldsymbol{\alpha}_{K}^{0}) &
		= O_{P}(T^{-1/2})\label{ssp2}%
		\end{align}
		for $l\neq k, \; l=1,\ldots, K$, where $({\boldsymbol{\ddot{\beta}}},
		\boldsymbol{\ddot{\alpha}})$  minimise the \textsc{c-lasso} objective, in \eqref{minimisation}.
	\end{theorem}
	
	Similarly to \textsc{c-lasso}, one can use alternative estimators for
	$\boldsymbol{{\alpha}}$. An estimator to be considered is the \emph{k-means}.
	Let $Q_{1,NT}={(NT)}^{-1}\sum_{i=1}^{N}\sum_{i=1}^{T}\frac{1}{2}\left(
	y_{it}-\boldsymbol{\beta}_{i}^{\prime}\boldsymbol{x}_{it}\right)  ^{2}$. The
	second methodology  involves the minimisation of:
	\begin{align}
	Q_{2,iNT,\lambda}\left(  \boldsymbol{{\beta}}\right)  =Q_{1,NT}+\frac{\lambda
	}{N}\sum_{i}\prod_{k=1}^{K}\left\Vert \boldsymbol{\beta}_{i}-f^{KM}%
	(\boldsymbol{\beta})\right\Vert _{2},\label{minimisation2}
	\end{align}
	where $f^{KM}(\boldsymbol{\beta})=\arg\min_{\boldsymbol{\alpha}}\sum
	_{i}\left\Vert \boldsymbol{\beta}_{i}-\boldsymbol{\alpha}_{k}\right\Vert
	_{2}^{2}$ is the \emph{k-means} solution. Then ${\boldsymbol{\widehat{\beta}}%
	}=\arg\min_{\boldsymbol{\beta}\in\mathbb{R}^{N\times p}}Q_{2,iNT,\lambda
	}\left(  \boldsymbol{{\beta}}\right)  $ and $\boldsymbol{\widehat{\alpha}%
	}=f^{KM}(\boldsymbol{\widehat{\beta}})$. Theorem \ref{consistencyeta} has similar
	asymptotic properties to the \textsc{c-lasso}, to Theorem \ref{SSP_cons}, but
	with slightly slower convergence rates.
	
	\begin{theorem}
		\label{consistencyeta} Consider Assumptions \ref{AssonX}--\ref{clust}, the
		linear model in \eqref{mod} and the minimisation in \eqref{minimisation2}.
		Then, for a tuning parameter $\lambda=o(1)$, we write
		\begin{align}
		\left\| {\boldsymbol{\widehat{\beta}}}_{i} - \boldsymbol{\beta}^{0}%
		_{i}\right\| _{2} & =O_{P}(T^{-1/2}+\lambda),\\
		\Vert\boldsymbol{\widehat{\alpha}} _{k}-\boldsymbol{\alpha} _{k}^{0}\Vert_{2}
		& =O_{P}\left( T^{-1/2} +N^{-1/2}\right) ,\label{ssp3}\\
		(\boldsymbol{\widehat{\alpha}}_{1}, \ldots, \boldsymbol{\widehat{\alpha}}_{K})
		- ( \boldsymbol{\alpha}_{1}^{0}, \ldots, \boldsymbol{\alpha}_{K}^{0})  &
		=O_{P}\left( T^{-1/2}+N^{-1/2}\right) ,
		\end{align}
		where by Assumption \ref{tuneparameters} \eqref{Asskm12},  $ \ln{T}^{9}/T = o(N^{1/2} ) $  and $(\boldsymbol{\widehat{\beta}},
		\widehat{\boldsymbol{\alpha}})$ are centres obtained through the \emph{k-means
			Lasso} iterative procedure, outlined in \eqref{minimisation2}.
	\end{theorem}
	
	\begin{remark}
		\label{consistentbeta} Theorem \eqref{consistencyeta} establishes point-wise
		convergence of the estimated slope parameter $\boldsymbol{\widehat{\beta}}%
		_{i}$. The asymptotic rate which guarantees consistency for
		$\boldsymbol{\widehat{\alpha}}$ can be $O_{P}(T^{-1/2})$ which is the rate
		using \textsc{c-lasso} in the case where $N$ is of larger order of magnitude
		than $T$, i.e. $T=o(N)$.
		The latter, while sub-optimal, is of interest as
		clustering is a simple method in terms of
		asymptotic results and can deal with empirical datasets with a  large cross-section
	\end{remark}

	\subsection{An alternative estimator for the grouping centres}
	
	\label{alternativeGroup} In this section we consider a feasible way to
	identify the cluster structure. We devise a simple method, where we obtain the estimates of the unit-specific parameters with \textsc{ols} 
	\[
	\boldsymbol{\widetilde{\beta}}_{i}=\left(  \boldsymbol{x}_{i}^{\prime
	}\boldsymbol{x}_{i}\right)  ^{-1}\boldsymbol{x}_{i}^{\prime}\boldsymbol{y}%
	_{i}.
	\]
	and   minimise ${Q}_{N}(\boldsymbol{G}_{k},\boldsymbol{\alpha}%
	)=\frac{1}{N}\sum_{i=1}^{N}\left(  \boldsymbol{\beta}_{i}-\boldsymbol{\alpha
	}_{k}\right)  ^{2}
	$ with respect to $\boldsymbol{\alpha}$, for $i=1,\ldots,N,\;k=1,\ldots,K$.
	
	It becomes clear that $\arg\min_{\boldsymbol{\alpha}} {Q}_{N}(G_{k},
	\boldsymbol{\alpha}) $ imposes an infeasible problem as there has to be
	knowledge of $\boldsymbol{\beta}_{i}$. Therefore, we consider a unit-by-unit
	estimation to obtain $\boldsymbol{\widehat{\beta}}_{i}$, using \eqref{inf}.
	Substituting \eqref{mod} in \eqref{inf}, we have that
	\begin{align}
	\boldsymbol{\widetilde{\beta}}_{i}=\boldsymbol{\beta}_{i}+\left(  \boldsymbol{x}%
	_{i}^{\prime}\boldsymbol{x}_{i}\right)  ^{-1}\boldsymbol{x}_{i}^{\prime
	}\boldsymbol{\varepsilon}_{i}=\boldsymbol{\beta} _{i}+\boldsymbol{\psi}_{i}=
	\boldsymbol{\alpha}_{k}+\boldsymbol{\eta} _{i}+\boldsymbol{\psi} _{i}\label{astt}
	\end{align}
	where $\boldsymbol{x}_{i}=(x_{i1},\ldots,x_{iT})^{\prime}$,
	$\boldsymbol{\varepsilon}_{i}=(\varepsilon_{i1},\ldots,\varepsilon
	_{iT})^{\prime}$ and $\boldsymbol{\psi}_{i}=\left(  \boldsymbol{x}_{i}%
	^{\prime}\boldsymbol{x}_{i}\right)  ^{-1}\boldsymbol{x}_{i}^{\prime
	}\boldsymbol{\varepsilon}_{i}$. Then, a feasible \emph{k-means} loss can be
	constructed, $\widehat{Q}_{N}(G_{k}, \boldsymbol{\alpha} )$, where the
	unit-by-unit estimates can be clustered instead of $\boldsymbol{\beta}_{i}$,
	\begin{align}
	\widehat{Q}_{N}(G_{k}, \boldsymbol{\alpha} )=\arg\min_{\boldsymbol{\alpha}%
	}\frac{1}{N}\sum_{i=1}^{N}\left( \boldsymbol{\widetilde{\beta}}_{i}-
	\boldsymbol{\alpha}_{k}\right)  ^{2}\label{A}%
	\end{align}
	Note that, minimising \eqref{A}, the grouping structure relies on the
	estimation of ${{\boldsymbol{\beta}}}_{i}^{\ast}$. We showcase, with the
	following example, that ${Q}_{N}(\cdot)$ can be minimised with respect to some
	grouping centre $\boldsymbol{\alpha}= \left(\boldsymbol{\alpha}_{j, 1}, \ldots, \boldsymbol{\alpha}_{j, K}			\right)'$, $j=1, \ldots, p$. 
	
	In the following theorem we show consistency of the estimated centres obtained by
	clustering $\boldsymbol{ \widehat{\beta}}$, in \eqref{inf}, through
	\emph{k-means}, we call this method \emph{feasible k-means}
	
	\begin{theorem}
		\label{Feasible Kmeans} Consider Assumptions \ref{AssonX}--\ref{Assonalpha},
		and \ref{clust} then the following holds
		\begin{align}
		\Vert\boldsymbol{\widetilde{\alpha}} _{k} -\boldsymbol{\alpha} _{k}^{0}\Vert_{2}
		& =O_{p}(N^{-1/2}), \label{aahat}
		\end{align}
		where $\boldsymbol{\widetilde{\alpha}} _{k} $ is the minimiser of \eqref{A}.
	\end{theorem}
	
	Naturally, Theorem \ref{Feasible Kmeans} relies heavily on Assumption
	\ref{clust}, where $( \boldsymbol{ {\alpha}}^{0}_{1}, \ldots, \boldsymbol{
		{\alpha}}^{0}_{K})^{\prime}$ is a unique minimiser of the population
	$Q(\cdot)_{N}$, therefore, $(\boldsymbol{\widetilde{\alpha}} _{1}, \ldots,
	\boldsymbol{\widetilde{\alpha}} _{K})^{\prime}$ minimises $\widehat{Q}_{N}%
	(G_{k}, \boldsymbol{\alpha} )$.
	
	\subsection{Classification under group heterogeneity}\label{Class_consistency}
	
	In this section we discuss the classification consistency of the different
	estimating techniques.
	
	\begin{definition}
		\label{def} Let estimators for the true regression coefficient of unit $i$ be
		given by $\boldsymbol{\widehat{\beta}}_{i}$, $i=1,\ldots,N$, and for the set of
		cluster centres by $\boldsymbol{\widehat{\alpha}}_{k}$, $k=1,\ldots,K.$ Consider
		$\boldsymbol{G}_{k}^{0}\subset\{1,2,\ldots,N\}$, $\cup_{k=1}^{K}%
		\boldsymbol{G}_{k}^{0}=\{1,2,\ldots,N\}$, and $\boldsymbol{G}_{k}^{0}%
		\cap\boldsymbol{G}_{j}^{0}=\varnothing$ for any $j\neq k$, $j=1,\ldots,K$.
		Then we define $\boldsymbol{\widehat{G}}_{k}\subseteq\{1,2,\ldots,N\}$ as the
		set of units whose associated estimated centre,  $\boldsymbol{\widehat{\alpha
		}}_{k}$ satisfies $p\lim\boldsymbol{\widehat{\alpha}}_{k}=\boldsymbol{\alpha
		}_{k}^{0}$, where a unit $i$ is associated with centre
		$\boldsymbol{\widehat{\alpha}}_{k}$ if $\arg\min_{j}\left\Vert
		\boldsymbol{\widehat{\beta}}_{i}-\boldsymbol{\widehat{\alpha}}_{j}\right\Vert
		=\boldsymbol{\widehat{\alpha}}_{k}$.
	\end{definition}
	
	Due to the presence of $\boldsymbol{\eta}_{i}$ in the generating process in
	\eqref{model}, the probabilities of events of the form $\mathcal{I}%
	_{1,i}=\{i\in\boldsymbol{\widehat{G}}_{k}|i\notin\boldsymbol{G}_{k}^{0}\}$,
	and $\mathcal{I}_{2i}=\{i\notin\boldsymbol{\widehat{G}}_{k}|i\in
	\boldsymbol{G}_{k}^{0}\}$, respectively, become difficult to control. The
	literature usually defines classification consistency as the property where
	$\Pr(\cup_{i}\mathcal{I}_{1,i})\rightarrow0$ and $\Pr(\cup_{i}\mathcal{I}%
	_{2,i})\rightarrow0.$ In fact in our case it is possible to show the converse.
	In particular, we have the following Lemma
	\begin{lemma}
		\label{probability} Consider Assumptions \ref{AssonX}--\ref{clust}, the model
		in \eqref{mod} and the minimisation in \eqref{minimisation2}. Then,
		\begin{align}
		\lim_{N,T\rightarrow\infty}\Pr(\cup_{i}\mathcal{I}_{1,i}) &
		>0,\;and\label{p1}\\
		\lim_{N,T\rightarrow\infty}\Pr(\cup_{i}\mathcal{I}_{2,i}) &  >0.\label{p2}%
		\end{align}
	\end{lemma}
	
	\begin{remark}\label{ghat}
		In Lemma \ref{probability} we show that classification consistency is not
		guaranteed due to the underlying generating process of $\boldsymbol{ {\beta}%
		}_{i}$. Further we show that the probability of Type I and II errors is larger
		than 0 as opposed to approaching 0 as $(N,T)\to\infty$. It is further
		important to note that similarly to Lemma \ref{probability}, this result can
		be shown using alternative estimators such as the \textsc{c-lasso} and the
		\emph{feasible k-means}.
	\end{remark}
	\section{Heterogeneity Inference}
	
	\label{FeasibleEstimation} The possibility of the presence of both a group
	structure as well as within group heterogeneity necessitates an exploration of
	data dependent methods for heterogeneity inference. Two issues arise. First
	determining whether there is heterogeneity within clusters and the second is, under heterogeneity,  how to determine the number of clusters. It is important to have the
	right sequencing in this investigation. We propose to start  with a test of
	the following null hypothesis
	\begin{align}
	H_{0}: \boldsymbol{\beta}_{i} &  =\boldsymbol{\alpha}_{{k}}, 
	\text{ against }\\
	H_{1}: \boldsymbol{\beta}_{i}& \neq\boldsymbol{\alpha}_{{k}}\; \forall\; i.\label{hypothesis}
	\end{align}
	Under $H_{0}$, methods such as \textsc{c-lasso} can be used to determine the
	number of clusters and obtain consistent estimates of their centres. These methods
	can then be used as discussed in Section \ref{tests} below. If the null is not
	rejected one reverts to existing methods and find ways to determine the number
	of clusters under heterogeneity. Existing methods associated to \emph{k-means}
	cluster analysis can be used and are discussed in Section \ref{gap_sec}. In
	the following discussion we keep matters simple by not allowing for more
	refined features such as weak heterogeneity (defined as heterogeneity that is
	asymptotically negligible) or heterogeneity in a proportion of clusters
	(although we discuss cluster specific tests below). All such extensions are
	possible but beyond the scope of the current paper.
	
	\subsection{Testing for the existence of heterogeneity}\label{tests}
	
	The main interest in this section is to verify the existence
	of different types of heterogeneity, i.e. cross-sectional, as well as within
	group. We follow the two-step method described in Section
	\ref{alternativeGroup} to extract feasibly both the group and unit-specific
	parameter and show that using the unit-by-unit estimate as a preliminary
	estimator of $\boldsymbol{\beta}_{i}$ can lead to identifiable groups via \emph{k-means},
	while the dimensionality of the cross-sectional parameters is allowed to diverge.
	
	We consider the model in \eqref{mod} and simplify our analysis so that $\boldsymbol{\beta} = ({\beta}_{1}, \ldots, {\beta}_{N})^{\prime}$, and $\boldsymbol{\alpha} = ({\alpha}_{1}, \ldots,{\alpha} _{k})$ are $N\times 1$ and $K\times 1$ vectors respectively.
	To investigate the existence  of heterogeneity in the cross section, and, consequently,  within a group ${G}_k$  we construct test statistics under the  null hypothesis in \eqref{hypothesis}.   Our methodology proceeds in distinct steps; We first obtain the estimates of $\widetilde{\beta}_{i}$ through  \textsc{ols} on
	$\boldsymbol{{y}}_{i}|\boldsymbol{x}_{i}$ and uncover the grouping structure using \emph{k-means} on ${\widetilde{\beta}}_{i}$ to  obtain ${\widetilde{\alpha}}_k,$ the minimiser of  \eqref{A}.  We then obtain   $\boldsymbol{\widehat{\varepsilon}}_{i}=\boldsymbol{{y}}_{i} -\boldsymbol{x}_{i}{\widetilde{\alpha}}_k $ where $i\in \widehat G_{k}$.  Note that here it is sensible to assume that $\widehat G_{k}\to_P  G_{k}^0$. The latter is clear under $H_0$, contrary, under $H_1$, where the results of Lemma \ref{probability} verify an alternative statement.
	Then under the Assumption that
	$E({\varepsilon}_{it}|{x}_{it})=0$, and $E({\eta}_{it}|{x}_{it})=0$, $T^{-1/2}
	\sum_{i} T^{-1}\boldsymbol{x}_{i}^{\prime}{\varepsilon}_{it} \sim
	N(0,\boldsymbol{\sigma}_{x\varepsilon})$, where $\boldsymbol{\sigma
	}_{x\varepsilon} = N^{-1} \sum_{i} T^{-1}\boldsymbol{x}_{i}^{\prime
	}{\varepsilon}_{it} {\varepsilon}_{it} ^{\prime}\boldsymbol{x}_{i}$. Recall from \eqref{astt} that
	\begin{align}
	\widetilde{\beta}_{i} = \beta_{i} + \left(  \boldsymbol{x}_{i}^{\prime
	}\boldsymbol{x}_{i} \right) ^{-1} \boldsymbol{x}_{i}^{\prime}%
	\boldsymbol{{\varepsilon}}_{i},\label{inf}%
	\end{align}
	where $\boldsymbol{{\varepsilon}}_{i}= ( \boldsymbol{{\varepsilon}}_{i1},
	\ldots, \boldsymbol{{\varepsilon}}_{iT} )^{\prime}$. A feasible solution to
	\eqref{inf} would be to replace $\boldsymbol{{\varepsilon}}_{i}$ with
	$\boldsymbol{\widehat{\varepsilon}}_{i} .$ One can then  construct the cross-sectional
	heterogeneity test statistic,
	\begin{align}
	\widehat{S}^{2}_{N} & = \left( \frac{1}{\sqrt{N}} \sum_{i=1}^{N} \frac{ \widehat{t}_{i}^2(\beta)-1  }{\widehat\sigma_{i}}\right) ^{2}
	\label{xiT}, \; \text{and }\;
	\widehat{t}_{i}(\beta)={\boldsymbol{\widehat{\varepsilon}}_{i}
		^{\prime}\boldsymbol{x}_{i}}{\widehat{\sigma}_{i} \left(  \boldsymbol{x}%
		_{i}^{\prime}\boldsymbol{x}_{i}\right)  ^{-1/2}},
	\end{align}
	where $\widehat{\sigma}_{i} =\frac{1}{T-1}\sum_{i}  \boldsymbol{\widehat{\varepsilon}}_{i} \boldsymbol{\widehat{\varepsilon}}_{i}
	^{\prime}$ a
	consistent estimate of $\sigma_i=E\frac{1}{T}\boldsymbol{{\varepsilon}}_{i} \boldsymbol{{\varepsilon}}_{i}
	^{\prime}$.
	
	Further, a poolability-type  test would rely on the $R^{2}$ statistic,
	where one uses the residuals resulting from unit-by-unit \textsc{ols}, see e.g.
	\eqref{inf}, to asses the fit against a pooled \textsc{ols} regression, e.g. $
	\widehat{\beta}_{\textsc{ols}} = ( \sum_{i=1}^{N} \sum_{t=1}^{T} \boldsymbol{x}%
	_{it} ^{\prime}\boldsymbol{x}_{it} ) ^{-1} (  \sum_{i=1}^{N}
	\sum_{t=1}^{T} \boldsymbol{x}_{it} ^{\prime}y_{it} ),
	$
	Further, under standard assumptions of independence, see e.g Assumption
	\ref{Assonalpha}, the null hypothesis  in \eqref{hypothesis} will be rejected once the
	unit-by-unit \textsc{ols} reports larger $R^{2}$ than the corresponding $R^{2}$ of the
	pooled \textsc{ols}, which indicates that treating the data as heterogeneous increases
	the overall panel fit.
	
	To that end we construct the $R^{2}$ statistic, resulting from the auxiliary
	regressions of $\boldsymbol{\widehat{\varepsilon}}_{{i}}|\boldsymbol{x}%
	_{i}$, such that
	\begin{align}
	\widehat{R}^{2}_{N} & =  \left(\frac{1}{\sqrt{N}}\frac{\sum_{i=1}^N T\widehat{r}_{i}^2(\beta) -1 }{\hat{\sigma}_{i, R}}  \right)^2
	\label{xiR}, \; \text{and }\;  \widehat{r}_{i}  =\left( \widehat\gamma(\beta) \boldsymbol{x}_i ' \boldsymbol{x}_i \widehat\gamma(\beta )  (\boldsymbol{\widehat{v}}_{i}(\beta)\boldsymbol{\widehat{v}}_{i}(\beta ) ')\right)^{1/2}
	\end{align}
	where $
	\boldsymbol{\widehat{v}}_{i}\boldsymbol{(}\beta\boldsymbol{)=\widehat{\varepsilon}}_{i}-\widehat{\gamma}\left(  \beta\right)  \boldsymbol{x}_{i}%
	$
	and
	$
	\widehat{\gamma}\left(  \beta\right)  ={\boldsymbol{\widehat{\varepsilon}}_{i} ^{\prime}\boldsymbol{x}_{i}}({\boldsymbol{x}_{i}^{\prime
		}\boldsymbol{x}_{i}})^{-1},%
	$ lastly, ${\sigma_{i, R}}$ can be defined similarly to ${\sigma_{i}}$.  Under $H_0$, both $\widehat{R}^{2}_{N} , \; \widehat{S}^{2}_{N}$ are asymptotically $\chi_1^2$ distributed, while consistent under the alternative.  The latter is the key finding of the following theorem.
	
	Then, we have the following theorem giving the asymptotic distribution of
	$S_{N}$ and $R_{N}$, defined in \eqref{xiT} and \eqref{xiR}.
	\begin{theorem}
		\label{statisticsTheorem} Consider Assumptions \ref{AssonX}, \ref{Assonalpha}
		and Assumption \ref{clust}. Then, under the null
		\begin{align}
		\widehat{S}^{2}_{N} \overset{d}{\to} \chi_{1}^{2},\quad\widehat{R}^{2}_{N}
		\overset{d}{\to} \chi_{1}^{2},\label{Statement1_test}
		\end{align}
		where $\widehat{S}^{2}_{N}$,  and
		$\widehat{R}^{2}_{N} $ are defined in \eqref{xiT} and \eqref{xiR} respectively.
		
	\end{theorem}
	
	\begin{remark}
		While testing for cross-sectional heterogeneity is an important first step towards uncovering a grouping structure, it is  also of interest to investigate the existence of heterogeneity within a group $k$.  More specifically,  in
		\eqref{xiT}--\eqref{xiR} we can substitute  $\boldsymbol{\widehat{\varepsilon}}_{{i,{k}}}$ instead of $\boldsymbol{\widehat{\varepsilon}%
		}_{{i}}$  such that 
		\begin{align}
		\widehat{S}_{\widehat{K}}^{2} & = \left( \frac{1}{\sqrt{\widehat N_{k}}} \sum_{i, \; i\in \widehat G_k}
		\frac{ \widehat{t}_{k}^2({\widetilde{\alpha}}_k) -1 }{\widehat{\sigma}_{k}%
		}\right) ^{2} ,\label{xiT_K}\quad \widehat{R}^{2}_{\widehat{K}}    =\frac{1}{\sqrt{\widehat N}_{k}} \sum_{i, \; i\in \widehat G_k}
		\frac{ (T\widehat{r}_{k}^{2}-1 ) }{\widehat{\sigma}^{2} _{\widehat{r}_{k}, }}
		, \; \text{$ i=1, \ldots,  N$,}
		\end{align}
		where $\widehat{t}_{k}({\widetilde{\alpha}}_k) = {\boldsymbol{\widehat{\varepsilon}}_{i} %
			^{\prime}\boldsymbol{x}_{i}}{\widehat{\sigma}_{k}({\widetilde{\alpha}})\left(  \boldsymbol{x}%
			_{i}^{\prime}\boldsymbol{x}_{i}\right)  ^{-1/2}},$  where ${\widetilde{\alpha}}_{k}$ is the minimiser of \eqref{A}.  Further, $\widehat{\sigma}_{k}=\frac{1}{T-1}\sum_{i\in G_{k}}
		\boldsymbol{\widehat{\varepsilon}}_{{i}%
		}\boldsymbol{\widehat{\varepsilon}}_{i}^{\prime},
		$ and  $\widehat{r}^{2}_{k} = \sum_{ i\in G_{k}}({\widetilde{\alpha}}_{k}\boldsymbol{x}_{i} ^{\prime}\boldsymbol{x}_{i}{\widetilde{\alpha}}_{k})^{\prime}({\boldsymbol{\widehat{\varepsilon}%
			}_{{i}} \boldsymbol{\widehat{\varepsilon}}_{{i}}^{\prime}%
		})^{-1} $ and $\widehat{\sigma}_{\widehat{r},k} $ can be defined similarly to  $\widehat{\sigma}_{k} $ for $1\leq i\leq N.$ Note that using Lemma \ref{probability} under $H_0 $,  $\widehat{G}_k\to_P G_k^{0}$ and $\widehat{N}_k\to_P N_k$ and the test-statitsics follow $\chi_1^2$.
		The statements in  \eqref{xiT_K}  can be shown following a similar analysis to show Theorem \ref{statisticsTheorem}.
	\end{remark}
	\begin{remark}
		Under $H_1$ the test statistics  \eqref{xiT}, \eqref{xiR},
		as well as their group-specific counterparts,  \eqref{xiT_K}  are consistent.  The statement in \eqref{Statement1_test} can
		accommodate a reasonably larger dimensionality  in terms of the number of covariates, $p$, as long as the clustering remains the same across covariates.
	\end{remark}
	
	\subsection{Determination of the number of groups under heterogeneity}
	
	\label{gap_sec}
	
	Allowing for different types of heterogeneity, the identifiability of groups
	becomes a difficult task. First, because in real datasets there is no a-priory
	knowledge of the necessity of grouping, the number of groups formed or the
	grouping structure. Second, in their majority grouping methods require
	spherical groups, this way allowing only for a degree of heterogeneity, or are
	density based techniques, which in turn require the knowledge of the empirical
	distribution for each group. Further, versatile clustering techniques, such as
	\emph{k-means}, appear effective, but in cases of no-grouping at all appear
	inappropriate as it forces grouping. We address that issue by estimating
	the number of groups using techniques known in the statistical literature. The
	most commonly used is the \textit{Gap Statistic}, see e.g. \cite{gap}, and can
	be constructed   as follows,
	\begin{align}
	\label{gap_in}\text{Gap}_{N}(k) = E  & \left[ \log(W_{k}) \right]  -
	\log(W_{k}),\quad 
	W_{k}= \sum_{k=1}^{K} \frac{1}{2N_{k}} D_{k},  & \; D_{k} ={ \sum_{i, j\in
			G_{k},}} \Vert \boldsymbol{\beta}_i-\boldsymbol{\beta}_j\Vert _2^2,
	\end{align}
	for ${i\neq j }= 1, \ldots, N$ and $k=1, \ldots, K$, where $\boldsymbol{d}$ is the square euclidean distance from a centre in group
	$k$, $W_{k}$ is the pooled within cluster sum of squares around the cluster
	means, $E\left[ \log(W_{k}) \right] $ denotes the expectation under a sample
	of size $N$ is drawn from a reference distribution and $N_{k}= |G_{k}|.$  Under
	Assumption \ref{clust} and well separated groups the number of clusters 
	${K}^{\ast }$ maximises $\text{Gap}_{N}(k), $ defined in \eqref{gap_in},  such that
	\begin{align}
	\text{Gap}_{N}(k) \geq \text{Gap}_{N}(k+1) - {N^{-1/2}\sum_{i=1}^N   (\log{W}_{k+1, i}) }\label{gap}
	\end{align}
	We consider a feasible alternative to maximise \eqref{gap_in} with respect to $\hat{K}$, such that  for ${i\neq j },$
	\begin{align}
	\widehat{W}_{k}= \sum_{k=1}^{K} \frac{1}{2\hat{N}_{k}}  { \sum_{i, j\in
			G_{k},}} \Vert \boldsymbol{\widehat{\beta}}_i-\boldsymbol{\widehat{\beta}}_j\Vert _2^2,\label{wk}
	\end{align} 
	where $\boldsymbol{\widehat{\beta}}_i$ an estimator of $\boldsymbol{\beta}_i$. Note that  $\boldsymbol{\widehat{\beta}}_i$ can be a mazimiser of either $Q_{iNT,\lambda}\left(  \boldsymbol{{\beta}}\right)$, $Q_{2,iNT,\lambda}\left(  \boldsymbol{{\beta}}\right)  $, or it can maximise the unit-wise least squares objective, such that $\arg\min_{\boldsymbol{\beta}_i\in \mathbb{R}^p}Q_{1,NT}, \; Q_{1,NT}={(NT)}^{-1}\sum_{i=1}^{N}\sum_{i=1}^{T}\frac{1}{2}\left(
	y_{it}-\boldsymbol{\beta}_{i}^{\prime}\boldsymbol{x}_{it}\right)  ^{2}$, denoted  \eqref{minimisation}, \eqref{minimisation2} and \eqref{inf} respectively. 
	Consider Assumptions \ref{AssonX}, \ref{Assonalpha}, and  \ref{clust}, then $\log\widehat{W}_{k} \to _P \log  W_{k} $ where ${W}_k$ is defined in \eqref{gap_in} and $\widehat{W}_k$   in \eqref{wk}.
	
	The gap statistic is an empirical tool to identify feasibly the number of clusters taking into account the within-cluster dispersion, given various clustering algorithms. In Section \ref{MC} we explore the small sample properties of the method using simulations, and report the  frequency of selecting the correct numbers of groups, where it is known. Further, as the findings of \cite{gap} suggest, the gap statistic outperforms other methods in the literature, even in the case of no grouping, i.e. $K=1$. 
	
	\section{Simulation study}

\label{MC} In this section we explore the finite sample properties of the
different estimators explored in Section \ref{sec1} and \ref{tests} and make comparisons. A detailed description of the methods considered can be found in
\ref{MC1}. Further, this section is divided in two sub-sections: In Section
\ref{MC1} we explore the small sample properties of the different methods in
terms of mean squared error and miss-classification rate, and in Section \ref{MC2}
we focus  on the performance of the tests proposed in Section \ref{tests}.

	\subsection{Parameter estimation  (misclassification) error }
	
	\label{MC1} To evaluate the finite-sample performance of the classification
	and estimation procedure, we consider 1000 simulations of model \eqref{mod},
	where the number of groups assumed are $K=2$, $\boldsymbol{G}_{k}$ is a
	membership vector for group $k=1,\ldots, K$, the simulated true centres are
	$\boldsymbol{\alpha} =(0.5, 2)^{\prime}$ and $\eta_{i}\sim N(0, 0.2)$. Further, $x_{it}\sim_{\text{i.i.d}} N(0,1)$,  $\varepsilon_{it}\sim_{\text{i.i.d}} N(0,1)$ and
	$(\beta_{1},\ldots, \beta_{N/2})^{\prime}\in\boldsymbol{G}_{1}, \;
	(\beta_{N/2+1}, \ldots, \beta_{N})^{\prime}\in\boldsymbol{G}_{2},$ where
	$N=[20,50,100,200], \; T=[50,100,200,500]$.
	
	We consider various different methods and evaluate the consistency of the
	grouped parameters $\beta_{i}$, and individual group centres, as well as the
	quality of grouping under homogeneous and heterogeneous parameters. Namely we
	consider the \textsc{c-lasso} of \cite{su2016identifying} reported as SSP, the
	\textsc{c-lasso} estimator using \emph{k-means} to obtain the group centres,
	reported as Km
	and the \textsc{c-lasso}  without imposing within class homogeneity, reported as \textsc{H-SSP}.
	The penalty parameter for the penalisation throughout methods was
	carried out via cross validation scheme, from a parameter grid, $\lambda
	\in[0.125, 0.25, 0.5, 1, 2]T^{-1/3} $. We carried out a k-fold cross
	validation scheme without re-shuffling, where $k=10$.
	
	To asses the consistency of the individual parameters we report the average
	mean squared error (MSE) across units under both the null and the alternative,
	e.g. \eqref{hypothesis}. We report these findings in Table \ref{betas}. In
	Table \ref{alphas} we report the average MSE of the centres of the first group
	to save space, e.g. $1000^{-1} \sum_{i=1}^{1000} (\alpha_{1,i}-
	\widehat{\alpha}_{1,i})$, where the estimation of the individual $K$ centres
	is carried out using four different methods, descried below.
	
	To quantify the quality of grouping we use a dissimilarity measure, Rand
	Index, e.g. \cite{rand1971objective}. The Rand index, ranges between 0 and 1,
	with 0 indicating that the two data groupings do not agree on any pair of
	points and 1 indicating that the data groupings are exactly identical to some
	permutation. More specifically, given  the following partitions
	$\boldsymbol{G}= \{\boldsymbol{G}_{1}, \ldots, \boldsymbol{G}_{r}
	\}$,   of $\boldsymbol{\beta} $ into $r$ subsets and
	$\boldsymbol{\widehat{G}}=\{ \boldsymbol{\widehat{G}}_{1}, \ldots,
	\boldsymbol{\widehat{G}}_{s} \}$, a partition of $\boldsymbol{\beta} $ into
	$s$ subsets, where $1<r, s\leq N$, we write the following measure
	\[
	RI = \frac{a+b}{{\binom{N }{2 }}}\equiv\frac{TP+TN}{TP+TN+FP+FN},
	\]
	where $N$ is the number of units to be clustered, TP is the number of true
	positives, TN is the number of true negatives, FP is the number of false
	positives, and FN is the number of false negatives. Further, $a=
	|\boldsymbol{B}^{*}|, \; \boldsymbol{B}^{*} = \{ (\beta_{i},\beta_{j}) |
	\beta_{i},\beta_{j} \in\boldsymbol{G}, \; \beta_{i},\beta_{j} \in
	\boldsymbol{\widehat{G}}\}$ and $b = |\boldsymbol{B}^{*}|, \; \boldsymbol{B}%
	^{*} = \{ (\beta_{i}, \beta_{j} ) | \beta_{i}\in\boldsymbol{I_{1}},\beta
	_{j}\in\boldsymbol{G}_{2}, \;\beta_{i}\in\boldsymbol{\widehat{G}}_{1}%
	,\beta_{j}\in\boldsymbol{\widehat{G}}_{2} \}$, where the former declares the number
	of pairs of elements in $\{\beta_{1}, \ldots, \beta_{N}\}$ that are in the
	same subset in $\boldsymbol{G}$ and in the same subset in
	$\boldsymbol{\widehat{G}}$, and the latter declares the number of pairs of
	elements in $\{\beta_{1}, \ldots, \beta_{N}\}$ that are in different subsets
	in $\boldsymbol{G}$ and in different subsets in $\boldsymbol{\widehat{G}}$. We
	report these findings in Table \ref{RI} both under homogenous and
	heterogeneous groups.
	
	Further, since heterogeneity is allowed within a group, we consider additional
	metrics to ensure that the grouping is not forced but indeed latent structures
	can be found in the cross section. We then estimate the number of groups that
	can be found using the unit-by-unit (UE) estimate, ${\beta}^{\ast}_i$ obtained in \eqref{inf}, reporting the frequency of selecting the
	true number of groups, $K_{0}=2$ throughout 1000 replications, through the gap
	statistic, e.g. \cite{gap}. We report these findings in Table \ref{gap} both
	under homogenous and heterogeneous coefficients.
	
	\begin{table}[ht!]
		\centering
		\resizebox{\textwidth}{!}{
			\begin{tabular}{ l l c c c c c  |c c c c c}	
				\toprule
				& \multicolumn{6}{c|}{$\beta_{i} = \alpha_{{i\in G_k}} $} & \multicolumn{5}{c}{$\beta_{i} = \alpha_{{i\in G_k}} + \eta_{i}$} \\ \toprule
				&$T/N$& {20}& {50}& {100}& {200}& {500}& {20}& {50}& {100}& {200}& {500}\\\midrule
				{SSP}&50& 0.0020 & 0.0008 & 0.0004 & 0.00021&0.0020&0.1829&0.1885&0.1846&0.1855&0.1710\\
				{Km}&&0.0149&0.0148&0.0147&0.0147&0.0146&0.0881&0.0887&0.0885&0.0886&0.0887\\
				{\textsc{H-SSP}}&& 0.0116 & 0.0109 &0.0107 &0.0107 &0.0107 &0.0889  &  0.0889  &  0.0885  &  0.0888 &   0.0887 \\
				\midrule
				{SSP}&100&0.0010 & 0.0004 & 0.0002 & 0.0001 &{0.0001}&0.1439&0.1444&0.1483&0.1442&0.1462\\
				{Km}&&0.0072&0.0070&0.0069&0.0069&0.0068&0.0818&0.0813&0.0811&0.0811&0.0808\\
				\textsc{H-SSP}&&  0.0072   & 0.0070 &0.0069 &0.0069 &   0.0069&0.0820 &   0.0815 &   0.0811   & 0.0811& 0.0809\\
				\midrule
				{SSP}&200&0.0005 & 0.0002 & 0.0001 & 0.0001&0.0001&0.1203&0.1252&0.1226&0.1256&	0.1280	\\
				{Km}&&0.0033&0.0033&0.0033&0.0032&0.0033&0.0770&0.0776&0.0777&0.0780&0.0778\\
				\textsc{H-SSP} && 0.0034   & 0.0033 &   0.0033   & 0.0033  &  0.0033& 0.0771  &  0.0777  &  0.0778   & 0.0781  &  0.0778\\
				\midrule
				{SSP}&500&0.0002&0.0001&0.0000&0.0000&0.0000& 0.1122&0.1139&0.1146&0.1163& 0.1158\\
				{Km}&&0.0013&0.0012&0.0012&0.0012& 0.0012&0.0775&0.0771&0.0768&0.0768&0.0767 \\
				\textsc{H-SSP}&&   0.0013   & 0.0012 &0.0012 &0.0012 &0.0012&0.0776 &0.0772  &  0.0768& 0.0768  &  0.0767
				\\
				\bottomrule
			\end{tabular}
		}\caption{MSE of $\protect\widehat{\beta}_{i, j} $, $i=1,\ldots, N$ and
			$j=1,\ldots, M$ the total number of methods examined (four), throughout 1000
			replications of model \eqref{mod} under the null, (left panel) and under the
			alternative, e.g. \eqref{hypothesis} (right panel), where $\eta_{i}\sim
			N(0,0.2)$.}%
		\label{betas}%
	\end{table}
	
	\begin{table}[ht!]
		\resizebox{\textwidth}{!}{
			\begin{tabular}{@{}cccccccccccc@{}}
				\toprule
				&    & \multicolumn{5}{c}{$\beta_{i}=\alpha_{{i\in G_k}}$}     & \multicolumn{5}{c}{$\beta_{i}=\alpha_{{i\in G_k}}+\eta_{i}$} \\ \midrule
				& $T/N$ & 20  & 50  & 100 & 200 & \multicolumn{1}{c|}{500} & 20  & 50    & 100   & 200   & 500   \\ \midrule
				SSP  & 50 & 0.0020 & 0.0008 & 0.0004 & 0.0002 & \multicolumn{1}{c|}{0.0001} & 0.0058 & 0.0022   & 0.0013   & 0.0006   & 0.0002   \\
				Km   &    & 0.0022 & 0.0009 & 0.0004 & 0.0003 & \multicolumn{1}{c|}{0.0001} & 0.0060 & 0.0024   & 0.0014   & 0.0007   & 0.0003   \\
				\textsc{H-SSP}  && 0.0026  &  0.0011 &0.0005 &0.0003 & \multicolumn{1}{c|}{0.0001} &0.0084 &0.0034 &0.0020 &0.0010 &0.0005\\
				\midrule
				SSP  & 100   & 0.0010 & 0.0004 & 0.0002 & 0.0001 & \multicolumn{1}{c|}{0.0001} & 0.0055 & 0.0021   & 0.0010   & 0.0005   & 0.0002   \\
				Km   &    & 0.0010 & 0.0004 & 0.0002 & 0.0001 & \multicolumn{1}{c|}{0.0001} & 0.0057 & 0.0021   & 0.0011   & 0.0006   & 0.0003   \\
				\textsc{H-SSP}&& 0.0013 &0.0005 &0.0003 &0.0001 & \multicolumn{1}{c|}{0.0001 }   &0.0076 &0.0030 &0.0015 &0.0008 &0.0004\\
				\midrule
				SSP  & 200   & 0.0005 & 0.0002 & 0.0001 & 0.0001 & \multicolumn{1}{c|}{0.00020}    & 0.0044 & 0.0018   & 0.0010   & 0.0004   &  	0.0008   \\
				Km   &    & 0.0005 & 0.0002 & 0.0001 & 0.0001 & \multicolumn{1}{c|}{0.00022}    & 0.0045 & 0.0018   & 0.0010   & 0.0005   &   	0.0008   \\
				\textsc{H-SSP} &&0.0006   &0.0003 &0.0001  &  0.0001 & \multicolumn{1}{c|}{0.00022} &0.0067 &0.0027 &0.0014 &0.0007 &0.0003 \\
				\midrule
				SSP  & 500   & 0.0002 & 0.0001 & 0.0000 & 0.0000 & \multicolumn{1}{c|}{0.0000}    & 0.0043 & 0.0018   & 0.0009   & 0.0004   &   0.0016   \\
				Km   &    & 0.0002 & 0.0001 & 0.0000 & 0.0000 & \multicolumn{1}{c|}{0.0000}    & 0.0043 & 0.0018   & 0.0009   & 0.0004   &   0.0018   \\
				\textsc{H-SSP}&&0.0002 &0.0001 &0.0000 &0.0000 & \multicolumn{1}{c|}{0.0000}   &0.0062 &0.0025 &0.0013 &0.0006 &0.0003\\
				\bottomrule
			\end{tabular}
		}\caption{MSE of $\protect\widehat{\alpha}_{k, j} $, see note in Table
			\ref{betas}. }%
		\label{alphas}%
	\end{table}As it is expected, in Table \ref{betas}, the MSE decreases as the
	sample size increases, while SSP reports a higher MSE compared to all other
	methods as both $N,T$ increase in the cases of heterogeneity (see right panel
	of Table \ref{betas}), while it also decreases with an increase of $T$. The
	latter is due to the fact that SSP does not allow  within group heterogeneity ,
	forcing the cross-sectional parameters to be penalised down to their
	corresponding group centre, while heterogeneity persists within a group, as
	the finding of Table \ref{tab1} indicate. $Km$ reports the smallest MSE across
	methods in cases of heterogeneity,
	In the cases of homogeneity (left panel of Table
	\ref{betas}), SSP is the best performing method, as it is expected, while all
	methods report smaller MSEs as $T$ increases.
	
	In Table \ref{alphas} all methods reported appear equivalent, while as $T$
	increases the error approaches zero. In Table \ref{RI} we report the average
	RI (see \eqref{RI}) throughout 1000 replications. As it is expected, under the
	null (see left panel of Table \ref{RI}), all methods perform are equivalently,
	i.e. most methods report average accuracy around or equal 100\%. In the case
	both when generated under the alternative hypothesis, i.e. heterogeneity across
	units, SSP reports the smallest accuracy throughout samples, but as the sample
	size $T$ increase, accuracy increases as well. The latter is a common finding
	across methods.
	
	Table \ref{gap} reports the average frequency of selecting the correct number
	of groups via the gap statistic. Table \ref{gap} compliments the results of
	Table \ref{RI}, verifying that (1) clustering the cross sectional units is
	necessary, and (2) while clustering is necessary, through Table \ref{RI} we
	verify that accuracy of grouping is on average high throughout samples.
	Finding (1) can be observed both generating under the null and the alternative
	as the average frequency of selecting two groups is close to 1 and increases
	with the sample size $T$.
	\begin{table}[ht!]%
		\begin{tabular}
			[c]{lcccccc|ccccc}%
			\toprule &  & \multicolumn{5}{c|}{$\beta_{i}=\alpha_{{i\in G_{k}}}$} &
			\multicolumn{5}{c}{$\beta_{i}=\alpha_{{i\in G_{k}}}+\eta_{i}$}\\
			\midrule & $T/N$ & 20 & 50 & 100 & 200 & 500 & 20 & 50 & 100 & 200 & 500\\
			\midrule SSP & 50 & 0.990 & 0.990 & 0.990 & 0.990 & 0.990 & 0.900 & 0.880 &
			0.880 & 0.880 & 0.880\\
			Km &  & 1 & 1 & 1 & 1 & 1 & 1 & 1 & 1 & 1 & 1\\
			F-Km &  & 1 & 1 & 1 & 1 & 1 & 1 & 1 & 1 & 1 & 1\\
			\textsc{H-SSP} &  & 1 & 1 & 1 & 0.999 & 0.996 & 0.995 & 0.996 & 0.995 & 0.996 &
			0.996\\
			\midrule SSP & 100 & 1 & 1 & 1 & 1 & 1 & 0.920 & 0.910 & 0.910 & 0.910 &
			0.910\\
			Km &  & 1 & 1 & 1 & 1 & 1 & 1 & 1 & 1 & 1 & 1\\
			F-Km &  & 1 & 1 & 1 & 1 & 1 & 1 & 1 & 1 & 1 & 1\\
			\textsc{H-SSP} &  & 1 & 1 & 1 & 1 & 1 & 0.997 & 0.998 & 0.998 & 0.999 & 0.998\\
			\midrule SSP & 200 & 1 & 1 & 1 & 1 & 1 & 0.930 & 0.930 & 0.930 & 0.930 &
			0.924\\
			Km &  & 1 & 1 & 1 & 1 & 1 & 1 & 1 & 1 & 1 & 0.999\\
			F-Km &  & 1 & 1 & 1 & 1 & 1 & 1 & 1 & 1 & 1 & 0.999\\
			\textsc{H-SSP} &  & 1 & 1 & 1 & 1 & 1 & 0.999 & 0.999 & 0.999 & 0.999 & 0.999\\
			\midrule SSP & 500 & 1 & 1 & 1 & 1 & 1 & 0.940 & 0.940 & 0.940 & 0.930 &
			0.934\\
			Km &  & 1 & 1 & 1 & 1 & 1 & 1 & 1 & 1 & 1 & 0.999\\
			F-Km &  & 1 & 1 & 1 & 1 & 1 & 1 & 1 & 1 & 1 & 0.999\\
			\textsc{H-SSP} &  & 1 & 1 & 1 & 1 & 1 & 0.999 & 0.999 & 0.999 & 0.999 & 0.999\\
			\bottomrule 
		\end{tabular}
		\bigskip\caption{Average Rand index (RI) through 1000 replications of 1000
			replications of model \eqref{mod} under the null e.g. \eqref{hypothesis} (left
			panel) and under the alternative (right panel), where $\eta_{i}\sim N(0,0.2)$.
		}%
		\label{RI}%
	\end{table}\begin{table}[h]
		\centering
		\begin{tabular}
			[c]{lccccc|cccccl}%
			\toprule & \multicolumn{5}{c|}{$\beta_{i} = \alpha_{{i\in G_{k}}} + \eta_{i}$}
			& \multicolumn{5}{c}{$\beta_{i} = \alpha_{{i\in G_{k}}} $} & \\
			\midrule $T/N$ & {20} & {50} & {100} & {200} & {500} & {20} & {50} & {100} &
			{200} & {500} & \\
			\midrule {50} & 0.954 & 0.998 & 1 & 1 & 1 & 0.956 & 0.999 & 1 & 1 & 1 & \\
			\midrule {100} & 0.961 & 1 & 1 & 1 & 1 & 0.950 & 0.997 & 1 & 1 & 1 & \\
			\midrule {200} & 0.963 & 0.998 & 1 & 1 & 1 & 0.953 & 0.998 & 1 & 1 & 1 & \\
			\midrule {500} & 0.967 & 0.997 & 1 & 1 & 1 & 0.962 & 0.999 & 1 & 1 & 1 & \\
			\bottomrule 
		\end{tabular}
		\caption{Frequency of selecting $K_{0}=2$ groups via the gap statistic,
			$\eta_{i}\sim N(0,0.2).$}%
		\label{gap}%
	\end{table}
	
	\subsection{Testing for the existence of heterogeneity}
	
	\label{MC2} We simulate the model in \eqref{mod} and take the one-way within
	transformation, see  \eqref{mod}, with $x_{it}\sim_{\text{i.i.d}} N(0,1)$, $\varepsilon_{it}\sim_{\text{i.i.d}} N(0,1)$, $\boldsymbol{\alpha} =(0.5, 2)^{\prime}$, $\eta_{i}\sim N(0,0.2).
	$  We generate under the null hypothesis of
	\eqref{hypothesis}, to report the size of the test in \eqref{xiT} on the right
	panel of Table \ref{tab1} and further power of the test can be reported by
	generating under the alternative, see the left panel of Table \ref{tab1}. The
	reported quantities of Table \ref{tab2} can be similarly described for the test
	in \eqref{xiR} where we assess the goodness of fit under clustered effects.
	\newline\indent In Table \ref{tabkh} we report the average frequency of
	selecting $K=2$ groups when the true number of groups is $K_{0}=2$ throughout
	1000 replications of model \eqref{mod}, while in Table \ref{tabkh1} we report
	the average frequency of selecting $K=1$ groups when the true number of groups
	is $K_{0}=1$. Further we report four different measures of identifying the number
	of latent groups in a vector of estimates \eqref{inf}, the maximum number of
	clusters considered is $K_{\max}=5$. Namely, we consider the gap statistic,
	e.g. \cite{gap}, reported as 'gap', the Silhouette statistic e.g. \cite{sil},
	reported as 'Sil', the Calinski-Harabasz statistic, e.g. \cite{ch} reported as
	'CH' and the Davies-Bouldin, e.g. \cite{db} reported as 'DB'. \newline%
	\indent We are interested in investigating the existence of heterogeneity and
	for that matter the existence of groups in two separate cases; First, we
	investigate the existence of heterogeneity in the cross-section, using the
	unit-by-unit \textsc{ols} estimator in \eqref{inf} (reported as \emph{CS} in Tables
	\ref{tab1}--\ref{tab2}) and second we investigate the existence of
	heterogeneity after grouping using the \emph{feasible k-means} estimator
	(reported as \emph{WG} in Tables \ref{tab1}--\ref{tab2}) minimising
	\eqref{A}. All the tests are carried out at $5\%$ level of significance. In
	Table \ref{tab1} there is clear indication that asymptotically, as $T$
	increases, the test in \eqref{xiT} rejects the null for both types of
	heterogeneity frequently, showing that indeed heterogeneity still persists
	both before and more importantly after grouping. Similarly to Table
	\ref{tab1}, in Table \ref{tab2} the goodness of fit both before and after
	grouping is appears better as $T$ increases, showing that grouping is
	necessary even when heterogeneity persists within each group. The latter is an
	interesting finding coupled with the findings of Table \ref{tabkh}%
	--\ref{tabkh1}, because there is an implication that in the cases where the
	number of groups $K=2$, asymptotically the gap measure
	identifies the correct number of groups 100\% o the times, while when there is
	no grouping at all, the rate at which the gap statistic selects the correct
	number of groups is 100\% throughout $N,\; T$. Therefore, there is clear
	indication of grouping structure in the cross-section, and the findings of
	Tables \ref{tab1}--\ref{tab2} show that  heterogeneity clearly exists within
	the different groups as well.
	
	\begin{table}[ht!]
		
		\centering
		\resizebox{0.8\linewidth}{!}{
			\begin{tabular}{@{}lcccccc|ccccc@{}}
				\toprule
				&                            & \multicolumn{5}{c|}{$\beta_{i} = \alpha_{k} + \eta_{i}$} & \multicolumn{5}{c}{$\beta_{i} = \alpha_{k} $} \\ \midrule
				&   {$N/T$} & 100       & 200       & 500       & 1000      & 2000     & 100     & 200     & 500     & 1000   & 2000   \\ \midrule
				CS &{100}   & 1      & 1      & 1      & 1      & 1     & 0.14    & 0.14    & 0.15    & 0.13   & 0.15   \\
				WG &  {}      & 0.86      & 0.87      & 0.90      & 0.91      & 0.99     & 0.15    & 0.13    & 0.16    & 0.13   & 0.15   \\ \midrule
				CS &  {200}   & 1      & 1      & 1      & 1      & 1     & 0.11    & 0.11    & 0.11    & 0.10   & 0.11   \\
				WG &  {}      & 0.91      & 0.92      & 0.92      & 0.93      & 0.98     & 0.10    & 0.11    & 0.11    & 0.10   & 0.12   \\ \midrule
				CS &  {500}   & 1      & 1      & 1      & 1      & 1     & 0.06    & 0.07    & 0.07    & 0.05   & 0.07   \\
				WG &  {}      & 0.95      & 0.94      & 0.93      & 0.95      & 0.96     & 0.05    & 0.07    & 0.07    & 0.07   & 0.08   \\ \midrule
				CS &  {1000}  & 1      & 1      & 1      & 1      & 1     & 0.04    & 0.04    & 0.06    & 0.04   & 0.05   \\
				WG &  {}      & 0.96      & 0.95      & 0.96      & 0.96      & 0.97     & 0.04    & 0.05    & 0.03    & 0.06   & 0.06   \\ \midrule
				CS &  {2000}  & 1      & 1      & 1      & 1      & 1     & 0.02    & 0.02    & 0.05    & 0.04   & 0.04   \\
				WG &  {}      & 0.98      & 0.97      & 0.97      & 0.96      & 0.96     & 0.02    & 0.04    & 0.03    & 0.03   & 0.04   \\ \bottomrule
		\end{tabular}}
		\caption{Size and power of the statistic  throughout 1000
			replications, generating under the null (right-hand panel) and alternative (right-hand panel) hypothesis, the number of groups is $K_{0}=2$. CS stands for cross-sectional, and WG stands for within group.}		\label{tab1}
	\end{table}
	\begin{table}[ht!]
		
		\centering
		\resizebox{0.80\textwidth}{!}{
			\begin{tabular}{@{}lccccccccccc@{}}
				\toprule
				&  & \multicolumn{5}{c}{$\beta_{i} = \alpha_{{i\in G_k}} + \eta_{i}$} & \multicolumn{5}{c}{$\beta_{i} = \alpha_{{i\in G_k}} $} \\ \midrule
				& N/T  & 100  & 200  & 500  & 1000 & \multicolumn{1}{c|}{2000} & 100  & 200  & 500  & 1000 & 2000 \\ \midrule
				CS  & 100  & 1 & 1 & 1 & 1 & \multicolumn{1}{c|}{1} & 0.06 & 0.04 & 0.08 & 0.13 & 0.07 \\
				WG &   & 0.17 & 0.57 & 0.98 & 1 & \multicolumn{1}{c|}{1} & 0.03 & 0.09 & 0.03 & 0.04 & 0.04 \\ \midrule
				CS  & 200  & 1 & 1 & 1 & 1 & \multicolumn{1}{c|}{1} & 0.08 & 0.1  & 0.12 & 0.06 & 0.04 \\
				WG &   & 0.13 & 0.37 & 0.95 & 1 & \multicolumn{1}{c|}{1} & 0.06 & 0.11 & 0.05 & 0.08 & 0.07 \\\midrule
				CS  & 500  & 1 & 1 & 1 & 1 & \multicolumn{1}{c|}{1} & 0.05 & 0.09 & 0.07 & 0.09 & 0.01 \\
				WG &   & 0.1  & 0.24 & 0.73 & 1 & \multicolumn{1}{c|}{1} & 0.07 & 0.05 & 0.05 & 0.05 & 0.04 \\ \midrule
				CS  & 1000 & 1 & 1 & 1 & 1 & \multicolumn{1}{c|}{1} & 0.04 & 0.07 & 0.04 & 0.05 & 0.04 \\
				WG &   & 0.13 & 0.13 & 0.59 & 0.97 & \multicolumn{1}{c|}{1} & 0.08 & 0.04 & 0.06 & 0.09 & 0.04 \\ \midrule
				CS  & 2000 & 1 & 1 & 1 & 1 & \multicolumn{1}{c|}{1} & 0.06 & 0.07 & 0.04 & 0.03 & 0.06 \\
				WG &   & 0.11 & 0.08 & 0.46 & 0.8  & \multicolumn{1}{c|}{1} & 0.07 & 0.02 & 0.05 & 0.04 & 0.06 \\ \bottomrule
			\end{tabular}
		}
		\caption{Size and power of the statistic in \eqref{xiR} throughout 1000
			replications, generating under the null and alternative hypotheses in
			\eqref{hypothesis}, the number of groups is $K_{0}=2$.}%
		\label{tab2}%
	\end{table}
	
	\begin{table}[ht!]
				\resizebox{\textwidth}{!}{\begin{tabular}{@{}lcccccccccccccccc@{}}
				\toprule
				\multicolumn{17}{l}{$\beta_i=\alpha_{{i\in G_k}}+\eta_{i}$, $\eta_{i}\sim i.i.d$} \\ \midrule
				N/T & 100 &  &  & \multicolumn{1}{c|}{} & 200 &  &  & \multicolumn{1}{c|}{} & 500 &  &  & \multicolumn{1}{c|}{} & 1000 &  &  &  \\ \midrule
				& Gap & Sil & CH & \multicolumn{1}{c|}{DB} & Gap & Sil & CH & \multicolumn{1}{c|}{DB} & Gap & Sil & CH & \multicolumn{1}{c|}{DB} & Gap & Sil & CH & DB \\ \midrule
				100 & 0.763 & 1&0& \multicolumn{1}{c|}{0.955} & 0.830 & 0.998 & 0.002 & \multicolumn{1}{c|}{0.964} & 0.851 & 0.999&0& \multicolumn{1}{c|}{0.959} & 0.849 & 1 & 0.001 & 0.973 \\
				200 & 0.932 & 1&0& \multicolumn{1}{c|}{0.999} & 0.942 & 1 & 0.001 & \multicolumn{1}{c|}{0.999} & 0.955 & 1&0& \multicolumn{1}{c|}{0.999} & 0.974 & 1 & 0.001 & 1 \\
				500 & 0.995 & 1&0& \multicolumn{1}{c|}{1} & 0.998 & 1&0& \multicolumn{1}{c|}{1} & 0.999 & 1&0& \multicolumn{1}{c|}{1} & 0.999 & 1&0& 1 \\
				1000 & 1 & 1&0& \multicolumn{1}{c|}{1} & 1 & 1&0& \multicolumn{1}{c|}{1} & 1 & 1&0& \multicolumn{1}{c|}{1} & 1 & 1&0& 1 \\ \midrule
				\multicolumn{17}{l}{$\beta_{i}= a_{{i\in G_k}}$} \\ \midrule
				N/T & 100 &  &  & \multicolumn{1}{c|}{} & 200 &  &  & \multicolumn{1}{c|}{} & 500 &  &  & \multicolumn{1}{c|}{} & 1000 &  &  &  \\ \midrule
				\multicolumn{1}{l}{} & Gap & Sil & CH & \multicolumn{1}{c|}{DB} & Gap & Sil & CH & \multicolumn{1}{c|}{DB} & Gap & Sil & CH & \multicolumn{1}{c|}{DB} & Gap & Sil & CH & DB \\ \midrule
				100 & 1 & 1 & 0.456 & \multicolumn{1}{c|}{1} & 1 & 1 & 0.490 & \multicolumn{1}{c|}{1} & 1 & 1 & 0.466 & \multicolumn{1}{c|}{1} & 1 & 1 & 0.485 & 1 \\
				200 & 1 & 1 & 0.652 & \multicolumn{1}{c|}{1} & 1 & 1 & 0.653 & \multicolumn{1}{c|}{1} & 1 & 1 & 0.700 & \multicolumn{1}{c|}{1} & 1 & 1 & 0.705 & 1 \\
				500 & 1 & 1 & 0.905 & \multicolumn{1}{c|}{1} & 1 & 1 & 0.922 & \multicolumn{1}{c|}{1} & 1 & 1 & 0.906 & \multicolumn{1}{c|}{1} & 1 & 1 & 0.914 & 1 \\
				1000 & 1 & 1 & 0.984 & \multicolumn{1}{c|}{1} & 1 & 1 & 0.989 & \multicolumn{1}{c|}{1} & 1 & 1 & 0.993 & \multicolumn{1}{c|}{1} & 1 & 1 & 0.990 & 1 \\ \bottomrule
		\end{tabular}} \caption{ Frequency of selecting $2$ groups when the true number of clusters
			is $K_{0}=2.$ }%
		\label{tabkh}%

	\end{table}
	
	\begin{table}[ht!]
		\resizebox{\textwidth}{!}{\begin{tabular}{@{}ccccccccccccccccc@{}}
				\toprule
				\multicolumn{17}{l}{$\beta_i=a_{k}+\eta_{i}$, $\eta_{i}\sim i.i.d$} \\ \midrule
				N/T & 100 &  &  & \multicolumn{1}{c|}{} & 200 &  &  & \multicolumn{1}{c|}{} & 500 &  &  & \multicolumn{1}{c|}{} & 1000 &  &  &  \\ \midrule
				& Gap & Sil & CH & \multicolumn{1}{c|}{DB} & Gap & Sil & CH & \multicolumn{1}{c|}{DB} & Gap & Sil & CH & \multicolumn{1}{c|}{DB} & Gap & Sil & CH & DB \\ \midrule
				100 & 1 & 0 & 0 & \multicolumn{1}{c|}{0} & 1 & 0 & 0 & \multicolumn{1}{c|}{0} & 1 & 0 & 0 & \multicolumn{1}{c|}{0} & 1 & 0 & 0 & 0 \\
				200 & 1 & 0 & 0 & \multicolumn{1}{c|}{0} & 1 & 0 & 0 & \multicolumn{1}{c|}{0} & 1 & 0 & 0 & \multicolumn{1}{c|}{0} & 1 & 0 & 0 & 0 \\
				500 & 1 & 0 & 0 & \multicolumn{1}{c|}{0} & 1 & 0 & 0 & \multicolumn{1}{c|}{0} & 1 & 0 & 0 & \multicolumn{1}{c|}{0} & 1 & 0 & 0 & 0 \\
				1000 & 1 & 0 & 0 & \multicolumn{1}{c|}{0} & 1 & 0 & 0 & \multicolumn{1}{c|}{0} & 1 & 0 & 0 & \multicolumn{1}{c|}{0} & 1 & 0 & 0 & 0 \\ \midrule
				\multicolumn{17}{l}{$\beta_{i}= a_{k}$} \\ \midrule
				N/T & 100 &  &  & \multicolumn{1}{c|}{} & 200 &  &  & \multicolumn{1}{c|}{} & 500 &  &  & \multicolumn{1}{c|}{} & 1000 &  &  &  \\ \midrule
				& Gap & Sil & CH & \multicolumn{1}{c|}{DB} & Gap & Sil & CH & \multicolumn{1}{c|}{DB} & Gap & Sil & CH & \multicolumn{1}{c|}{DB} & Gap & Sil & CH & DB \\ \midrule
				100 & 1 & 0 & 0 & \multicolumn{1}{c|}{0} & 1 & 0 & 0 & \multicolumn{1}{c|}{0} & 1 & 0 & 0 & \multicolumn{1}{c|}{0} & 1 & 0 & 0 & 0 \\
				200 & 1 & 0 & 0 & \multicolumn{1}{c|}{0} & 1 & 0 & 0 & \multicolumn{1}{c|}{0} & 1 & 0 & 0 & \multicolumn{1}{c|}{0} & 1 & 0 & 0 & 0 \\
				500 & 1 & 0 & 0 & \multicolumn{1}{c|}{0} & 1 & 0 & 0 & \multicolumn{1}{c|}{0} & 1 & 0 & 0 & \multicolumn{1}{c|}{0} & 1 & 0 & 0 & 0 \\
				1000 & 1 & 0 & 0 & \multicolumn{1}{c|}{0} & 1 & 0 & 0 & \multicolumn{1}{c|}{0} & 1 & 0 & 0 & \multicolumn{1}{c|}{0} & 1 & 0 & 0 & 0 \\ \bottomrule
		\end{tabular}}\caption{ Frequency of selecting $1$ group when the true
			number of clusters is $K_{0}=1.$ }%
		\label{tabkh1}%
	\end{table}

	\section{Empirical study}
	
	\label{Empirics} In this section we discuss the results of the empirical
	analysis, and provide a detailed discussion of the datasets we used, in the Appendix.

	In Table \ref{tab_empirics} we report the statistics in \eqref{xiT} and
	\eqref{xiR} respectively, with \emph{p-values} and report statistical significance at  different levels using   $(*)$. Note that
	${***}$ signifies 1\% significant, ${**} $ signifies 5\% significant, and ${*}
	$ signifies 10\% significant.
	Note that the datasets used in the empirical analysis have different number of
	variables, hence the critical value for the $\chi_{d}^{2}$ with which the
	statistics \eqref{xiT} and \eqref{xiR} are compared, changes, where $d$ are
	the degrees of freedom. Namely, for $d=1: c=3.841$, $d=2: c=5.991$, $d=3:
	c=7.815$, $d=4: c=9.488$.
	
	Across  datasets we use the   feasible \emph{k-means} estimator based
	on the unit-wise estimation in \eqref{inf} and estimate the number of clusters
	using the gap statistic, see e.g. \cite{gap}. The optimal number of clusters
	is selected from the range ${K} \in\{ 1, \ldots, 20 \}$ to maximise the
	criterion in \eqref{gap_in}. Further we obtain the solution of \eqref{A} based
	on the optimal number of clusters $\widehat{K}$ and use \eqref{xiT} to
	evaluate whether there is remainder heterogeneity collectively across groups
	and asses the fit of the proposed clustering method using \eqref{xiR}.
	
	In our empirical application we use eleven datasets, in which for the means of clustering we have
	standardised both $\boldsymbol{X}$ and $\boldsymbol{y}$, the number of
	variables for each dataset is indicated with $p$. Throughout the datasets, we observe 
	substantial heterogeneity across countries was observed in all the
	macroeconomic variables, it is important then to construct groups that allow
	for within group heterogeneity and at the same time, keeping the groups
	identifiable. It is observable that heterogeneity within groups  exists at most datasets,
	within groups. The only dataset that indicates near-homogeneous groups is the
	dataset of the Italian regions, where the F-statistic reported is
	statistically significant at 10\% level of significance. Further, the
	\textit{"US MSA's"," Gravity"} and \textit{"Gasoline Demand"} datasets report
	one group. This finding for the \textit{"US MSAs"}, at least, can appear odd
	because the Metropolitan housing areas are 381, and there are grounds for
	grouping, as the housing market corresponds to different incomes, while for
	the rest of the two a similar hypothesis can be adopted since income is
	different across different groups of countries. In these datasets strong
	heterogeneity (to noisy to find a signal, i.e. groups) in the cross-section is
	observed, indicative of no distinct grouping. Therefore the gap statistic
	fails to produce optimal number of groups since it is based on the hypothesis
	that the groups while heterogeneous, are clearly distinct.
	\begin{table}[ht!]
		\centering
		\resizebox{0.65\textwidth}{!}{
			\begin{tabular}{@{}lllc@{}}
				\toprule
				\multicolumn{1}{l}{  \textbf{Production functions}} &${\widehat{S}^2_{N, \widehat{K}}}$ &${\widehat{R}^2_{N, \widehat{K}}}$\\\midrule
				OECD, $p=2, \; \widehat{K}=2$   & 6.769$^{*}$ & 36.187$^{***}$  &  \\
				\midrule
				EU, $p=2$, $ \widehat{K}=3$      & 14.954$^{***}$  & 85.019 $^{***}$ &  \\
				\midrule
				ITA, 	$ p = 1, \; \widehat{K}=4$   & 2.081$^{*}$  & 26.889$^{***}$  &  \\
				\midrule
				US States, $p=2$, $ \widehat{K}=3$     & 17.355$^{**}$   & 27.005$^{***}$   &  \\
				\midrule
				UNIDO, 	$p=1, \;  \widehat{K}=3$  & 4.812$^{**}$ & 15.696$^{***}$ &  \\
				\midrule
				Production R\&D, $p=3$, $ \widehat{K}=2$     & 11.475$^{***}$  & 155.059$^{***}$ &  \\
				\midrule
				Gravity, $p=6$, $ \widehat{K}=1$   & 24.337$^{***}$   & 78.541$^{***}$   &  \\
				\midrule
				Gasoline Demand, $p=2$, $ \widehat{K}=1$   & 20.776$^{***}$   & 59.853 $^{***}$  &  \\
				\midrule
				\multicolumn{3}{l}{  \textbf{Income elasticity}}\\\midrule
				US States, 	$ p=1, \widehat{K}=4$          & 12.927$^{***}$  & 29.087$^{***}$  &  \\
				US MSAs, $p=1, \; \widehat{K}=1$            & 142.939$^{***}$ & 263.817$^{***}$&  \\
				Saving's rate,$p=4$, $ \widehat{K}=2$  & 10.835$^{**}$  & 181.976$^{***}$  &  \\
				\bottomrule
			\end{tabular}
		}
		\caption{Statistics and corresponding significance  of \eqref{xiT}, \eqref{xiR} under the null.   In cases where the test fails to reject the null,  each group is homogeneous.}
		\label{tab_empirics}
	\end{table}
	\subsection{Within-group analysis}
	
	\label{wgA} We test for heterogeneity within each group, using \eqref{xiT_K}, across the eleven different datasets considered in Table
	\ref{tab_empirics}. First, we note that across datasets heterogeneity has been
	observed within groups. It is interesting to point out that our test reports
	different levels of significance across different variables and groups,
	implying borderline homogeneity for some groups, if the test reports
	\textit{p-values} $\approx10\%$. While our findings do not suggest borderline
	homogeneity, the findings of the \textit{"Savings"} dataset are worth
	discussing. Specifically, we find that in the first group of ${S}_{t}$ (the
	ratio of savings to GDP), the \textit{p-value} is $0.0003<<1\%$, while in the
	second group a similar result applies. Regarding the second variable ${ I}%
	_{t}$ (the CPI-based inflation rate) the \textit{p-value} of the first group
	is $0.00026$, and for the second group it is $0.00014<<1\%$, while for
	${GDP}_{t}$ (the per capita GDP growth rate) similar results are found. These
	findings are indicative of heterogeneity within each group for the different
	variables, as the test results appear significant at 1\%, with
	\textit{p-values} significantly lower than the nominal rate. For the third
	variable, ${R}_{t}$ (the real interest rate), \textit{p-values} for the first
	and second group are $0.04$, $0.04<5\%$, which indicate that our tests report
	statistically significant results at the $5\%$ level of significance. The
	latter indicates that interest rates across countries (economic units) can be
	indicative of two groups and further those two groups appear less
	heterogeneous than the rest.
	
	We use the same analysis in the remainder of the datasets and find, across these datasets, that all groups rejected  the null hypothesis at $1\%$ level of significance.
	
	\section{Discussion}
	
	\label{Discussion} We explore the idea of different types of heterogeneity in
	large panel data models where the number of units are  allowed to diverge faster
	than the sample size. We follow the work of \cite{su2016identifying}, and
	use several approaches to identify and estimate latent grouping structures in
	panel data, developing panel penalized profile likelihood methods for
	classification and estimation. More importantly, we argue that, often,
	cross-sectional heterogeneity is not alleviated through grouping, creating a
	degree of bias and further identification issues in the model. In this paper,
	we explore the asymptotic properties of such a model, and emphasize on the
	identification of heterogeneous components within a group. We devise a testing
	procedure that identifies heterogeneity, for which the statistic is
	asymptotically chi-squared under the hypothesis of homogeneity, while it is
	consistent otherwise.
	
	These techniques combined, provide a general approach to grouping and
	estimating panel models with unknown groups, heterogeneity within and across
	groups, and an unknown number of groups. We use known in the literature
	statistics, e.g. \cite{gap}, to identify the number of groups, while the
	method is forcing neither grouping, nor within group heterogeneity.
	
	Simulations show that the different approaches considered have good
	finite-sample performance and can be implemented in empirical datasets.
	Further, the two tests considered have good size and power under the null
	hypothesis. We use eleven empirical applications which reveal the advantages
	of data-determined identification of latent grouping structures in empirical
	panel modeling, while most importantly our tests show that the different
	groups identified across datasets are heterogeneous.
	
	This work can be extended in a number of different directions, which we leave for future research.
	First, it may be appealing to consider a more general framework that
	allows the number $(K)$ of groups to grow with the sample size. The
	theoretical framework used in this paper can allow $K$ to grow with $N$ at a
	slower rate. Second, an extension to panel models with endogeneity will raise
	new statistical and computational challenges. Lastly, and in our view, a more
	appealing issue is considering mixed panel models, allowing for more
	heterogeneity in the sense of noise, while grouping structures can be
	considered as signals. It is then natural to devise different testing
	procedures to identify both the heterogeneous components and the signals, such
	as LM-type of tests for the former and simple t-tests for the latter. This
	framework though welcomes issues such as different degrees of bias to control,
	while joint estimation procedures can come with an additional mathematical and
	computational challenge, empirically though this framework is more realistic
	and general.

	\newpage
	\appendix	
	\setcounter{page}{1}
	
	\setcounter{equation}{0}
	\renewcommand \theequation{\thesection.\arabic{equation}}
	\setcounter{lemma}{0}
	\renewcommand \thelemma{\thesection.\arabic{lemma}}
	
	\section*{Appendix}
	\section{Theoretical Appendix}
	
	\label{Appendix_1} Appendix \ref{Appendix_1} provides technical proofs of the  results the main paper. It is organised as follows: Section \ref{MAIN} provides proofs of 
	Theorems  \ref{SSP_cons}--\ref{Feasible Kmeans}, Lemma \ref{probability}, and Theorem \ref{statisticsTheorem} of the main paper.
	Section \ref{Appendix_2} presents auxiliary technical lemmas, used to show the results in Section \ref{MAIN}, and Section \ref{AUX_proofs} provides technical proofs for the  of auxiliary  lemmas presented in Section \ref{Appendix_2}
	
	Appendix \ref{EmpiricalApp} provides   descriptions of the empirical datasets, and their specifications, used in Section \ref{Empirics}
	of the main paper.
	\bigskip 
	
	\noindent{\textbf{{\large Notation}}}\newline For any vector $\boldsymbol{x}%
	\in\mathbb{R}^{n}, $ we denote the $\ell_{p}$-, and $\ell_{\infty}$ as
	$\left\|  \boldsymbol{x} \right\| _{p}=\left( \sum_{i=1}^{n} | x_ {i}
	|^{p}\right) ^{1/p}$, $\lVert\boldsymbol{x} \rVert_{\infty} = \max
	_{i=1,\ldots,n}|x_{i}|$.  Further, $\boldsymbol{1}\{\cdot\}$ denotes an indicator function and  $\boldsymbol{0}_{p\times 1}$ denotes a zero valued $p\times 1$ vector. We use "$\to_{P}$" to denote convergence in
	probability. For two deterministic sequences $a_{n}$ and $b_{n}$ we define
	asymptotic proportionality, "$\asymp$", by writing $a_{n} \asymp b_{n}$ if
	there exist constants $0 < a_{1} \leq a_{2}$ such that $a_{1}b_{n} \leq a_{n}
	\leq a_{2}b_{n}$ for all $n \geq1$. For any set $A$, $|A|$ denotes its
	cardinality, while $A^{c}$ denotes its complement.

	\subsection{Proofs of main results}\label{MAIN}
	In this section we provide  proofs of the  Theorems and Lemmas presented in the main paper, namely Theorems  \ref{SSP_cons}--\ref{Feasible Kmeans}, Lemma \ref{probability}, and Theorem \ref{tests}, in that ordering.  
	\subsubsection*{Proof of Theorem \ref{SSP_cons}}
	The proof proceeds in two parts: In the first we show argument \eqref{s1}  and in the second we show \eqref{s2}-\eqref{ssp2}.
	Let $\boldsymbol{Q}_{NT,i}(\boldsymbol{	\beta}_i)=(1/2)(NT)^{-1}  \sum_{i=1}^{N} \sum_{t=1}^{T} ({y}_{it} - \boldsymbol{	\beta}_i'\boldsymbol{{x}}_{it})^2$ and $\boldsymbol{Q}_{iNT,\lambda}(\boldsymbol{	\beta}_i, \boldsymbol{\alpha})=\boldsymbol{Q}_{NT,i}(\boldsymbol{	\beta}_i) + \frac{\lambda}{N}\sum_{i=1}^{N}\prod_{k=1}^{K} \Vert\boldsymbol{\beta}_i - \boldsymbol\alpha_{{k}} \Vert_2 $.   Further, let, $\boldsymbol{b}_i = {\boldsymbol{\ddot{\beta}}}_{i} - \boldsymbol{\beta}_i^0$ and $\boldsymbol{\ddot{b}}_i = \boldsymbol{\beta}_{ i} - {\boldsymbol{\ddot{\beta}}}_{i}$.
	Then, for $i=1,\ldots, N$, we have
	\begin{align}
	\boldsymbol{Q}_{NT,i}({\boldsymbol{\ddot{\beta}}}_{i}) -\boldsymbol{Q}_{NT,i}(\boldsymbol{	\beta}^0_i) &= \boldsymbol{\ddot{b}}_i '\frac{1}{T}\sum_{t=1}^T \boldsymbol{ x }_{it} '\boldsymbol{ x }_{it} \boldsymbol{\ddot{b}}_i - \boldsymbol{\ddot{b}}_i '\frac{1}{T}\sum_{t=1}^T \boldsymbol{ x }_{it} '{ \varepsilon}_{it}
	\label{ineq_A11}.
	\end{align}
	Under Assumption \ref{AssonX}, 	for each $i=1,\ldots, N$ we have that  $\frac{1}{T}\sum_{t=1}^T \boldsymbol{ x }_{it} '\boldsymbol{ x }_{it}= O_P(1)$ and $\frac{1}{T}\sum_{t=1}^T \boldsymbol{ x }_{it} '{ \varepsilon}_{it} = O_P(T^{-1/2})$.
	%
	We proceed with the multiplicative part of $\boldsymbol{Q}_{iNT,\lambda}(\boldsymbol{	\beta}_i, \boldsymbol{\alpha})$, by the reverse triangle inequality, we have
	{ \small \begin{align}
		\left \vert \prod_{k=1}^{K}  \left\Vert   {\boldsymbol{\ddot{\beta}}}_{i}- \boldsymbol{	\alpha} _{k} \right\Vert_2 - \prod_{k=1}^{K}  \left\Vert   \boldsymbol{	{\beta}}^0_{i} - \boldsymbol{	\alpha} _{k} \right\Vert_2  \right\vert
		&\leq \left \vert 		\prod_{k=1}^{K-1}  \left\Vert  {\boldsymbol{\ddot{\beta}}}_{i}- \boldsymbol{	\alpha} _{k} \right\Vert_2 \left(	  \left\Vert {\boldsymbol{\ddot{\beta}}}_{i}- \boldsymbol{	\alpha} _{k} \right\Vert_2-  \left\Vert   \boldsymbol{	{\beta}}^0_{i} - \boldsymbol{	\alpha} _{k} \right\Vert_2	\right) \right\vert   \nonumber\\
		&+\left \vert 		\prod_{k=1}^{K-2}  \left\Vert  {\boldsymbol{\ddot{\beta}}}_{i}- \boldsymbol{	\alpha} _{k} \right\Vert_2  \left\Vert   \boldsymbol{	{\beta}}^{0}_{i} - \boldsymbol{	\alpha} _{k} \right\Vert_2 \right.\nonumber \\
		&\quad \left. \left(  \left\Vert   {\boldsymbol{\ddot{\beta}}}_{i}- \boldsymbol{	\alpha}_{K-1} \right\Vert_2 - \left\Vert   \boldsymbol{	{\beta}}^0_{i} - \boldsymbol{	\alpha}_{{K-1}} \right\Vert_2	\right) \right\vert     \nonumber \\
		&+ \left \vert 		\prod_{k=1}^{K-3}  \left\Vert   {\boldsymbol{\ddot{\beta}}}_{i}- \boldsymbol{	\alpha} _{k} \right\Vert_2  \left\Vert   \boldsymbol{	{\beta}}^{0}_{i} - \boldsymbol{	\alpha} _{k} \right\Vert_2 \right.\nonumber \\
		&\quad \left. \left(  \left\Vert  {\boldsymbol{\ddot{\beta}}}_{i}- \boldsymbol{	\alpha}_{{K-2}} \right\Vert_2 - \left\Vert   \boldsymbol{	{\beta}}^0_{i} - \boldsymbol{	\alpha}_{{K-2}} \right\Vert_2	\right) \right\vert  + \cdots \nonumber\\
		& + \left \vert 		\prod_{k=2}^{K}  \left\Vert   \boldsymbol{	{\beta}}^{0}_{i} - \boldsymbol{	\alpha} _{k} \right\Vert_2  \left(	 \left\Vert  {\boldsymbol{\ddot{\beta}}}_{i} - \boldsymbol{	\alpha}_{1} \right\Vert_2 - \left\Vert   \boldsymbol{	{\beta}}^{0}_{i} - \boldsymbol{	\alpha}_{1} \right\Vert_2 	\right) \right\vert \nonumber \\
		&\leq \ddot{C}_{i, NT}(\boldsymbol{	\alpha}) \left \Vert {\boldsymbol{\ddot{\beta}}}_{i}-   \boldsymbol{	{\beta}}^{0}_{i}	  \right\Vert_{2}=\ddot{C}_{i, NT}(\boldsymbol{	\alpha}) \left \Vert	\boldsymbol{\ddot	{b}}_{i} 	  \right\Vert_{2} , \label{multiplicative}
		\end{align}}
	where
	{\begin{align}
		\ddot{C}_{i, NT}(\boldsymbol{	\alpha})  &=   		\prod_{k=1}^{K-1}  \left\Vert  {\boldsymbol{\ddot{\beta}}}_{i} - \boldsymbol{	\alpha} _{k} \right\Vert_2 +	\prod_{k=1}^{K-2}  \left\Vert  {\boldsymbol{\ddot{\beta}}}_{i} - \boldsymbol{	\alpha} _{k} \right\Vert_2 \left\Vert   \boldsymbol{	{\beta}}^0_{i} - \boldsymbol{	\alpha} _{k} \right\Vert_2 +	\cdots+\prod_{k=2}^{K}  \left\Vert   \boldsymbol{	{\beta}}^0_{i} - \boldsymbol{	\alpha} _{k} \right\Vert_2
		\label{heterog}
		\end{align}}
	by implication of  \eqref{minimisation}, $N^{-1}\ddot{C}_{i, NT}(\boldsymbol{	\alpha})   = O_P(1)$. Further, by \eqref{ineq_A11}--\eqref{heterog}, and the fact that $\boldsymbol{Q}_{iNT,\lambda}({\boldsymbol{\ddot{\beta}}}_{i}, \boldsymbol{\ddot{\alpha}})\leq \boldsymbol{Q}_{iNT,\lambda}(\boldsymbol{	{\beta}}^{0}_i, \boldsymbol{\ddot{\alpha}})$ we have that
	\begin{align}
	\boldsymbol{\ddot{b}}_i '\frac{1}{T}\sum_{t=1}^T \boldsymbol{ x }_{it} '\boldsymbol{ x }_{it} \boldsymbol{\ddot{b}}_i   &\leq   \frac{2}{T}\sum_{t=1}^T \boldsymbol{ x }_{it} '{ \varepsilon}_{it} + \ddot{C}_{i, NT}(\boldsymbol{	\alpha}) \left \Vert	\boldsymbol{\ddot	{b}}_{i} 	  \right\Vert_{2} ,\quad \text{then}  \\
	\left \Vert	\boldsymbol{\ddot	{b}}_{i} 	  \right\Vert_{2}^2 \left\Vert\boldsymbol{	Q}_{xx, i}\right\Vert_2&\leq  \left\Vert 2 \boldsymbol{	Q}_{x\varepsilon, i} \right\Vert _2 + \ddot{C}_{i, NT}(\boldsymbol{	\alpha}) \left \Vert	\boldsymbol{\ddot	{b}}_{i} 	  \right\Vert_{2},
	\end{align}
	where for each $i=1,\ldots, N$  $\boldsymbol{	Q}_{xx, i} = \frac{1}{T}\sum_{t=1}^T \boldsymbol{ x }_{it} '\boldsymbol{ x }_{it} =O_P(1)$, $\boldsymbol{	Q}_{x\varepsilon, i} =\frac{1}{T}\sum_{t=1}^T \boldsymbol{ x }_{it} '{ \varepsilon}_{it}   = O_P(T^{-1/2})$, then, using  the arguments in \eqref{heterog} and the results of Lemma \ref{betaAUX},   it follows that
	\begin{align}
	\left \Vert	\boldsymbol{\ddot	{b}}_{i} 	  \right\Vert_{2}  = O_P\left(	T^{-1/2}	+ \lambda \right),
	\end{align}
	where $\lambda $ follows assumption \ref{assKm1} where $T/N \to 0 $ as $T,N\to \infty$, completing the proof of  claim \eqref{s1} of Theorem \ref{SSP_cons}.  \newline
	\indent	It remains to show mean square convergence of ${\boldsymbol{\ddot{\beta}}}- \boldsymbol{{\beta}}^0$. To that end, let   $\boldsymbol{	\beta} = \boldsymbol{	\beta}^0 + T^{-1/2} \boldsymbol{v} $, where $\boldsymbol{v} = (\boldsymbol{v}_1, \ldots, \boldsymbol{v}_N)$ is a $p\times N$ matrix and each $\boldsymbol{	v}_i, \; i=1,\ldots, N$ a $p\times 1$ vector of  deviations from each $\boldsymbol{	{\beta}}^0_j, \; j=1,\ldots, p. $ We wish to show that here is local minimum $\{ {\boldsymbol{\ddot{\beta}}},  \boldsymbol{	\ddot{\alpha}} \}$  such that $N^{-1} \sum_{i=1}^N   \lVert \boldsymbol{\ddot	{b}}\rVert^2_2 = O_P(T^{-1})$, which holds regardless of the property of $\boldsymbol{\ddot{\alpha}}. $  Using the basic inequality in  \eqref{ineq_A11} and since $ \boldsymbol{Q}_{iNT,\lambda}(\boldsymbol{	\beta}_i, \boldsymbol{\ddot{\alpha}}) - \boldsymbol{Q}_{iNT,\lambda}(\boldsymbol{	\beta}^0, \boldsymbol{\alpha}^0) =  	 \boldsymbol{Q}_{iNT,\lambda}(\boldsymbol{	\beta}^0+T^{-1/2} \boldsymbol{v}  , \boldsymbol{\ddot{\alpha}}) - \boldsymbol{Q}_{iNT,\lambda}(\boldsymbol{	\beta}^0, \boldsymbol{\alpha}^0)$,  we have that
	\begin{align}
	&T	[ \boldsymbol{Q}_{iNT,\lambda}(\boldsymbol{	\beta}^0+T^{-1/2} \boldsymbol{v}  , \boldsymbol{\ddot{\alpha}}) - \boldsymbol{Q}_{iNT,\lambda}(\boldsymbol{	\beta}^0, \boldsymbol{\alpha}^0)] \nonumber\\
	&=  \frac{1}{N} \sum_{i}^N \boldsymbol{v}_i' \boldsymbol{	Q}_{xx, i} \boldsymbol{v}_i  -\frac{\sqrt{T}}{2N}  \sum _{i=1}^N \boldsymbol{v}_i '\boldsymbol{	Q}_{x\varepsilon, i}   + \frac{\lambda \sqrt{T}}{N} \sum_{i}^N \prod _{k=1}^K \left\Vert	\boldsymbol{	{\beta}}^0_i  + T^{-1/2} \boldsymbol{v}_i- \boldsymbol{	\ddot{\alpha}} _{k}	 \right\Vert_2 \nonumber\\
	&\geq  \frac{ c_{\boldsymbol{	Q}_{xx, i}}}{N}  \sum _{i=1}^N \left\Vert  \boldsymbol{v}_i \right\Vert ^2_2 - \left[		\frac{1}{N}	\sum_{i=1}^N \left\Vert  \boldsymbol{v}_i \right\Vert ^2_2   	\right]^{1/2} \left[\frac{T}{N}	\sum_{i=1}^N \left\Vert \boldsymbol{	Q}_{x\varepsilon, i}  \right\Vert _2^2	\right]^{1/2}  \equiv D_{1} + D_2,		 
	\end{align}
	where $c_{\boldsymbol{	Q}_{xx, i}} = \boldsymbol{	Q}_{x\varepsilon, i}  = O_P(1),$ and $\frac{T}{N}	\sum_{i=1}^N \left\Vert \boldsymbol{	Q}_{x\varepsilon, i}  \right\Vert _2^2	= O_P(1)$,  $D_1$ dominates $D_2$,   that is $ \boldsymbol{Q}_{iNT,\lambda}(\boldsymbol{	\beta}^0+T^{-1/2} \boldsymbol{v}  , \boldsymbol{\ddot{\alpha}}) > \boldsymbol{Q}_{iNT,\lambda}(\boldsymbol{	\beta}^0, \boldsymbol{\alpha}^0)$.  The latter has as a consequence $\Vert \boldsymbol{\ddot	{b}}_i \Vert _2^2= O_P(T^{-1})$.  To further clarify, if $T	[ \boldsymbol{Q}_{iNT,\lambda}(\boldsymbol{	\beta}^0+T^{-1/2} \boldsymbol{v}  , \boldsymbol{\ddot{\alpha}}) - \boldsymbol{Q}_{iNT,\lambda}(\boldsymbol{	\beta}^0, \boldsymbol{\alpha}^0)] >0 $ holds then
	\begin{align}
	P \left(	\inf_{N^{-1}\sum_{i=1}^N \Vert 	\boldsymbol{	v}_i	\Vert^2_2 = L }	\boldsymbol{Q}_{iNT,\lambda}(\boldsymbol{	\beta}^0+T^{-1/2} \boldsymbol{v}  , \boldsymbol{\ddot{\alpha}}) > \boldsymbol{Q}_{iNT,\lambda}(\boldsymbol{	\beta}^0, \boldsymbol{\alpha}^0)	\right)\geq 1- c_*, \label{MSC}
	\end{align}
	where $c_*>0$, and $L = L(c_*)$ a finite, large and positive constant such that \eqref{MSC} holds for  large enough $N,T$.
	\newline
	We proceed to show \eqref{s2} of Theorem \ref{SSP_cons}.  By \eqref{multiplicative} we have that
	\begin{align}
	\ddot{C}_{i, NT}(\boldsymbol{	\alpha}) &\leq 	\prod_{k=1}^{K-1}  \left\Vert   {\boldsymbol{\ddot{\beta}}}_{i}- \boldsymbol{\beta}^0_{i} \right\Vert_2 +\left\Vert   {\boldsymbol{\ddot{\beta}}}_{i} - \boldsymbol{	\alpha} _{k} \right\Vert_2 \nonumber\\
	&\quad  +	\prod_{k=1}^{K-2}   \left[\left\Vert  {\boldsymbol{\ddot{\beta}}}_{i} - \boldsymbol{\beta}^0_{i} \right\Vert_2+\left\Vert {\boldsymbol{\ddot{\beta}}}_{i}- \boldsymbol{	\alpha} _{k} \right\Vert_2\right] \left\Vert   \boldsymbol{	{\beta}}^0_{i} - \boldsymbol{	\alpha} _{k} \right\Vert_2 +	\cdots+\prod_{k=2}^{K}  \left\Vert   \boldsymbol{	{\beta}}^0_{i} - \boldsymbol{	\alpha} _{k} \right\Vert_2\nonumber\\
	&= \sum_{j=0}^{K-1} \left\Vert  {\boldsymbol{\ddot{\beta}}}_{i}- \boldsymbol{\beta}^0_{i} \right\Vert_2 ^{j} + \prod_{k=1}^j c_{kj} \left\Vert \boldsymbol{\beta}^0_{i} - \boldsymbol{	\alpha} _{k}\right\Vert_2^{K-1-j}\nonumber\\
	&\leq C_{K}(\boldsymbol{	\alpha})   \sum_{j=0}^{K-1} \left\Vert  {\boldsymbol{\ddot{\beta}}}_{i}- \boldsymbol{\beta}^0_{i} \right\Vert_2 ^{j}\leq  C_{K}(\boldsymbol{	\alpha}) \left(	1+2 		 \left \Vert	\boldsymbol{\ddot	{b}}_{i} 	  \right\Vert_{2}  	\right),
	\label{alphaa}
	\end{align}
	where $c_{kj}\in \mathbb{Z}^{+}$, and for a group $k=1,\ldots, K$,  $\boldsymbol{\beta}_{i}^0 = \boldsymbol{\alpha}^0 _{k} + \boldsymbol{	\eta} _{k}$ we have that
	\begin{align}
	C_{K}(\boldsymbol{	\alpha}) & \leq \max_{i} \max_{1 \leq j \leq k\leq K-1} \prod_{k=1}^{j} c_{kj} \left\Vert \boldsymbol{\beta}^0_{i} - \boldsymbol{	\alpha}_{k}\right\Vert_2^{K-1-j}\nonumber\\
	&   \leq \max_{1\leq l\leq K} \max_{1 \leq j \leq k\leq K-1}  \left\Vert (\boldsymbol{\alpha}^0_{l} - \boldsymbol{	\alpha} _{k} )  +  \boldsymbol\eta_{i\in G_l}\right\Vert_2^{K-1-j}  ,
	\end{align}
	Let $P_{NT}(\boldsymbol{\beta}, \boldsymbol{\alpha}) = N^{-1}\sum_{i=1}^{N}  \prod_{k=1}^{K} \Vert	\boldsymbol{\beta}_i - \boldsymbol{	\alpha} _{k} \Vert _2$, then  by \eqref{multiplicative} and \eqref{heterog}   we have that
	\begin{align}
	\left\vert 		P_{NT}\left(	{\boldsymbol{\ddot{\beta}}}, \boldsymbol{	\alpha}	\right) -	P_{NT}\left(	\boldsymbol{	{\beta}}^0, \boldsymbol{	\alpha}	\right) 	\right\vert
	& \leq   C_{K}(\boldsymbol{	\alpha}^{0})N^{-1}\sum_{i=1}^N \left\Vert \boldsymbol{\ddot{b}}_i \right\Vert _2 + 2 C_{K}(\boldsymbol{	\alpha}^{0})N^{-1}\sum_{i=1}^N \left\Vert \boldsymbol{\ddot{b}}_i \right\Vert _2^2\label{abtilde}\\
	&=  O_P\left(\frac{1}{\sqrt{T}} \right) + O_P\left(\frac{1}{\sqrt{N}T}\right)  = O_P\left(\frac{1}{\sqrt{T}}\right).
	\label{A.7}
	\end{align}
	where $0<N_k/N  <1$.   Note  that for cluster $k= 1,\ldots, K$, $\boldsymbol\beta_{i}^{0}=\boldsymbol{\alpha} _{k}^{0}+\boldsymbol{\eta}_i$.  Then, we will proceed by showcasing the desired result for two groups and generalise to $K$ groups.   For two groups, we have
	\begin{align}
	P_{NT}\left(	\boldsymbol{	{\beta}}^0, \boldsymbol{	\alpha}^0	\right) &=	\frac{1}{N}\sum_{i}\left\Vert \boldsymbol{\beta}_{i}^{0}-\boldsymbol{\alpha}_{ 1}^{0}\right\Vert_2
	\left\Vert\boldsymbol{\beta}_{i}^{0}-\boldsymbol{\alpha}_{ 2}^{0}\right\Vert_2 \nonumber \\
	& =\frac{1}{N_1}\sum_{i\in{G}_{1}}\left\Vert \boldsymbol{\beta}_{i}^{0}-\boldsymbol{\alpha}_{ 1}^{0}\right\Vert_2
	\left\Vert\boldsymbol{\beta}_{i}^{0}-\boldsymbol{\alpha}_{ 2}^{0}\right\Vert_2+
	\frac{1}{N_2}\sum_{i\in {G}_{2}}\left\Vert \boldsymbol{\beta}_{i}^{0}-\boldsymbol{\alpha}_{ 1}^{0}\right\Vert_2
	\left\Vert\boldsymbol{\beta}_{i}^{0}-\boldsymbol{\alpha}_{ 2}^{0}\right\Vert_2 \nonumber\\
	& \leq \frac{1}{N_1}\sum_{i\in {G}_{1}}\left\Vert \boldsymbol{\eta}_{ 1}\right\Vert_2
	\left\Vert\boldsymbol{\eta}_{ 1}+(\boldsymbol{	{\alpha}}_{ 1}^{0}-\boldsymbol{	{\alpha}}_{ 2}^{0})\right\Vert_2
	+		 \frac{1}{N_2}\sum_{i\in {G}_{2}}\left\Vert \boldsymbol{\eta}_{ 2}\right\Vert_2
	\left\Vert\boldsymbol{\eta}_{ 2}+(\boldsymbol{	{\alpha}}_{ 2}^{0}-\boldsymbol{	{\alpha}}_{ 1}^{0})\right\Vert _2.
	\end{align}%
	Consequently,
	\begin{align}
	P_{NT}\left(	\boldsymbol{	{\beta}}^0, \boldsymbol{\widetilde{	\alpha}}	\right)& = \frac{1}{N}\sum_{i}\left\Vert \boldsymbol{\beta}_{i}^{0}-\boldsymbol{\ddot{\alpha}}_{ 1}\right\Vert_2
	\left\Vert  \boldsymbol{\beta}_{i}^{0}-\boldsymbol{\ddot{\alpha}}_{ 2}\right\Vert_2\nonumber \\
	& =	\frac{1}{N_1}\sum_{i\in G_1}\left\Vert \boldsymbol{\beta}_{i}^{0}-\boldsymbol{\ddot{\alpha}}_{ 1}\right\Vert
	\left\Vert  \boldsymbol{\beta}_{i}^{0}-\boldsymbol{\ddot{\alpha}}_{ 2}\right\Vert +
	\frac{1}{N_2}\sum_{i \in G_2}\left\Vert \boldsymbol{\beta}_{i}^{0}-\boldsymbol{\ddot{\alpha}}_{ 1}\right\Vert
	\left\Vert  \boldsymbol{\beta}_{i}^{0}-\boldsymbol{\ddot{\alpha}}_{ 2}\right\Vert_2\nonumber \\
	& =\frac{1}{N_1}\sum_{i\in {G}_{1}}\left\Vert \boldsymbol{\eta}_{i}+\left(
	\boldsymbol{	{\alpha}}_{1}^{0}-\boldsymbol{\ddot{\alpha}}_{ 1}\right)  \right\Vert_2 \left\Vert\boldsymbol{\eta}_{i}+\left(  \boldsymbol{{\alpha}}_{1}^{0}-\boldsymbol{\ddot{\alpha}}_{2}\right)  \right\Vert_2 \nonumber\\
	& \quad +\frac{1}{N_2}\sum_{i\in{G}_{2}}\left\Vert \boldsymbol{	\eta}_{i}+\left(
	\boldsymbol{\alpha}_{2}^{0}-\boldsymbol{\ddot{\alpha}}_{1}\right)  \right\Vert_2 \left\Vert \boldsymbol{	\eta}_{i}+\left(  \boldsymbol{\alpha}_{2}^{0}-\boldsymbol{\ddot{\alpha}}_{2}\right)  \right\Vert_2.
	\end{align}
	Then,
	\begin{align}
	P_{NT}\left(	\boldsymbol{	{\beta}}^0, \boldsymbol{{	\alpha}}^0	\right)  - 			P_{NT}\left(	\boldsymbol{	{\beta}}^0, \boldsymbol{\widetilde{	\alpha}}	\right) &\leq
	\frac{1}{N_1}\sum_{i\in {G}_{1}}\left\Vert \boldsymbol {	\eta}_{i}\right\Vert_2
	\left\Vert  \boldsymbol {	\eta}_{i}+(\boldsymbol{\alpha}_{1}^{0}-\boldsymbol{\alpha}_{2}^{0})\right\Vert_2\nonumber \\
	&\quad  - \left\Vert
	\boldsymbol {	\eta}_{i}+\left( \boldsymbol{	\alpha}_{1}^{0}-\boldsymbol{	\ddot{\alpha}}_{1}\right)  \right\Vert_2
	\left\Vert	\boldsymbol{	\eta}_{i}+\left(   \boldsymbol{	\alpha}_{1}^{0}-\boldsymbol{	\ddot{\alpha}}_{2}\right)
	\right\Vert_2 \nonumber \\
	&\quad+			\frac{1}{N_2}\sum_{i\in {G}_{2}}\left\Vert \boldsymbol{	\eta}_{i}\right\Vert_2
	\left\Vert  \boldsymbol{	\eta}_{i}+(\boldsymbol{\alpha}_{1}^{0}-\boldsymbol{\alpha}_{2}^{0})\right\Vert_2\nonumber\\
	&\quad  -\left\Vert
	\boldsymbol{	\eta}_{i}^{0}+\left( \boldsymbol{	\alpha}_{1}^{0}-\boldsymbol{	\ddot{\alpha}}_{1}\right)  \right\Vert_2
	\left\Vert	\boldsymbol{	\eta}_{i}+\left(   \boldsymbol{	\alpha}_{1}^{0}-\boldsymbol{	\ddot{\alpha}}_{2}\right)
	\right\Vert_2 \nonumber\\
	& =\frac{1}{N_1}\sum_{i\in {G}_{1}}\left\Vert \boldsymbol{\eta}_{i}\right\Vert_2
	\left\Vert \boldsymbol{\eta}_{i}+(\boldsymbol{\alpha}_{1}^{0}-\boldsymbol{\alpha}_{ 2}^{0})\right\Vert_2\nonumber\\
	&\quad -\left\Vert	\boldsymbol{\eta}_{ i}+\left( \boldsymbol{ \alpha}_{1}^{0}-\boldsymbol{\ddot{\alpha}}_{1}\right)  \right\Vert_2   			\left\Vert 	\boldsymbol{\eta}_{i }+\left(  	\boldsymbol{\alpha}_{ 1}^{0}-\boldsymbol{\alpha}_{ 2}^{0}\right)  +\left(
	\boldsymbol{\alpha}_{2}^{0}-	\boldsymbol{\ddot{\alpha}}_{ 2}\right)  \right\Vert_2  \nonumber \\
	& \quad +\frac{1}{N_2}\sum_{i\in {G}_{2}}\left\Vert \boldsymbol{\eta}_{ i}\right\Vert
	\left\Vert \boldsymbol{\eta}_{i}+(\boldsymbol{\alpha}_ {2}^{0}-\boldsymbol{\alpha}_{ 1}^{0})\right\Vert_2\nonumber\\
	&\quad -\left\Vert	\boldsymbol{\eta}_{i }+\left( \boldsymbol{ \alpha}_{2}^{0}-\boldsymbol{\ddot{\alpha}}_{2}\right)  \right\Vert_2  		
	\left\Vert 	\boldsymbol{\eta}_{i} + \left(  	\boldsymbol{\alpha}_ {2}^{0}-\boldsymbol{\alpha}_{ 2}^{0}\right)  +\left(
	\boldsymbol{\alpha}_{2}^{0} -	\boldsymbol{\ddot{\alpha}}_{ 1}\right)  \right\Vert_2\label{2groups} \leq 0 .
	\end{align}
	It is then clear that generalising \eqref{2groups} for a finite number of groups $K$, we have that
	{\small\begin{align}
		P_{NT}\left(	\boldsymbol{	{\beta}}^0, \boldsymbol{{	\alpha}}^0	\right)  - 			P_{NT}\left(	\boldsymbol{	{\beta}}^0, \boldsymbol{\widetilde{	\alpha}}	\right) &=
		\frac{1}{N_1}  \sum_{i \in G_1} \prod_{k=1}^{K}  \boldsymbol{V}_{1,k} \left(	\boldsymbol{	\alpha}^0 + \Delta\left (\boldsymbol{	{\alpha}}, \boldsymbol{\widetilde	{\alpha}} \right)	, \boldsymbol{	\eta}\right) \left\Vert 	\boldsymbol{	H}^{(1)}(\boldsymbol{	{\alpha}}, \boldsymbol{\widetilde	{\alpha}})	+ \boldsymbol{	\eta}_{i } 	\right\Vert_2\nonumber\\
		&\quad	+\cdots +	\frac{1}{N_K}  \sum_{i \in G_K} \prod_{k=1}^{K}  \boldsymbol{V}_{K,k} \left(	\boldsymbol{	\alpha}^0 + \Delta(\boldsymbol{	{\alpha}}, \boldsymbol{\widetilde	{\alpha}}),  \boldsymbol{	\eta}	\right) \left\Vert 	\boldsymbol{	H}^{(1)}(\boldsymbol{	{\alpha}}, \boldsymbol{\widetilde	{\alpha}})	+ \boldsymbol{	\eta}_{i} 	\right\Vert_2,\label{groups}
		\end{align}}
	where for a group $k\leq K$,
	\begin{align} \boldsymbol{V}_{k,k} \left(	\boldsymbol{	\alpha}^0 + \Delta(\boldsymbol{	{\alpha}}, \boldsymbol{\widetilde	{\alpha}})\right)
	&=\left\Vert \boldsymbol{\eta}_{i}\right\Vert_2
	\left\Vert \boldsymbol{\eta}_{i}+(\boldsymbol{\alpha}_{ k}^{0}-\boldsymbol{{\alpha}}_{{k+1}}^{0})\right\Vert_2   -  \left\Vert
	\boldsymbol{	\eta}_{i}+\left( \boldsymbol{	\alpha}_{k}^{0}-\boldsymbol{	\ddot{\alpha}}_{k}\right)  \right\Vert_2 \label{aa2}
	\end{align}
	and $\boldsymbol{	H}^{(k )}(\boldsymbol{	{\alpha}}, \boldsymbol{\widetilde	{\alpha}})	 = \left(
	\boldsymbol{\alpha}_{k}^{0}-	\boldsymbol{\ddot{\alpha}}_{k}\right) $. Further,
	by implication of  \eqref{A.7}, $P_{NT}\left(	\boldsymbol{	{\beta}}^0, \boldsymbol{{	\alpha}}^0	\right)  - 			P_{NT}\left(	\boldsymbol{	{\beta}}^0, \boldsymbol{\widetilde{	\alpha}}	\right) = O_P(T^{-1/2})$, implying that $(\boldsymbol{\ddot{\alpha}}_{1}, \ldots, \boldsymbol{\ddot{\alpha}}_{ K})$  is consistent up to a permutation, i.e. $(\boldsymbol{\ddot{\alpha}}_{1}, \ldots, \boldsymbol{\ddot{\alpha}}_{K}) - ( \boldsymbol{\alpha}_{1}^{0}, \ldots, \boldsymbol{\alpha}_{K}^{0})= O_P(T^{-1/2})$,  which completes the proof of \eqref{ssp2} and therefore completing the proof of the theorem.  	\qed		
	
	\subsubsection*{Proof of Theorem \ref{consistencyeta}}
	To show the results of  Theorem \ref{consistencyeta} we follow a similar  line of arguments as in Theorem \ref{SSP_cons}, with the only difference that now $\boldsymbol{	\widehat{\alpha}}$ is obtained through \emph{k-means}.
	To show that $(\boldsymbol{\widehat{\alpha}}_{1}, \ldots, \boldsymbol{\widehat{\alpha}}_{K})$  is consistent up to a permutation, i.e. $(\boldsymbol{\widehat{\alpha}}_{1}, \ldots, \boldsymbol{\widehat{\alpha}}_{K}) - ( \boldsymbol{\alpha}_{1}^{0}, \ldots, \boldsymbol{\alpha}_{ K}^{0})= O_P(T^{-1/2} + N^{-1/2})$, we follow the analysis in \eqref{groups} with the difference that $\left(
	\boldsymbol{\alpha}_{1}^{0}-	\boldsymbol{\widehat{\alpha}}_{ 1}\right)=O_P(N^{-1/2})$, which arises by implication of Theorem C of \cite{pollard1982central}.\qed
	
	\subsubsection*{Proof of Theorem \ref{Feasible Kmeans}}
	The proof proceeds in two steps. The first is to show statement of interest in Theorem  \ref{Feasible Kmeans} under the null hypothesis in  \ref{hypothesis}, and consequently, under the alternative.
	We start by showing \eqref{aahat} under $H_0: \boldsymbol{	{\beta}}_i=\boldsymbol{\alpha}_k$. Then, we construct the mean group estimator of $\boldsymbol{\beta}_i$, for $i\in \boldsymbol{	G}_k$, considering further that $\boldsymbol{	\widehat{G}}_k\to_{P} \boldsymbol{G}^0_k$, so that for $1\leq i\leq N$ we have that
	$\boldsymbol{\widetilde{\alpha}}_k = \arg \min_{\boldsymbol{\alpha}}  N^{-1} \sum_{i} (\boldsymbol{\widetilde{\beta}}_i -\boldsymbol{	{\alpha}}_k	)$, where $\boldsymbol{\widetilde{\beta}}$ is defined in \eqref{astt}.
	Then under $H_0$,  Assumption \ref{tuneparameters} \eqref{a21}, the fact that $N_k/N\in (0,1)$ and  by using Lemma \ref{KSS}  we have that
	\begin{align}
	{\sqrt{N_kT}} \left(	\boldsymbol{\widetilde{\alpha}}_k - \boldsymbol{\alpha}_k^0\right) =  \frac{1}{\sqrt{N_k}} \sum_{i\in \boldsymbol{G}_k}  \frac{1}{\sqrt{N}}\frac{\frac{\sum_{j=1}^{N}\boldsymbol{x}_j '\boldsymbol{\varepsilon}_j  }{\sqrt{T}}}{\sum_{j=1}^N\frac{\boldsymbol{x}_j '\boldsymbol{x}_j }{T} } \to N(\boldsymbol{0}, \Sigma_{\boldsymbol{x\varepsilon}}) ,
	\end{align}
	where $\Sigma_{\boldsymbol{x\varepsilon}} = N^{-1}\lim_{N\to \infty } \sum_i  \Sigma_{\boldsymbol{\varepsilon}, i}\Sigma_{\boldsymbol{xx}, i}^{-1}$ and $\Sigma_{\boldsymbol{\varepsilon}} = E({T^{-1}\boldsymbol{\varepsilon}_i}{\boldsymbol{\varepsilon}_i'})$.  It follows that  $	\boldsymbol{\widetilde{\alpha}} - \boldsymbol{\alpha}^0= {\sqrt{N N_kT}}^{-1} $, which is slightly sub-otpimal compared to   $\sqrt{NT}^{-1}. $
	
	The, under $H_1$, the  proof of \eqref{aahat} follows  from the results of Lemma \ref{Aux1} and Lemma \ref{LemmaB.2} of Appendix  \ref{Appendix_2}.\qed

	\subsubsection*{Proof of Lemma \ref{probability}}
	The proof follows similar steps as the proof of Theorem 2.2 of \cite{su2016identifying}.  To show \eqref{p1}--\eqref{p2} we make the following statements:   By consistency of $\boldsymbol{\widehat{\beta}}_{i}$ and $\boldsymbol{\widehat{\alpha}}_k$, the fact that $\min_{1\leq k\leq j\leq K}\Vert \boldsymbol{\alpha}_k^0 - \boldsymbol{\alpha}^{0}_j\Vert _2 \geq c_a, \; c_a\geq 0 $ and Assumption \ref{assKm1} \eqref{a21}
	\[\boldsymbol{\widehat{\beta}}_i-\boldsymbol{\widehat{\alpha}}_l \to_P \boldsymbol{{\alpha}}^0_k- \boldsymbol\alpha_l^0 + \boldsymbol{\eta}_i \neq 0,\; \forall \; i\in G_k^0.\]  Consequently,
	\begin{equation}
	\widehat{c}_{k,i}\equiv \prod_{l=1, \; l\neq k}^K \Vert \boldsymbol{\widehat{\beta}}_i-\boldsymbol{\widehat{\alpha}}_l  \Vert _2 \to_P c_k^0 \equiv \prod_{l=1, \; l\neq k}^K\Vert \boldsymbol{{\alpha}}^0_k-\boldsymbol{\widehat{\alpha}}_l +\boldsymbol\eta_{i}  \Vert _2,\label{ckhat}
	\end{equation}
	where the latter is similar to the first term of $ \boldsymbol{V}_{k,k} \left(	\boldsymbol{	\alpha}^0 + \Delta(\boldsymbol{	{\alpha}}, \boldsymbol{\widehat	{\alpha}})\right)$ in Theorem \ref{consistencyeta}. Recall the generating process of $\boldsymbol{\beta}_i^0$ in \eqref{model}, then $\Vert \boldsymbol{\widehat{\beta}}_i-\boldsymbol{\widehat{\alpha}}_k  \Vert _2\neq 0 $ for some $i\in \boldsymbol{G}_k^0$. Taking the first order conditions with respect to $\boldsymbol{\beta}_i$ for the minimisation \eqref{minimisation2}, we have that
	\begin{align}
	\boldsymbol{0}_{p\times 1}&=  \frac{1}{\sqrt{T}}  [- (y_{it}- \boldsymbol{\beta}_{i}' \boldsymbol{x}_{it} ) \boldsymbol{x}_{it}] +\sqrt{T} \lambda \sum_{j=1}^{K} \frac{\boldsymbol{\widehat{\beta}}_{i}	-\boldsymbol{\widehat{\alpha}}_{j}	}{\Vert  \boldsymbol{\widehat{\beta}}_{i}	-\boldsymbol{\widehat{\alpha}}_{j}	 \Vert_2} \prod_{l=1, \; l\neq j}^{K} \Vert  \boldsymbol{\widehat{\beta}}_{i}	-\boldsymbol{\widehat{\alpha}}_{l}	 \Vert_2\\
	&=  \frac{1}{ \sqrt{T}} \sum_{t=1}^{T} \left(\boldsymbol{x}_{it}	 (\boldsymbol{\beta}_{i}-\boldsymbol{\beta}_{i}^0) \boldsymbol{x}_{it} \right) -  \frac{1}{ \sqrt{T}}\sum_{t=1}^{T} \boldsymbol{x}_{it}\varepsilon_{it} \\
	&+\sqrt{T}\frac{\lambda \widehat{c}_{k,i}}{	\Vert  \boldsymbol{\widehat{\beta}}_{i}	-\boldsymbol{\widehat{\alpha}}_{k} \Vert _2 	} \boldsymbol{\iota}_{p}\left( \boldsymbol{\widehat{\beta}}_{i} - \boldsymbol{\widehat{\alpha}}_k  \right)+  \left[\frac{1}{{T}} \sum_{t=1}^{T} E\left( \boldsymbol{x}_{it} \boldsymbol{x}_{it}'\right) \right]\sqrt{T}\left( \boldsymbol{\widehat{\beta}}_{i} - \boldsymbol{\widehat{\alpha}}_k  \right)\\
	&+\sqrt{T}\left [\frac{1}{{T}}	\sum_{t=1}^T  \boldsymbol{x}_{it}\left(	\boldsymbol{\beta}_{i} -\boldsymbol{\beta}^0_{i} \right) \boldsymbol{x}_{it}	\right] + \frac{1}{{T}}	\sum_{t=1}^T E\left( \boldsymbol{x}_{it} \boldsymbol{x}_{it}'\right) \sqrt{T} \left( \boldsymbol{\widehat{\alpha}}_{k} - \boldsymbol{\alpha}^0_k 	\right)\\
	&+\sqrt{T} \lambda \sum_{j=1, \; j\neq k}^{K} \frac{\boldsymbol{\widehat{\beta}}_{i}	-\boldsymbol{\widehat{\alpha}}_{j}	}{\Vert  \boldsymbol{\widehat{\beta}}_{i}	-\boldsymbol{\widehat{\alpha}}_{j}	 \Vert_2} \prod_{l=1, \; l\neq j}^{K} \Vert  \boldsymbol{\widehat{\beta}}_{i}	-\boldsymbol{\widehat{\alpha}}_{l}	 \Vert_2\\
	&=  B_{i,1}+  B_{i,2}+  B_{i,3}+  B_{i,4}+B_{i,5}+  B_{i,6}+B_{i,7}.  \label{forP}
	\end{align}
	It is clear that $ \frac{\boldsymbol{\widehat{\beta}}_{i}	-\boldsymbol{\widehat{\alpha}}_{j}	}{\Vert  \boldsymbol{\widehat{\beta}}_{i}	-\boldsymbol{\widehat{\alpha}}_{j}	 \Vert_2}  $ in term $B_{i, 7}$ is not asymptotically vanishing because $\Vert  \boldsymbol{\widehat{\beta}}_{i}	-\boldsymbol{\widehat{\alpha}}_{l}	 \Vert_2 \neq 0 $, in the case of homogeneous groups the term would be
	$   \Vert	\frac{\boldsymbol{\widehat{\beta}}_{i}	-\boldsymbol{\widehat{\alpha}}_{j}	}{\Vert  \boldsymbol{\widehat{\beta}}_{i}	-\boldsymbol{\widehat{\alpha}}_{j}	 \Vert_2}  \Vert _2\leq 1$.
	The analysis in \eqref{forP} is useful to show statement \eqref{p1} below:
	\begin{align}
	P\left( \cup_{i\notin \boldsymbol{\widehat{G}}_k} i\notin \boldsymbol{\widehat{G}}_k , i\in \boldsymbol{G}_k^0  \right) &= P\left( 	-	B_{i,1}-B_{i,2} = B_{i,3}+  B_{i,4}+B_{i,5}+  B_{i,6}+B_{i,7} \right) \\
	&\leq  P\left( \left\vert B_{i,1}+B_{i,2} \right\vert \leq \left\vert B_{i,3}+  B_{i,4}+B_{i,5}+  B_{i,6}+B_{i,7}\right\vert \right) \label{math1}.
	\end{align}
	We make use of the argument in   \eqref{ckhat} and substitute $\left( \boldsymbol{\widehat{\beta}}_{i} - \boldsymbol{\widehat{\alpha}}_{k} \right)$ to   \eqref{math1}, such that   \eqref{math1} becomes
	{\small\begin{align}
		&P\left( \left\vert \left( \boldsymbol{\widehat{\beta}}_{i} - \boldsymbol{\widehat{\alpha}}_{k} \right)'(B_{i,1}+B_{i,2}) \right\vert \leq \left\vert \left( \boldsymbol{\widehat{\beta}}_{i} - \boldsymbol{\widehat{\alpha}}_{k} \right)'(B_{i,3}+  B_{i,4}+B_{i,5}+  B_{i,6}+B_{i,7})\right\vert \right) .\label{probs}
		\end{align}}
	We first analyse the following
	{\begin{align}
		& P\left(\max_{i\in G_k^0}\left \Vert \widehat{c}_{k,i} - {c}_{k}^0 \right\Vert_2   \geq	 \frac{{c}_{k}^0}{2}  \right)\nonumber\\
		&  =  P\left (\left\Vert \prod_{l=1, \; l\neq k}^K \Vert \boldsymbol{\widehat{\beta}}_i-\boldsymbol{\widehat{\alpha}}_l  \Vert _2 - \prod_{l=1, \; l\neq k}^K\Vert \boldsymbol{{\alpha}}^0_k-\boldsymbol{\widehat{\alpha}}_l +\boldsymbol\eta_{i}  \Vert _2 \right\Vert_2 \geq \frac{1}{2} \prod_{l=1, \; l\neq k}^K\Vert \boldsymbol{{\alpha}}^0_k-\boldsymbol{\widehat{\alpha}}_l +\boldsymbol\eta_{i}  \Vert _2 \right)\nonumber \\
		&=  P\left (\max_{i\in G_k^0}\left\Vert \prod_{l=1, \; l\neq k}^K \Vert (\boldsymbol{\widehat{\beta}}_i - \boldsymbol{{\beta}}^0_i) + \left(\boldsymbol{\alpha}^0_l - \boldsymbol{\widehat{\alpha}}_l\right)   +\boldsymbol{\eta}_i\Vert _2 - \prod_{l=1, \; l\neq k}^K\Vert \boldsymbol{{\alpha}}^0_k-\boldsymbol{\widehat{\alpha}}_l +\boldsymbol\eta_{i}  \Vert _2 \right\Vert_2 \right.\nonumber\\
		&\quad \quad \quad  \quad \left.\geq \frac{1}{2} \prod_{l=1, \; l\neq k}^K\Vert \boldsymbol{{\alpha}}^0_k-\boldsymbol{\widehat{\alpha}}_l +\boldsymbol\eta_{i}  \Vert _2 \right),\nonumber
		\end{align}
	}
	where, by implication of \eqref{abtilde}--\eqref{aa2}, $P\left(\max_{i\in G_k^0}\left \Vert \widehat{c}_{k,i} - {c}_{k}^0 \right\Vert_2   \geq	 \frac{{c}_{k}^0}{2}  \right) =o(1). $ Further, we consider the following intersection of events
	{\small \begin{align}
		\mathcal{E}_{k,i,T}& \equiv \left\{ 		\max_{i\in G_k^0}\left \Vert \widehat{c}_{k,i} - {c}_{k}^0 \right\Vert_2   \geq	 \frac{{c}_{k}^0}{2} 	\right\} \cap  \left\{ 		\max_{i\in G_k^0}\left \Vert \frac{1}{{T}} \sum_{t=1}^{T}  E\left( \boldsymbol{x}_{it} \boldsymbol{x}_{it}'\right)  \right\Vert_2 \left\Vert  \boldsymbol{\widehat{\beta}}_{i}- \boldsymbol{\beta}^0_{i}\right\Vert_2  \geq	\sqrt{T} {c }_1	\right\}\nonumber  \\
		& \quad \cap  \left\{   \max_{i\in G_k^0}\left\Vert    B_{i,5}  \right\Vert_2 \leq T(\ln{T})^{\nu+3} c_2 \right\} \cap \left\{   \max_{i\in G_k^0}\left\Vert    B_{i,6}  \right\Vert_2 \leq T(\sqrt{T}+\sqrt{N}) (\ln{T})^{\nu} c_3 \right\} \nonumber \\
		& \quad \cap  \left\{   \max_{i\in G_k^0}\left\Vert    B_{i,7}  \right\Vert_2 \leq \sqrt{T}  c_4 \right\} ,\label{events}
		\end{align}}
	for some $c_1, c_2, c_3>0$, $c_4=\left(T^{-1}(\ln T)^3+\lambda\right)(\ln T)^\nu, \; \nu>0$.  We condition \eqref{probs} on $\mathcal{E}_{k,i,T}$, and note that uniformly on $i\in \boldsymbol{G}^0_k$, $\left\Vert    B_{i,7}  \right\Vert_2 $ is an asymptotically non-vanishing term, due to $\Vert	\boldsymbol{\widehat{\beta}}_i - \boldsymbol{\widehat{\alpha}}_k \Vert_2\neq 0$.  Then, considering Assumption \ref{AssonX}-\ref{clust},  the results of Theorem \ref{consistencyeta}, and Lemma \ref{betaAUX},  it becomes  clear that \eqref{probs}$\geq 0$ uniformly on $i\in \boldsymbol{G}_k^0$. Note that in the case of  $T=o(N)$ the probability of the third event in \eqref{events} is $o(1)$.  The result of \eqref{p1} follows.
	We continue to show statement  \eqref{p2}:
	{ \begin{align}
		P\left(\cup_{k=1}^K i\in \boldsymbol{\widehat{G}}_k , i\notin \boldsymbol{G}_k^0  \right)\nonumber
		&\leq \sum_{k=1}^K\sum_{i\notin \boldsymbol{{G}}^0_k} P\left(i\in \boldsymbol{\widehat{G}}_k , i\notin \boldsymbol{G}_k^0  \right) \\
		&=  \sum_{l=1, l\neq k}^K \sum_{i\notin \boldsymbol{{G}}^0_k} P\left(i\notin \boldsymbol{\widehat{G}}_k , i \notin \boldsymbol{G}_{l}^0  \right)
		= \sum_{l=1, l\neq k}^K \sum_{i\notin \boldsymbol{{G}}^0_k} P\left(i\in  \boldsymbol{\widehat{G}}_k , i \in  \boldsymbol{G}_{l}^0  \right)
		\end{align}}
	Following the analysis on \eqref{p1}, $\sum_{l=1, l\neq k}^K \sum_{i\notin \boldsymbol{{G}}^0_k} P\left(i\in  \boldsymbol{\widehat{G}}_k , i \in  \boldsymbol{G}_{l}^0  \right) \geq 0 $ for  sufficiently large $(N,T)$.  The results follows, completing the proof of the Lemma.\qed
	
	\subsubsection*{Proof of Theorem \ref{statisticsTheorem}}
	
	We begin by showing  statement \eqref{Statement1_test} of the theorem.
	To show \eqref{Statement1_test}, we need to first show that $\widehat{S}_{N}\to N(0,1)$ under the null  in \eqref{hypothesis}.
	To show show that $\widehat{S}_{N}\to N(0,1)$, we first need to show that
	\begin{align}
	{S}^{*}_{N}= \frac{1}{\sqrt{N}} \sum_{i=1}^{N} \frac{ {t}_{i}^2(\beta)-1  }{\widehat\sigma_{i}} , \quad \widehat{S}_{N}-{S}^{*}_{N}=o_P(1).  \label{op1}
	\end{align}
	Let $\widehat\sigma_{i}=T^{-1} {\boldsymbol{\widehat{\varepsilon}}_{i}^{\prime}\boldsymbol{\widehat{\varepsilon}}_{i}%
	} $ and  $\sigma_{i}=E\left( T^{-1}  {\boldsymbol{\varepsilon}_{i}^{\prime}\boldsymbol{\varepsilon}_{i}%
	}\right)  $.  Consider a general estimator $\widehat{\beta}$ for $\beta^0$.
	We  start the analysis with the nominator.    The test statistic of interest is
	$
	\frac{1}{\sqrt{N}}\sum_{i}t_{i}(\widehat{\beta})
	$
	where
	$
	t_{i}(\beta)={\boldsymbol{\varepsilon}_{i}\boldsymbol{(}\beta\boldsymbol{)}%
		^{\prime}\boldsymbol{x}_{i}}{\widehat{\sigma}_{i}(\beta)\left(  \boldsymbol{x}%
		_{i}^{\prime}\boldsymbol{x}_{i}\right)  ^{-1/2}},
	$
	$
	\boldsymbol{\varepsilon}_{i}\boldsymbol{(}\beta\boldsymbol{)=y}_{i}-\beta\boldsymbol{x}%
	_{i}
	$, %
	$
	\widehat{\sigma}_{i}(\beta)=T^{-1}{\boldsymbol{\widehat{v}}_{i}\boldsymbol{(}%
		\beta\boldsymbol{)}^{\prime}\boldsymbol{\widehat{v}}_{i}\boldsymbol{(}%
		\beta\boldsymbol{)}}%
	$
	, $
	\boldsymbol{\widehat{v}}_{i}\boldsymbol{(}\beta\boldsymbol{)=\varepsilon}_{i}\boldsymbol{(}%
	\beta\boldsymbol{)-}\widehat{\gamma}\left(  \beta\right)  \boldsymbol{x}_{i}%
	$
	and
	$
	\widehat{\gamma}\left(  \beta\right)  ={\boldsymbol{\varepsilon}_{i}\boldsymbol{(}%
		\beta\boldsymbol{)}^{\prime}\boldsymbol{x}_{i}}({\boldsymbol{x}_{i}^{\prime
		}\boldsymbol{x}_{i}})^{-1},%
	$ and $\widehat{\beta}$  an estimator of $\beta_{0}.$
	We have that
	\[
	\frac{1}{\sqrt{N}}\sum_{i}t_{i}(\widehat{\beta})=\frac{1}{\sqrt{N}}\sum_{i}\left(
	t_{i}(\widehat{\beta})-t_{i}(\beta^{0})\right)  +\frac{1}{\sqrt{N}}\sum_{i}%
	t_{i}(\beta^{0})
	\]
	Then looking at the first term we have that for $1\leq i\leq N$
	\begin{align}
	\frac{1}{\sqrt{N}}\sum_{i}\left(  t_{i}(\widehat{\beta})-t_{i}(\beta^{0})\right)
	&  =\frac{1}{\sqrt{N}}\sum_{i}\frac{\left(  \boldsymbol{\varepsilon}_{i}\boldsymbol{(}%
		\widehat{\beta}\boldsymbol{)}-\boldsymbol{\varepsilon}_{i}\boldsymbol{(}\beta^{0}%
		\boldsymbol{)}\right)  ^{\prime}\boldsymbol{x}_{i}}{\left(  \boldsymbol{x}%
		_{i}^{\prime}\boldsymbol{x}_{i}\right)  ^{1/2}}=\\
	\frac{1}{\sqrt{N}}\sum_{i}\frac{\left(  \widehat{\beta}-\beta^{0}\right)
		\boldsymbol{x}_{i}^{\prime}\boldsymbol{x}_{i}}{\left(  \boldsymbol{x}%
		_{i}^{\prime}\boldsymbol{x}_{i}\right)  ^{1/2}}  &  =\left(  \frac{1}{\sqrt
		{N}}\sum_{i}\left(  \boldsymbol{x}_{i}^{\prime}\boldsymbol{x}_{i}\right)
	^{1/2}\right)  \left(  \widehat{\beta}-\beta^{0}\right)
	\end{align}
	But%
	\begin{align}
	\frac{1}{\sqrt{N}}\sum_{i}\left(  \boldsymbol{x}_{i}^{\prime}\boldsymbol{x}%
	_{i}\right)  ^{1/2}=\sqrt{NT}\left(  \frac{1}{N}\sum_{i}\left(  \frac
	{\boldsymbol{x}_{i}^{\prime}\boldsymbol{x}_{i}}{T}\right)  ^{1/2}\right)\label{xxi}
	\end{align}
	By Assumption \ref{AssonX} \eqref{Xt} --  \eqref{posdef}, it becomes apparent that
	\begin{align}
	\frac{1}{N}\sum_{i}\left(  \frac{\boldsymbol{x}_{i}^{\prime}\boldsymbol{x}%
		_{i}}{T}\right)  ^{1/2}=O_{p}\left(  1\right)
	\end{align}
	Consider that $\widehat{\beta}\to_{P}\beta^{0}$, so that  $\widehat{\beta}-\beta^{0}=$ $O_{p}\left(  \left(  \sqrt{NT}\right)
	^{-1}\right)  $, so the following holds
	\begin{align}
	\frac{1}{\sqrt{N}}\sum_{i}\left(  t_{i}(\widehat{\beta})-t_{i}(\beta^{0})\right)
	=O_{p}(1).
	\end{align}
	We further analyse the test:
	We focus on the Mean Group (MG)
	estimator of $\beta^{0}$, $\widehat{\beta}_{MG}$.  Then for $1\leq i\leq N$ we write that
	\begin{align}
	\widehat{\beta}_{MG}-\beta^{0}=\frac{1}{N}\sum_{i}\left(  \frac{\boldsymbol{y}%
		_{i}^{\prime}\boldsymbol{x}_{i}}{\boldsymbol{x}_{i}^{\prime}\boldsymbol{x}%
		_{i}}-\beta^{0}\right)  =\frac{1}{N}\sum_{i}\frac{\boldsymbol{x}_{i}^{\prime
		}\boldsymbol{\varepsilon}_{i}}{\boldsymbol{x}_{i}^{\prime}\boldsymbol{x}_{i}}%
	\end{align}
	So,
	\begin{align}
	\sqrt{NT}\left(  \widehat{\beta}_{MG}-\beta^{0}\right)  =\frac{1}{\sqrt{N}}\sum
	_{i}\frac{\left(  \boldsymbol{x}_{i}^{\prime}\boldsymbol{\varepsilon}_{i}\right)
		/\sqrt{T}}{\left(  \boldsymbol{x}_{i}^{\prime}\boldsymbol{x}_{i}/T\right)
	}\rightarrow N\left(  0,\sigma_{\varepsilon/x}^{2}\right)\label{ss}
	\end{align}
	where $\sigma_{\varepsilon/x}^{2}=\lim_{N\rightarrow\infty}\frac{1}{N}\sum_{i}%
	\frac{\sigma_{i\varepsilon}^{2}}{\sigma_{ix}^{2}}$. Further, reinserting the probability
	limit of $\widehat{\sigma}(\beta)$%
	\begin{align}
	p\lim\frac{1}{N}\sum_{i}\left(  \frac{\boldsymbol{x}_{i}^{\prime
		}\boldsymbol{x}_{i}}{\sigma_{iu}^{2}T}\right)  ^{1/2}=\lim_{N\rightarrow
		\infty}\frac{1}{N}\sum_{i}\frac{\sigma_{ix}}{\sigma_{iu}}=\sigma_{x/\varepsilon}%
	\end{align}
	So
	\begin{align}
	\frac{1}{\sqrt{N}}\sum_{i}\left(  t_{i}(\widehat{\beta}_{MG})-t_{i}(\beta^{0})\right)
	\rightarrow N\left(  0,\sigma_{x/\varepsilon}^{2}\sigma_{\varepsilon/x}^{2}\right)\label{sse}
	\end{align}
	Next, note that
	\begin{align}
	\frac{1}{\sqrt{N}}\sum_{i}t_{i}(\beta^{0})=\frac{1}{\sqrt{N}}\sum_{i}%
	\frac{\boldsymbol{u}_{i}\boldsymbol{(\beta^{0})}^{\prime}\boldsymbol{x}_{i}%
	}{\widehat{\sigma}_{i}(\beta^{0})\left(  \boldsymbol{x}_{i}^{\prime}%
		\boldsymbol{x}_{i}\right)  ^{1/2}}%
	\end{align}
	So
	\begin{align}
	&  \frac{1}{\sqrt{N}}\sum_{i}\left(  t_{i}(\widehat{\beta}_{MG})-t_{i}(\beta
	^{0})\right)  +\frac{1}{\sqrt{N}}\sum_{i}t_{i}(\beta^{0})\\
	&  =\frac{1}{\sqrt{N}}\sum_{i}\frac{\left(  \widehat{\beta}_{MG}-\beta^{0}\right)
		\boldsymbol{x}_{i}^{\prime}\boldsymbol{x}_{i}}{\left(  \boldsymbol{x}%
		_{i}^{\prime}\boldsymbol{x}_{i}\right)  ^{1/2}}+\frac{1}{\sqrt{N}}\sum
	_{i}t_{i}(\beta^{0})\\
	&  =\frac{1}{\sqrt{N}}\sum_{i}\frac{\left(  \frac{1}{N}\sum_{j}\frac
		{\boldsymbol{x}_{j}^{\prime}\boldsymbol{\varepsilon}_{j}}{\boldsymbol{x}_{j}^{\prime
			}\boldsymbol{x}_{j}}\right)  \boldsymbol{x}_{i}^{\prime}\boldsymbol{x}_{i}%
	}{\left(  \boldsymbol{x}_{i}^{\prime}\boldsymbol{x}_{i}\right)  ^{1/2}}%
	+\frac{1}{\sqrt{N}}\sum_{i}\frac{\boldsymbol{\varepsilon}_{i}{}^{\prime}\boldsymbol{x}%
		_{i}}{\widehat{\sigma}_{i}(\beta^{0})\left(  \boldsymbol{x}_{i}^{\prime
		}\boldsymbol{x}_{i}\right)  ^{1/2}}\\
	&  =\left(  \frac{1}{N}\sum_{i}\left(  \frac{\boldsymbol{x}_{i}^{\prime
		}\boldsymbol{x}_{i}}{T}\right)  ^{1/2}\right)  \left(  \frac{1}{\sqrt{N}}%
	\sum_{j}\frac{\frac{\boldsymbol{x}_{j}^{\prime}\boldsymbol{\varepsilon}_{j}}{\sqrt{T}}%
	}{\frac{\boldsymbol{x}_{j}^{\prime}\boldsymbol{x}_{j}}{T}}\right)  +\frac
	{1}{\sqrt{N}}\sum_{i}\frac{\frac{\boldsymbol{\varepsilon}_{i}{}^{\prime}\boldsymbol{x}%
			_{i}}{\sqrt{T}}}{\widehat{\sigma}_{i}(\beta^{0})\left(  \frac{\boldsymbol{x}%
			_{i}^{\prime}\boldsymbol{x}_{i}}{T}\right)  ^{1/2}}\\
	&  =\frac{1}{\sqrt{N}}\sum_{i}\left[  \left(  \frac{\frac{1}{N}\sum_{j}\left(
		\frac{\boldsymbol{x}_{j}^{\prime}\boldsymbol{x}_{j}}{T}\right)  ^{1/2}}%
	{\frac{\boldsymbol{x}_{i}^{\prime}\boldsymbol{x}_{i}}{T}}\right)  \left(
	\frac{\boldsymbol{x}_{i}^{\prime}\boldsymbol{\varepsilon}_{i}}{\sqrt{T}}\right)
	+\left(  \frac{1}{\widehat{\sigma}_{i}(\beta^{0})\left(  \frac{\boldsymbol{x}%
			_{i}^{\prime}\boldsymbol{x}_{i}}{T}\right)  ^{1/2}}\right)  \left(
	\frac{\boldsymbol{x}_{i}^{\prime}\boldsymbol{\varepsilon}_{i}}{\sqrt{T}}\right)  \right]
	\\
	&  =\frac{1}{\sqrt{N}}\sum_{i}\left[  \left[  \left(  \frac{\frac{1}{N}%
		\sum_{j}\left(  \frac{\boldsymbol{x}_{j}^{\prime}\boldsymbol{x}_{j}}%
		{T}\right)  ^{1/2}}{\frac{\boldsymbol{x}_{i}^{\prime}\boldsymbol{x}_{i}}{T}%
	}\right)  +\left(  \frac{1}{\widehat{\sigma}_{i}(\beta^{0})\left(  \frac
		{\boldsymbol{x}_{i}^{\prime}\boldsymbol{x}_{i}}{T}\right)  ^{1/2}}\right)
	\right]  \left(  \frac{\boldsymbol{x}_{i}^{\prime}\boldsymbol{\varepsilon}_{i}}{\sqrt
		{T}}\right)  \right]
	\end{align}
	It can be easily shown that
	{\small \begin{align}
		\frac{1}{\sqrt{N}}\sum_{i}\left[  \left(  \frac{1}{\widehat{\sigma}_{i}(\beta
			^{0})\left(  \frac{\boldsymbol{x}_{i}^{\prime}\boldsymbol{x}_{i}}{T}\right)
			^{1/2}}\right)  \left(  \frac{\boldsymbol{x}_{i}^{\prime}\boldsymbol{\varepsilon}_{i}%
		}{\sqrt{T}}\right)  \right]  -\frac{1}{\sqrt{N}}\sum_{i}\left[  \left(
		\frac{1}{\sigma_{i}\left(  \frac{\boldsymbol{x}_{i}^{\prime}\boldsymbol{x}%
				_{i}}{T}\right)  ^{1/2}}\right)  \left(  \frac{\boldsymbol{x}_{i}^{\prime
			}\boldsymbol{\varepsilon}_{i}}{\sqrt{T}}\right)  \right]  =o_{p}\left(  1\right)\label{ssig}
		\end{align}}
	Then, letting
	\begin{align}
	w_{Ti}=\left[  \left(  \frac{\frac{1}{N}\sum_{j}\left(  \frac{\boldsymbol{x}%
			_{j}^{\prime}\boldsymbol{x}_{j}}{T}\right)  ^{1/2}}{\frac{\boldsymbol{x}%
			_{i}^{\prime}\boldsymbol{x}_{i}}{T}}\right)  +\left(  \frac{1}{\sigma
		_{i}\left(  \frac{\boldsymbol{x}_{i}^{\prime}\boldsymbol{x}_{i}}{T}\right)
		^{1/2}}\right)  \right]  \left(  \frac{\boldsymbol{x}_{i}^{\prime
		}\boldsymbol{\varepsilon}_{i}}{\sqrt{T}}\right)  ,
	\end{align}
	we have
	\begin{align}
	& \frac{1}{\sqrt{N}}\sum_{i}\left[  \left[  \left(  \frac{\frac{1}{N}\sum
		_{j}\left(  \frac{\boldsymbol{x}_{j}^{\prime}\boldsymbol{x}_{j}}{T}\right)
		^{1/2}}{\frac{\boldsymbol{x}_{i}^{\prime}\boldsymbol{x}_{i}}{T}}\right)
	+\left(  \frac{1}{\sigma_{i}\left(  \frac{\boldsymbol{x}_{i}^{\prime
			}\boldsymbol{x}_{i}}{T}\right)  ^{1/2}}\right)  \right]  \left(
	\frac{\boldsymbol{x}_{i}^{\prime}\boldsymbol{\varepsilon}_{i}}{\sqrt{T}}\right)
	\right]  \\
	& =\frac{1}{\sqrt{N}}\sum_{i}w_{Ti}=\frac{1}{\sqrt{N}}\sum_{i}\left(
	w_{Ti}-E\left(  w_{Ti}\right)  \right)  +\frac{1}{\sqrt{N}}\sum_{i}E\left(
	w_{Ti}\right).
	\end{align}
	Another statistic is
	\begin{align}
	\frac{1}{\sqrt{N}}\left(\sum_{i}t_{i}(\widehat{\beta}_{MG})^{2}-1\right)=\frac{1}{\sqrt{N}}\sum
	_{i}\left(  t_{i}(\widehat{\beta}_{MG})^{2}-t_{i}(\beta^{0})^{2}\right)  +\frac
	{1}{\sqrt{N}}\sum_{i}t_{i}(\beta^{0})^{2}-1
	\end{align}
	Focusing on%
	\begin{align}
	&\frac{1}{\sqrt{N}}\sum_{i}\left(  t_{i}(\widehat{\beta}_{MG})^{2}-t_{i}(\beta^{0}%
	)^{2}\right)  \nonumber\\
	&  =\frac{1}{\sqrt{N}}\sum_{i}\left(  t_{i}(\widehat{\beta}_{MG}%
	)^{2}-t_{i}(\widehat{\beta}_{MG})t_{i}(\beta^{0})+t_{i}(\widehat{\beta}_{MG})t_{i}(\beta
	^{0})-t_{i}(\beta^{0})^{2}\right)   \nonumber
	\\
	&=  \frac{1}{\sqrt{N}}\sum_{i}\left(  t_{i}(\widehat{\beta}_{MG})^{2}-t_{i}(\widehat{\beta
	})t_{i}(\beta^{0})\right)  +\frac{1}{\sqrt{N}}\sum_{i}\left(  t_{i}(\widehat
	{\beta})t_{i}(\beta^{0})-t_{i}(\beta^{0})^{2}\right) \nonumber  \\
	&  =\frac{1}{\sqrt{N}}\sum_{i}t_{i}(\widehat{\beta}_{MG})\left(  t_{i}(\widehat{\beta
	}_{MG})-t_{i}(\beta^{0})\right)\nonumber\\
	&  +\frac{1}{\sqrt{N}}\sum_{i}t_{i}(\beta^{0})\left(
	t_{i}(\widehat{\beta}_{MG})-t_{i}(\beta^{0})\right)  \nonumber \\
	&  =\frac{1}{\sqrt{N}}\sum_{i}\left(  t_{i}(\widehat{\beta}_{MG})-t_{i}(\beta
	^{0})\right)  ^{2}+\frac{2}{\sqrt{N}}\sum_{i}t_{i}(\beta^{0})\left(
	t_{i}(\widehat{\beta}_{MG})-t_{i}(\beta^{0})\right)\label{analysis1}
	\end{align}
	By the analysis in \eqref{analysis1}%
	\begin{align}
	\frac{1}{\sqrt{N}}\sum_{i}\left(  t_{i}(\widehat{\beta}_{MG})-t_{i}(\beta^{0})\right)
	^{2}=o_{p}\left(  1\right).
	\end{align}
	Then, for $  1\leq i\leq N$ we write%
	\begin{align}
	&  \frac{1}{\sqrt{N}}\sum_{i}t_{i}(\beta^{0})\left(  t_{i}(\widehat{\beta}_{MG}%
	)-t_{i}(\beta^{0})\right) \nonumber \\
	&  =\frac{1}{\sqrt{N}}\sum_{i}\left(  \frac{\boldsymbol{1}}{\widehat{\sigma}%
		_{i}(\beta^{0})\left(  \boldsymbol{x}_{i}^{\prime}\boldsymbol{x}_{i}\right)
		^{1/2}}\right)  \boldsymbol{\varepsilon}_{i}\boldsymbol{(}\beta^{0}\boldsymbol{)}%
	^{\prime}\boldsymbol{x}_{i}\left(  t_{i}(\widehat{\beta}_{MG})-t_{i}(\beta^{0})\right)\nonumber
	\\
	&  =\frac{1}{\sqrt{N}}\sum_{i}\left(  \frac{\boldsymbol{1}}{\widehat{\sigma}%
		_{i}(\beta^{0})\left(  \boldsymbol{x}_{i}^{\prime}\boldsymbol{x}_{i}\right)
		^{1/2}}\right)  \boldsymbol{\varepsilon}_{i}\boldsymbol{(}\beta^{0}\boldsymbol{)}%
	^{\prime}\boldsymbol{x}_{i}\frac{\left(  \boldsymbol{\varepsilon}_{i}\boldsymbol{(}%
		\widehat{\beta}_{MG}\boldsymbol{)}-\boldsymbol{\varepsilon}_{i}\boldsymbol{(}\beta^{0}%
		\boldsymbol{)}\right)  ^{\prime}\boldsymbol{x}_{i}}{\left(  \boldsymbol{x}%
		_{i}^{\prime}\boldsymbol{x}_{i}\right)  ^{1/2}}\nonumber\\
	&  =\frac{\left(  \widehat{\beta}_{MG}-\beta^{0}\right)  }{\sqrt{N}}\sum_{i}%
	\boldsymbol{\varepsilon}_{i}\boldsymbol{(}\beta^{0}\boldsymbol{)}^{\prime}%
	\boldsymbol{x}_{i}\frac{\boldsymbol{x}_{i}^{\prime}\boldsymbol{x}_{i}}%
	{\widehat{\sigma}_{i}(\beta^{0})\left(  \boldsymbol{x}_{i}^{\prime}\boldsymbol{x}%
		_{i}\right)  }\nonumber\\
	&  =\frac{\left(  \widehat{\beta}_{MG}-\beta^{0}\right)  }{\sqrt{N}}\sum_{i}%
	\frac{\boldsymbol{\varepsilon}_{i}\boldsymbol{(}\beta^{0}\boldsymbol{)}^{\prime
		}\boldsymbol{x}_{i}}{\widehat{\sigma}_{i}(\beta^{0})}\nonumber\\
	&  =\sqrt{NT}\left(  \widehat{\beta}_{MG}-\beta^{0}\right)  \left(  \frac{1}{N}\sum
	_{i}\frac{\boldsymbol{\varepsilon}_{i}\boldsymbol{(}\beta^{0}\boldsymbol{)}^{\prime
		}\boldsymbol{x}_{i}}{\sqrt{T}\widehat{\sigma}_{i}(\beta^{0})}\right)
	=o_{p}\left(  1\right)
	\end{align}
	So
	\begin{align}
	\frac{1}{\sqrt{N}}\sum_{i}\left(t_{i}(\widehat{\beta}_{MG})^{2}-1\right)&=\frac{1}{\sqrt{N}}\sum
	_{i}\left(  t_{i}(\widehat{\beta}_{MG})^{2}-t_{i}(\beta^{0})^{2}\right)  +\frac
	{1}{\sqrt{N}}\sum_{i}t_{i}(\beta^{0})^{2}-1\nonumber\\
	&=\frac{1}{\sqrt{N}}\sum_{i}%
	t_{i}(\beta^{0})^{2}-1+o_{p}\left(  1\right)
	\end{align}
	Then,  for $i\neq j,  1\leq j\leq N$ we write%
	\begin{align}
	\frac{1}{\sqrt{N}}\sum_{i}(t_{i}(\beta^{0})^{2}-1)&=\frac{1}{\sqrt{N}}\sum
	_{i}t_{i}(\beta^{0})^{2}-E\left[  t_{i}(\beta^{0})^{2}\right]\nonumber  \\
	&+\frac{1}%
	{\sqrt{N}}\sum_{i}(E\left[  t_{i}(\beta^{0})^{2}\right]  -1)=A_{1}+A_{2}.
	\end{align}
	Then, each $t_i(\cdot)$ is independent from $t_j(\cdot), \; i\neq j, \; i,j=1,\ldots, N$, and the processes implicated in $t_i(\cdot) $ obey Assumption \ref{AssonX}.  Then by Lemma \ref{KSS}, $A_1$   is asymptotically normally distributed, while using the Laplace
	approximation method of \cite{lieberman1994laplace},
	$
	A_{2}=O\left(  {\sqrt{N}}/{T}\right)  .
	$
	As long as ${\sqrt{N}}/{T}=o(1)$, then $A_2$ is asymptotically negligible and so  $N^{-1/2}\sum
	_{i}^N(t_{i}(\beta^{0})^{2}-1)$ is asymptotically normally distributed.
	For both asymptotically normally distributed statistics the asymptotic
	variance can be consistently estimated by $(N^{-1})\sum_{i}t_{i}(\widehat
	{\beta})^{2}$ and $(N^{-1/2})\sum_{i}\left(  t_{i}(\widehat{\beta}_{MG}%
	)^{2}-1\right)  ^{2}$, respectively.
	\newline
	\indent The next step is to show that the denominator in the first statement  of \eqref{op1}:
	\begin{align}
	\widehat\sigma_{i}-\sigma_{i}
	&= \frac{1}{T}\sum_{i} \left(\boldsymbol{y}_i - \widehat\beta_{MG}\boldsymbol{x}_{i}\right)\left(\boldsymbol{y}_i - \widehat\beta_{MG}\boldsymbol{x}_{i}\right)' -  E\left[ \frac{1}{T} \sum_{i} \left(\boldsymbol{y}_i - \beta^0\boldsymbol{x}_{i}\right)\left(\boldsymbol{y}_i - \beta^0\boldsymbol{x}_{i}\right)'  \right]\nonumber\\
	&= \frac{1}{T}\sum_{i} \left( ( \beta^0- \widehat\beta_{MG})\boldsymbol{x}_{i} +\boldsymbol{\varepsilon}_i  \right)\left( ( \beta^0- \widehat\beta_{MG})\boldsymbol{x}_{i} +\boldsymbol{\varepsilon}_i  \right)'
	\\
	&	-  E\left[ \frac{1}{T} \sum_{i} \left(\boldsymbol{y}_i - \beta^0\boldsymbol{x}_{i}\right)\left(\boldsymbol{y}_i - \beta^0\boldsymbol{x}_{i}\right)'  \right]\label{var1}.
	\end{align}
	But  $( \beta^0- \widehat\beta_{MG}) = O_P((NT)^{-1/2})$, and
	\begin{align}
	\Var\left[T^{-1}   \left(\boldsymbol{x}_{i}'\boldsymbol{\varepsilon}_{i}	\right)\right]&	= E\left(T^{-2} \boldsymbol{x}_{i}^{\prime} \boldsymbol{\varepsilon}_i \boldsymbol{\varepsilon}_i^{\prime} \boldsymbol{x}_{i}\right) =E\left[T^{-2} E\left(\boldsymbol{x}_{i}^{\prime} \boldsymbol{\varepsilon}_i \boldsymbol{\varepsilon}_i^{\prime} \boldsymbol{x}_{i} \mid \boldsymbol{x}_i\right)\right]=\sigma_{\varepsilon,i}^2 T^{-1} E\left(\boldsymbol{x}_{i}^{\prime} \boldsymbol{x}_{i}\right) .\nonumber
	\end{align}
	But $T^{-1} \boldsymbol{x}_{i}'\boldsymbol{x}_{i}\to_P \lim_{T\to \infty } E( \boldsymbol{x}_{i}'\boldsymbol{x}_{i}) = \boldsymbol{	\Sigma}_{xx,i} $, so then $\Var[T^{-1}   \boldsymbol{x}_{i}'\boldsymbol{\varepsilon}_{i}  ]\to 0, $ as $T\to \infty$.  Therefore, $T^{-1}   \boldsymbol{x}_{i}'\boldsymbol{\varepsilon}_{i}   \to_P 0 $.  So then, it becomes clear that  $\widehat\sigma_{i}\to_{P} \sigma_{i} $ as $T\to \infty.$
	\newline We continue by showing the second statement of \eqref{Statement1_test}. Let
	\begin{align}
	R^{\ast}_N = N^{-1/2} \frac{\sum_{i=1}^N Tr_{i}^2(\beta) -1 }{\sigma^2_{i, R}},\label{stR}
	\end{align}
	where $r_{i}^2(\beta) =\widehat\gamma(\beta^0) \boldsymbol{x}_i ' \boldsymbol{x}_i \widehat\gamma(\beta^0)  (\boldsymbol{\widehat{v}}_{i}(\beta^0)\boldsymbol{\widehat{v}}_{i}(\beta^0) ') $, and
	$ \sigma^2_{i, R} = E[T^{-1}( \boldsymbol{y}_{i}-\beta\boldsymbol{x}%
	_{i})'( \boldsymbol{y}_{i}-\beta\boldsymbol{x}%
	_{i})]$.
	It is sufficient to show that
	\begin{align}
	R^{\ast}_N \sim N(0,1)  \text{ and } \widehat{R}_N - R^{\ast}_N =o_P(1).
	\end{align}
	We first analyse the nominator of $\widehat{R}_N - R^{\ast}_N$:
	\begin{align}
	\frac{1}{\sqrt{N}} \sum_{i} \left(r_{i}(\widehat{\beta}) - r_{i}(\beta^0) \right) + \frac{1}{\sqrt{N}}\sum_{i}r_{i}(\beta^0)=R_{i,1}+R_{i,2},
	\end{align}
	where $\widehat{\beta}$ is an estimator of $\beta^0$.  Looking at the first term we have that
	\begin{align}
	R_{i,1} &= \frac{1}{\sqrt{N}} \sum_{i}   \left [     \widehat\gamma(\widehat\beta) (\boldsymbol{x}_i ' \boldsymbol{x}_i )^{1/2} \widehat\sigma_{i,r}(\beta) ^{-1} (\boldsymbol{\widehat{v}}_{i}(\widehat{\beta})'\boldsymbol{\widehat{v}}_{i}(\widehat\beta))^{1/2}  \right.  \nonumber\\
	&\quad\left. - \widehat\gamma(\beta^0)   (\boldsymbol{x}_i ' \boldsymbol{x}_i )^{1/2} \sigma_{i,r}(\beta)^{-1}(\boldsymbol{{v}}_{i}(\beta^0)\boldsymbol{{v}}_{i}(\beta^0)')^{1/2}   \right] \nonumber\\
	&=  \frac{1}{\sqrt{N}} \sum_{i}    (\boldsymbol{x}_i ' \boldsymbol{x}_i )^{1/2}
	\left[		  (\boldsymbol{\widehat{v}}_{i}(\widehat{\beta})\boldsymbol{\widehat{v}}_{i}(\widehat\beta)')^{1/2}-  (\boldsymbol{{v}}_{i}(\beta^0)\boldsymbol{{v}}_{i}(\beta^0)')^{1/2}
	\right]
	( \widehat\gamma(\widehat\beta) - \widehat\gamma(\beta^0))  
	\end{align}
	where $ \widehat\sigma_{i,r}(\beta) = T^{-1}\boldsymbol{\widehat{v}}_{i}(\widehat{\beta})\boldsymbol{\widehat{v}}_{i}(\widehat\beta) $ and $\widehat\sigma_{i,r}(\beta) -\sigma_{i,r}(\beta) =o_P(1)$.  But, by \eqref{xxi}, ${{N}^{-1/2}} \sum_{i}    (\boldsymbol{x}_i ' \boldsymbol{x}_i )^{1/2} =\sqrt{NT} ( {{N}^{-1}} \sum_{i}    (T^{-1}\boldsymbol{x}_i ' \boldsymbol{x}_i )^{1/2} )$, so then it becomes clear that  $({{N}^{-1}} \sum_{i}    (T^{-1}\boldsymbol{x}_i ' \boldsymbol{x}_i )^{1/2} =O_P(1)$. Further, consider that $\widehat{\beta}\to_{P}\beta^{0}$, so that  $\widehat{\beta}-\beta^{0}=$ $O_{p}(  (  \sqrt{NT})
	^{-1})  $, so the following holds
	\begin{align}
	R_{i,1}=O_{p}(1).
	\end{align}
	We focus on the  MG
	estimator of $\beta^{0}$.  Then for $1\leq i\leq N$ we write that
	\begin{align}
	\widehat{\beta}_{MG}-\beta^{0}=\frac{1}{N}\sum_{i}\left(  \frac{\boldsymbol{y}%
		_{i}^{\prime}\boldsymbol{x}_{i}}{\boldsymbol{x}_{i}^{\prime}\boldsymbol{x}%
		_{i}}-\beta^{0}\right)  =\frac{1}{N}\sum_{i}\frac{\boldsymbol{x}_{i}^{\prime
		}\boldsymbol{\varepsilon}_{i}}{\boldsymbol{x}_{i}^{\prime}\boldsymbol{x}_{i}}%
	\end{align}
	So, similarly to the analysis in  \eqref{ss}--\eqref{sse},
	\begin{align}
	\frac{1}{\sqrt{N}} \sum_{i} \left(r_{i}(\widehat{\beta}_{MG}) - r_{i}(\beta^0) \right)\rightarrow N(0, \sigma_{x/\varepsilon,r}^2\sigma_{\varepsilon/x,r}^2),
	\end{align}
	where $\sigma_{x/\varepsilon,r}^2\sigma_{\varepsilon/x,r}^2$ can be  defined similarly to $\sigma_{x/\varepsilon}^2\sigma_{\varepsilon/x}^2$.
	Further, note that for $1\leq i\leq N$
	\begin{align}
	\frac{1}{\sqrt{N}} \sum_{i} r_{i}(\beta^0) = \frac{1}{\sqrt{N}} \sum_{i}   \widehat\gamma(\beta^0)   (\boldsymbol{x}_i ' \boldsymbol{x}_i )^{1/2}\frac{ (\boldsymbol{{v}}_{i}(\beta^0)\boldsymbol{{v}}_{i}(\beta^0)')^{1/2}}{\sigma_{i,r}(\beta) }.
	\end{align}
	Then,
	\begin{align}
	&\frac{1}{\sqrt{N}} \sum_{i}[r_{i}(\widehat\beta_{MG}) -  r_{i}(\beta^0) ]+\frac{1}{\sqrt{N}} \sum_{i} r_{i}(\beta^0)\nonumber\\
	&= \frac{1}{\sqrt{N}} \sum_{i}   \widehat\gamma(\widehat\beta_{MG})   (\boldsymbol{x}_i ' \boldsymbol{x}_i )^{1/2}\frac{ (\boldsymbol{{v}}_{i}(\widehat\beta_{MG})\boldsymbol{{v}}_{i}(\widehat\beta_{MG})')^{1/2}}{\sigma_{i,r}(\widehat\beta_{MG}) }\nonumber\\
	&\quad - \frac{1}{\sqrt{N}} \sum_{i}   \widehat\gamma(\beta^0)   (\boldsymbol{x}_i ' \boldsymbol{x}_i )^{1/2}\frac{ (\boldsymbol{{v}}_{i}(\beta^0)\boldsymbol{{v}}_{i}(\beta^0)')^{1/2}}{\sigma_{i,r}(\beta) }\nonumber\\
	&\quad +\frac{1}{\sqrt{N}} \sum_{i}   \widehat\gamma(\beta^0)   (\boldsymbol{x}_i ' \boldsymbol{x}_i )^{1/2}\frac{ (\boldsymbol{{v}}_{i}(\beta^0)\boldsymbol{{v}}_{i}(\beta^0)')^{1/2}}{\sigma_{i,r}(\beta) }\nonumber\\
	&=\frac{1}{\sqrt{N}} \sum_{i}   \widehat\gamma(\widehat\beta_{MG})    (\boldsymbol{x}_i ' \boldsymbol{x}_i )^{1/2} \left[	(\boldsymbol{{v}}_{i}(\widehat\beta_{MG})\boldsymbol{{v}}_{i}(\widehat\beta_{MG})')^{1/2}-  (\boldsymbol{{v}}_{i}(\beta^0)\boldsymbol{{v}}_{i}(\beta^0)')^{1/2} 	\right] \nonumber \\
	&\quad + \frac{1}{\sqrt{N}} \sum_{i}\left[  \widehat\gamma(\widehat\beta_{MG})  -  \widehat\gamma(\beta^0) 	\right]    (\boldsymbol{x}_i ' \boldsymbol{x}_i )^{1/2}(\boldsymbol{{v}}_{i}(\beta^0)\boldsymbol{{v}}_{i}(\beta^0)')^{1/2} \nonumber\\
	&= A_{T,i, 1}+A_{T,i, 2}+A_{T, i,3}.
	\end{align}
	where $A_{T, i,3}=\frac{1}{\sqrt{N}} \sum_{i} r_{i}(\beta^0) . $ But $   \widehat\gamma(\widehat\beta_{MG}) = \left(  \frac{1}{N}\sum_{j}\frac
	{\boldsymbol{x}_{j}^{\prime}\boldsymbol{\varepsilon}_{j}}{\boldsymbol{x}_{j}^{\prime
		}\boldsymbol{x}_{j}}\right) $ and
	$\boldsymbol{{v}}_{i}(\widehat\beta_{MG}) = (\beta^0 - \widehat\beta_{MG}) \boldsymbol{x}_i +\boldsymbol\varepsilon_{i}$, so then to aid the analysis, $A_{T,i, 1} $ becomes:
	{\small \begin{align}
		A_{T,i, 1}& = \frac{1}{\sqrt{N}} \sum_{i}   \left(  \frac{1}{N}\sum_{j}\frac
		{\boldsymbol{x}_{j}^{\prime}\boldsymbol{\varepsilon}_{j}}{\boldsymbol{x}_{j}^{\prime}\boldsymbol{x}_{j}}\right) (\boldsymbol{x}_i ' \boldsymbol{x}_i )^{1/2}\left[	 (\beta^0 - \widehat\beta_{MG}) \boldsymbol{x}_i \right]^{-1/2}  \nonumber  \\
		&=  \frac{1}{\sqrt{N}} \sum_{i}   \left(  \frac{1}{N}\sum_{j}\frac
		{T^{-1/2}\boldsymbol{x}_{j}^{\prime}\boldsymbol{\varepsilon}_{j}}{T^{-1}\boldsymbol{x}_{j}^{\prime}\boldsymbol{x}_{j}}\right) \left(\frac{1}{\sqrt{N}}\sum_{i} T^{-1}\boldsymbol{x}_i ' \boldsymbol{x}_i \right)^{1/2}\left[\frac{1}{\sqrt{N}}\sum_{i} 	T^{-1/2} (\beta^0 - \widehat\beta_{MG}) \boldsymbol{x}_i \right]^{-1/2}   .
		\end{align}}
	We then
	consider the terms  $A_{T,i, 1}+A_{T,i, 3}$
	\begin{align}
	w_{T,R,i}=\left[  \left(  \frac{\frac{1}{N}\sum_{j}\left(  {T}^{-1}{\boldsymbol{x}%
			_{j}^{\prime}\boldsymbol{x}_{j}}\right)  ^{1/2}}{T^{-1}{\boldsymbol{x}%
			_{i}^{\prime}\boldsymbol{x}_{i}}}\right)  +\left(  {\sigma
		_{i,r}^{-1}\left(  {T}^{-1}{\boldsymbol{x}_{i}^{\prime}\boldsymbol{x}_{i}}\right)
		^{1/2}}\right)  \right]  \left(  \frac{\boldsymbol{x}_{i}^{\prime
		}\boldsymbol{\varepsilon}_{i}}{\sqrt{T}}\right)  ,\label{at1}
	\end{align}
	so we have
	\begin{align}
	& \frac{1}{\sqrt{N}}\sum_{i}\left[  \left[  \left(  \frac{\frac{1}{N}\sum
		_{j}\left(  T^{-1}{\boldsymbol{x}_{j}^{\prime}\boldsymbol{x}_{j}}\right)
		^{1/2}}{T^{-1}{\boldsymbol{x}_{i}^{\prime}\boldsymbol{x}_{i}}}\right)
	+\left( {\sigma_{i,r}\left(  T^{-1}{\boldsymbol{x}_{i}^{\prime
			}\boldsymbol{x}_{i}}\right)  ^{-1/2}}\right)  \right]  \left(
	T^{-1/2}{\boldsymbol{x}_{i}^{\prime}\boldsymbol{\varepsilon}_{i}}\right)
	\right]  \\
	& =\frac{1}{\sqrt{N}}\sum_{i}w_{T,R,i}=\frac{1}{\sqrt{N}}\sum_{i}\left(
	w_{T,R,i}-E\left(  w_{T,R,i}\right)  \right)  +\frac{1}{\sqrt{N}}\sum_{i}E\left(
	w_{T,R,i}\right).
	\end{align}
	Lastly, by considering  Assumptions \ref{AssonX} and the proof of statement \eqref{s222} of Lemma \ref{KSS}, and similarly to \eqref{ssig}, it can be easily shown that
	\begin{align}
	&\frac{1}{\sqrt{N}}\sum_{i}\left[  \left( {\widehat{\sigma}_{i}(\beta
		^{0})\left(  {T}^{-1}{\boldsymbol{x}_{i}^{\prime}\boldsymbol{x}_{i}}\right)
		^{-1/2}}\right)  \left(  T^{-1/2}{\boldsymbol{x}_{i}^{\prime}\boldsymbol{\varepsilon}_{i}%
	}\right)  \right] \nonumber\\
	& -\frac{1}{\sqrt{N}}\sum_{i}\left[  \left(
	{\sigma_{i}\left(  T^{-1}{\boldsymbol{x}_{i}^{\prime}\boldsymbol{x}%
			_{i}}\right)  ^{-1/2}}\right)  \left(  T^{-1/2}{\boldsymbol{x}_{i}^{\prime
		}\boldsymbol{\varepsilon}_{i}}\right)  \right]  =o_{p}\left(  1\right).\label{ssig_R}
	\end{align}
	We further, focus on
	$\frac{1}{\sqrt{N}}  {\sum_{i=1}^N Tr_{i}^2(\widehat\beta_{MG}) -1 }	$, so
	\begin{align}
	\frac{1}{\sqrt{N}} {\sum_{i=1}^N r_{i}^2(\widehat\beta_{MG})  - 1}&= \left(\frac{1}{\sqrt{N}}  {\sum_{i=1}^N Tr_{i}^2(\widehat\beta_{MG})  }	 -\frac{1}{\sqrt{N}} {\sum_{i=1}^N Tr_{i}^2(\beta^{0})  }\right)	+\frac{1}{\sqrt{N}}  {\sum_{i=1}^N Tr_{i}^2(\beta^{0}) -1 }\nonumber	 \\
	&= B_{T,i,1}+B_{T,i,2}.
	\end{align}
	Following a similar analysis with \eqref{analysis1}, it  can be easily shown that  $B_{T,i,1}=o_P(1)$. So then
	\begin{align}
	\frac{1}{\sqrt{N}}  {\sum_{i=1}^N Tr_{i}^2(\widehat\beta_{MG})  - 1} &=\frac{1}{\sqrt{N}}  \sum_{i=1}^N \left[ Tr_{i}^2(\widehat\beta_{MG})  -T  r_{i}^2(\beta^{0}) \right] +	\frac{1}{\sqrt{N}} \sum_{i=1}^N	\left[ Tr_{i}^2(\beta^{0})  - 1\right] \nonumber\\
	&\underset{B_{T,i,1}=o_P(1)}{=} o_P(1) + N^{-1/2} \sum_{i=1}^N	\left[ Tr_{i}^2(\beta^{0})  - 1\right],
	\end{align}
	then considering $B_{T,i,2}$, we have
	\begin{align}
	\frac{1}{\sqrt{N}}  \sum_{i=1}^N	\left[ Tr_{i}^2(\beta^{0})  - 1\right] &=  \frac{1}{\sqrt{N}}  \sum_{i=1}^N	\left ( Tr_{i}^2(\beta^{0})   - E\left[Tr_{i}^2(\beta^{0})  \right] \right)\nonumber \\
	&+  \frac{1}{\sqrt{N}}  \sum_{i=1}^N	 \left( E\left[Tr_{i}^2(\beta^{0})  \right] - 1 \right)  = b_{T,i,1}+b_{T,i,2}.
	\end{align}
	By Lemma \ref{KSS}, $b_{T,i,1}\sim N(0,1)$, and using the Laplace approximation method of \cite{lieberman1994laplace}, $b_{T,i,2}=o(1)$. Then the LHS, $T^{-1}N^{-1/2} \sum_{i=1}^N	\left[ Tr_{i}^2(\beta^{0})  - 1\right] \sim N(0,1)$. The asymptotic variance of $ N^{-1/2}\sum_{i=1}^{N}\left[ Tr_{i}^2(\beta^{0})  - 1\right] $ can be consistently estimated by $N^{-1/2} \sum_{i=1}^N [Tr_{i}^2(\widehat\beta_{MG}) - 1]$ and similarly, the asymptotic variance of $ N^{-1}\sum_{i=1}^{N}\left[ Tr_{i}^2(\beta^{0})  \right] $ can be consistently estimated by $N^{-1} \sum_{i=1}^N [Tr_{i}^2(\widehat\beta_{MG}) ]$.  The remainder of the analysis  focuses on the variance estimation of the test statistic in \eqref{stR}, and follows from the analysis in  \eqref{var1}, therefore it is omitted.
	The result of the first statement of the Theorem follows, concluding the proof of  \eqref{Statement1_test}.\qed
	
	\subsection{Auxiliary Lemmas}
	
	\label{Appendix_2} In this section we provide technical Lemmas useful for the
	proofs of the main results.
	\begin{lemma}
		\label{KSS} Let $w_{i, T}$ and $\mu_{i, T}$ for $i=1, \ldots, N$ and $T=1,
		\ldots$, be arrays of vectors of random variables and constants such that
		$w_{i, T}-\mu_{i, T}$ is a spatial martingale difference array with $E\left[
		\left( w_{i, T}-\mu_{i, T}\right) \left( w_{i, T}-\mu_{i, T}\right) ^{\prime
		}\right] =\Sigma_{i, T}$, and $\sup_{i, T} E\left\| w_{i, T}\right\|
		^{2+\delta}<\infty$ for some $\delta>0$. Assume that $\Sigma=\lim_{N, T
			\rightarrow\infty} N^{-1} \sum_{i=1}^{N} \Sigma_{i, T}$ is positive definite
		and $\sup_{N, T} N^{-1} \sum_{i=1}^{N} \Sigma_{i, T}<\infty$. Then, as $N
		\rightarrow\infty$,
		\[
		N^{-1 / 2} \sum_{i=1}^{N}\left( w_{i, T}-\mu_{i, T}\right)  \rightarrow_{d}
		N(0, \Sigma) .
		\]
	\end{lemma}

	\begin{lemma}
		\label{betaAUX} Let $\boldsymbol{\widehat{S}}_{i}=T^{-1}\sum_{t=1}^{T}-(y_{it}
		- {\boldsymbol{ {\beta}}_{i}^{0}}^{\prime}\boldsymbol{ x}_{it})\boldsymbol{
			x}_{it}$, and $\boldsymbol{S}_{i}=T^{-1}\sum_{t=1}^{T}-(Ey_{it} -
		{\boldsymbol{ {\beta}}_{i}^{0}}^{\prime}E\boldsymbol{ x}_{it})E\boldsymbol{
			x}_{it}$, then for each $i=1,\ldots, N,$
		
		\begin{enumerate}[nolistsep]

			\item $\boldsymbol{S}_{i}= O_{P}(T^{-1/2})$, \label{AUX1}
			
			\item $\boldsymbol{\widehat{S}}_{i}-\boldsymbol{S}_{i} = O_{P}(T^{-1/2})$,
			\label{AUX2}
			
			\item $2N^{-1}\Vert\boldsymbol{\widehat{S}}_{i} \Vert_{2} ^{2} =O_{P}%
			(T^{-1}).$\label{AUX3}
		\end{enumerate}
	\end{lemma}
	
	\begin{lemma}
		\label{Aux1} Let Assumptions \ref{AssonX}--\ref{Assonalpha}, and Assumption
		\ref{clust} hold, then for $\boldsymbol{G}=\{ G_{1}, \ldots, G_{K} \}$ and
		$\boldsymbol{ {\alpha}}=(\boldsymbol{ {\alpha}}_{1}, \ldots, \boldsymbol{
			{\alpha}}_{K} )$, and $\boldsymbol\eta_{i} $ in a group $\boldsymbol G_{k}$, and for $\epsilon>0$ we
		have that
		{\small \begin{align}
			\lim_{N,T\to\infty}\sup_{i} P\left(  \sup_{i\in G_{k} } \left\Vert {Q}_N(
			\boldsymbol{G}, \boldsymbol{\alpha}) - \frac{1}{N} \sum_{i=1}^{N}
			\boldsymbol{\psi}_{i} ^{\prime}\boldsymbol{\psi}_{i}- \frac{1}{N_k} \sum
			_{k=1}^{K} \left(  \boldsymbol{ {\alpha}}_{k}^{0} - \boldsymbol{ {\alpha}}_{k}
			\right) ^{2} \right\Vert _{1} > C_{0} \left( \frac{\eta_{i}}{\sqrt{N}}\right)
			^{2} \right)  \leq\epsilon,\label{B.1}%
			\end{align}}
		where $\boldsymbol{\psi}_{i}=\left(  \boldsymbol{x}_{i}^{\prime}%
		\boldsymbol{x}_{i}\right)  ^{-1}\boldsymbol{x}_{i}^{\prime}%
		\boldsymbol{\varepsilon}_{i}$, and $i=1,\ldots,N$.
	\end{lemma}
	
	\begin{lemma}
		\label{LemmaB.2} Suppose Assumption \ref{AssonX}--\ref{Assonalpha}, and
		Assumption \ref{clust} hold, then for $\boldsymbol{G}=\{ G_{1}, \ldots, G_{K}
		\}$ and $\boldsymbol{ {\alpha}}=(\boldsymbol{ {\alpha}}_{1}, \ldots,
		\boldsymbol{ {\alpha}}_{ K} )$, and $\eta_{i} $ in a group $k$, and for
		$\epsilon>0$ we have that
		\begin{align}
		\sup_{i, i\in G_k} P\left(  \frac{1}{N_{k}}\sum_{i}\left(  \boldsymbol{\alpha}_{k}^{0}-
		\boldsymbol{\widetilde{\alpha}} _{k}\right)  > C \left( \frac{\eta
			_{i}}{\sqrt{N_{k}}}\right)  \right)  \leq\epsilon,
		\end{align}
		where $\boldsymbol{\widetilde{\alpha}} $ is the solution to \eqref{A},
		for $\epsilon>0$ a small and finite constant, $C>0$.
	\end{lemma}
	
	\subsection{Proofs of auxiliary Lemmas}\label{AUX_proofs}
	\subsubsection*{Proof of Lemma \ref{KSS}}
	By Theorem 12.11 of \cite{davidson1994stochastic}, if $\sup _{i, T} E\left\|w_{i, T}\right\|^{2+\delta}<\infty$, then we obtain the uniform integrability condition:
	\begin{align}
	\lim _{M \rightarrow \infty} \sup _{i, T} E\left(\left\|w_{i, T}-\mu_{i, T}\right\| I_{\left\{\left\|w_{i, T}-\mu_{i, T}\right\|>M\right\}}\right)=0\label{s111}
	\end{align}
	Together with $\sup _{N, T} N^{-1} \sum_{i=1}^N \Sigma_{i, T}<\infty$, this implies that the Lindeberg condition holds by Theorem 23.18 of \cite{davidson1994stochastic}. Then, by Theorem 23.16 of \cite{davidson1994stochastic}, it follows that
	\begin{align}
	\max _{i, T} N^{-1}\left(w_{i, T}-\mu_{i, T}\right) \rightarrow_p 0\label{s222}
	\end{align}
	Together with $\sup _{i, T} E\left\|w_{i, T}\right\|^{2+\delta}<\infty$, \eqref{s222} implies \eqref{s111} by Theorem 24.3 of \cite{davidson1994stochastic}.
	Lemma \ref{KSS} can be used to derive rates, specifically is used in Theorem \ref{statisticsTheorem}, noting that if $c_T \rightarrow_p c$ and $d_T \rightarrow_d d$, then $c_T d_T \rightarrow_d c d$, where $d$ is a normal variate. Further, if $c=0$, then $c_T d_T \rightarrow_p 0$, implying that $c_T d_T=O_p\left(c_T\right)$.\qed

	\subsubsection*{Proof of Lemma \ref{betaAUX}}
	We show \eqref{AUX1}. 	Let $U_i(y_{it}, \boldsymbol{x}_{it}; \boldsymbol{	{\beta}}^{0}) =-(y_{it} - {\boldsymbol{	{\beta}}_i^{0}}'\boldsymbol{	x}_{it})\boldsymbol{	x}_{it}$.  To show that $\boldsymbol{S}_{i}=T^{-1}\sum_{t=1}^T U_i(y_{it}, \boldsymbol{x}_{it}; \boldsymbol{	{\beta}}^{0}) = O_P(T^{-1/2})$, it is   sufficient to show that $\mathrm{Var}(\boldsymbol{\iota}_p ' \boldsymbol{S}_i) = O_P(T^{-1})$, where $\boldsymbol{\iota}_p $ is a $p\times 1$ non random and arbitrary vector with property $\|\boldsymbol{\iota}_p   \|_2=1$.  Note that $E(U_i(y_{it}, \boldsymbol{x}_{it}; \boldsymbol{	{\beta}}^{0})) = 0 $, and
	\begin{align}
	\mathrm{Var}(\boldsymbol{\iota}_p ' \boldsymbol{S}_i) &= \left(\frac{1}{T}\right)^2 \sum_{t=1}^{T}\sum_{s=1}^{T} \mathrm{Cov}\left[	\boldsymbol{\iota}_p ' T^{-1}\sum_{t=1}^T  U_i(y_{it}, \boldsymbol{x}_{it}; \boldsymbol{	{\beta}}^{0}) , T^{-1}\sum_{s=1}^T  U_i(y_{is}, \boldsymbol{x}_{is}; \boldsymbol{	{\beta}}^{0})'\boldsymbol{\iota}_p \right] \\
	&\underset{(1)}{\leq}   8\max_{i, t}\left[E\left(		\left|		\boldsymbol{\iota}_p ' T^{-1}\sum_{t=1}^T
	U_i(y_{it}, \boldsymbol{x}_{it}; \boldsymbol{	{\beta}}^{0})	\right|	^{q}	\right)	\right]^{\frac{2}{q}}	\frac{1}{T^2} \sum_{t=1}^{T}\sum_{s=1}^{T} a(|	t-s 	|	)^{1-2/q}\\
	&\underset{(2)}{\leq} 8\max_{i, t}\left[E\left(		\left|		\boldsymbol{\iota}_p ' T^{-1}\sum_{t=1}^T  U_i(y_{it}, \boldsymbol{x}_{it}; \boldsymbol{	{\beta}}^{0})	\right|	^{q}	\right)	\right]^{\frac{2}{q}} o_P(1) = O_P(T^{-1}),
	\end{align}
	where $(1)$ results from Corollary A.2 of \cite{hall2014martingale}, and $(2)$ because there is no serial correlation in $\{\boldsymbol{x}_{it}\}, \; \{ y_{it} \}$. The results follows.
	We continue to show \eqref{AUX2}. 	Further, we show that $ \boldsymbol{\widehat{S}}_{i}-  \boldsymbol{S}_{i}  = O_P(T^{-1/2})$.
	Then, we have that
	\begin{align}
	\left\Vert 	  \boldsymbol{\widehat{S}}_{i}-\boldsymbol{S}_{i} \right\Vert _2
	&\leq \left\Vert \frac{1}{T} \sum_{i=1}^T (y_{it} - {\boldsymbol{	{\beta}}_i^{0}}'\boldsymbol{	x}_{it})\boldsymbol{	x}_{it} -   (Ey_{it} - {\boldsymbol{	{\beta}}_i^{0}}'E\boldsymbol{	x}_{it})E\boldsymbol{	x}_{it}  \right\Vert _2 .\label{eqTay}
	\end{align}
	By Assumption \ref{AssonX}, \eqref{eqTay} is $O_P(T^{-1/2})$, completing the proof of this statement.
	We show \eqref{AUX3}.  It is clear that
	\begin{align}
	2N^{-1}\Vert \boldsymbol{\widehat{S}}_{i} \Vert_2 ^2 \leq  2N^{-1}\Vert \boldsymbol{{S}}_{i} \Vert_2 ^2+2N^{-1}\Vert \boldsymbol{\widehat{S}}_{i} - \boldsymbol{{S}}_{i} \Vert_2 ^2 = O_P(T^{-1})  + O_P(T^{-1}).
	\end{align}
	The first term results from the Markov inequality and the proof of statement \eqref{AUX1}.  While for  the second term, we use the decomposition of statement \eqref{AUX2}, in statement \eqref{AUX1} and can readily show the result, competing the proof.\qed

	\subsubsection*{Proof of Lemma \ref{Aux1}}
	Following \eqref{A}  we can write:
	\begin{align}
	{Q}_{N}(G_k, \boldsymbol{\alpha}) =&\frac{1}{N}\sum_{i} \boldsymbol\eta_{i}'\boldsymbol\eta_{i}+\frac{1}{N}%
	\sum_{i}\left( \boldsymbol\alpha _{k}^{0}- \boldsymbol\alpha _{k}\right) ^{2}+ \frac{2}{NT}\sum_{i}\sum_{t}\varepsilon_{it}\left(  \boldsymbol\alpha _{k}^{0}- \boldsymbol\alpha _{k}\right)   \\
	&+\frac{1}{N}\sum_{i}\boldsymbol\psi _{i}'\boldsymbol\psi _{i}+\frac{2}{N}\sum_{i} \boldsymbol\eta_{i}'\boldsymbol\psi _{i}+ \frac{2}{N}\sum_{i}\boldsymbol\psi _{i}\left( \boldsymbol\alpha_{k} ^{0}- \boldsymbol\alpha _{k}\right)= \sum_{j=1}^{6}B_j,   \label{fkm2}
	\end{align}
	$\boldsymbol\psi _{i}=\left( \boldsymbol{x}_{i}^{\prime }\boldsymbol{x}_{i}\right) ^{-1}\boldsymbol{x}_{i}^{\prime }\boldsymbol{\varepsilon}_{i}$, and $i=1,\ldots,N$.
	Substituting \eqref{fkm2} in \eqref{B.1} we have
	\begin{align}
	&P \left(\left\Vert  \frac{1}{N}\sum_{i} \boldsymbol\eta_{i}'\boldsymbol\eta_{i}+\frac{2}{NT}\sum_{i}\sum_{t}\varepsilon_{it}\left(  \boldsymbol\alpha _{k}^{0}- \boldsymbol\alpha _{k}\right)+\frac{2}{N}\sum_{i} \boldsymbol \eta_{i}\boldsymbol\psi _{i}\right. \right. \\
	& \quad \quad \left.\left.+ \frac{2}{N}\sum_{i}\boldsymbol\psi _{i}\left( \boldsymbol\alpha _{ k} ^{0}- \boldsymbol\alpha _{ k}\right) \right\Vert _1 	 	> C_0  \left(\frac{ \boldsymbol \eta_{i} '\boldsymbol \eta_{i}}{{N}}\right)	\right ) \\
	&\leq P\left(   \left\Vert  \frac{1}{N}\sum_{i}\boldsymbol\eta_{i}' \boldsymbol \eta_{i} \right\Vert _1 >\frac{C_0}{4}\left(\frac{ \boldsymbol \eta_{i} '\boldsymbol \eta_{i}}{{N}}\right)\right) \\
	&+ P\left(\left\Vert \frac{2}{NT}\sum_{i}\sum_{t}\varepsilon_{it}\left(  \boldsymbol\alpha _{k}^{0}- \boldsymbol\alpha _{k}\right)  \right\Vert _1 >\frac{C_0}{4} \left(\frac{ \boldsymbol \eta_{i} '\boldsymbol \eta_{i}}{{N}}\right)\right)   \\
	&+P\left(   \left\Vert 	\frac{2}{NT}\sum_{i}\sum_{t} \boldsymbol\eta_{i}'\boldsymbol\psi _{i}	\right\Vert _1>\frac{C_0}{4}\left(\frac{ \boldsymbol \eta_{i} '\boldsymbol \eta_{i}}{{N}}\right)\right)\\
	&+ P\left(   \left\Vert   \frac{2}{N}\sum_{i}\boldsymbol\psi _{i}\left( \boldsymbol\alpha _{k} ^{0}- \boldsymbol\alpha _{k}\right)  	\right\Vert _1>\frac{C_0}{4} \left(\frac{ \boldsymbol \eta_{i} '\boldsymbol \eta_{i}}{{N}}\right)\right) = C_{(1)} +C_{(2)}+C_{(3)}+C_{(4)}.
	\end{align}
	By Assumption \ref{AssonX}--\ref{Assonalpha}, and Assumption \ref{clust}, term $C_{(1)}$, $C_{(3)}$ are $o(1)$.  Further, by the fact that ${Q}_{N}(\boldsymbol G_k, \boldsymbol{\alpha}) \leq {Q}_{N}(\boldsymbol G^0_k, \boldsymbol{\alpha}^0)$, $\left(  {\boldsymbol{\alpha} }_{k}^{0}- \boldsymbol{\alpha}_{k}\right) < C_\alpha$, for a small positive constant $C_\alpha$. Then, we have that
	\begin{align}
	C_{(2)}\leq P\left(\left| \frac{2}{NT}\sum_{i}\sum_t \varepsilon_{it}\right|\geq\frac{C_0}{C_{\alpha}4} \left(\frac{ \boldsymbol \eta_{i} '\boldsymbol \eta_{i}}{{N}}\right)\right) &{=}  P\left(\left| \frac{2}{T}\sum_{i}\sum_t \varepsilon_{it}\widehat{\sigma}^{-2}_{\eta, i}\right|\geq\frac{NC_0}{4C_{\alpha}} \right), \nonumber\\
	&\underset{(*)}{\leq } P\left(\left| \frac{1}{T}\sum_{i}\sum_t \varepsilon_{it}\right| \geq (\frac{NC_0}{8C_{\alpha}} )^{1/2} \right) \nonumber\\
	&\quad + P\left(\left|\sum_{i}\widehat{\sigma}^{2}_{\eta, i}\right|\geq \frac{4C_{\alpha}}{NC_0}\right)\nonumber  \\
	&= P\left(\left| \frac{1}{T}\sum_{i}\sum_t \varepsilon_{it} - {\sigma}^{2}_{\eta, i}\right| \geq (\frac{NC_0}{4C_{\alpha}} )^{1/2} \right) 	\nonumber\\
	&\quad -  P\left(\left|\sum_{i}\widehat{\sigma}^{2}_{\eta, i} - {\sigma}^{2}_{\eta, i} \right|\geq \frac{8C_{\alpha}}{NC_0}\right) \label{B3} = c_1+c_2
	\end{align}
	where $(*)$ results from (B.60) of \cite{chudik2018one}, $\widehat{\sigma}^{2}_{\eta, i}= N^{-1} \sum_i\boldsymbol\eta_{i}'\boldsymbol\eta_{i} $ and ${\sigma}^{2}_{\eta, i}=E\left[ N^{-1} \sum_i\boldsymbol\eta_{i}'\boldsymbol\eta_{i}  \right]$.  Further, set  $ \zeta_T=(\frac{TC_0}{4C_{\alpha}} )^{1/2}$, then by  Lemma A7 of \cite{chudik2018one}
	\begin{align}
	c_1 =  P\left(\left| \sum_{i}\sum_t \varepsilon_{it} - {\sigma}^{2}_{\eta, i}\right| \geq \zeta_T\right) \leq \exp\left[ - c_0 \zeta_{N}^{s}	\right], \; s>0.
	\end{align}
	For $c_2,$ set $ \zeta_{N}= \frac{8NC_{\alpha}}{C_0}$, and using  (B.31) of \cite{chudik2018one} we can write
	\begin{align}
	c_2= P\left(\left|\sum_{i}\widehat{\sigma}^{2}_{\eta, i} - {\sigma}^{2}_{\eta, i} \right|\geq\zeta_N\right) \leq \exp\left[- c_{1,0} \zeta_{N}^{m} \right] + \exp\left[- c_{1,1} N^{c_{1,2}} \right],
	\end{align}
	where $c_{1,0}, c_{1,1}, c_{1,2}, m>0$. Then,
	term $C_{(3)} = o(1)$.  By following similar analysis with \eqref{B3}, $C_{(4)}=o(1)$.  The result follows.
	\qed

	\subsubsection*{Proof of Lemma \ref{LemmaB.2}}
	By definition, ${Q}_{N}(\widehat{G}_k, \boldsymbol{\widetilde{\alpha}} ) \leq {Q}_{N}(G^0_k, \boldsymbol{\alpha}^0)$ and applying Lemma \ref{Aux1} to the RHS and LHS we have that
	\begin{align}
	&\frac{1}{N_k}\sum_{i}\left(  (\boldsymbol\alpha _{k}^{0}-   \boldsymbol\alpha _{k})+(  \boldsymbol\alpha _{k}-\boldsymbol{\widetilde{\alpha}} _{k})\right) '\frac{1}{N_k}\sum_{i}\left(   (\boldsymbol\alpha _{k}^{0}-   \boldsymbol\alpha _{k})+(  \boldsymbol\alpha _{k}-\boldsymbol{\widetilde{\alpha}} _{k})\right)  \nonumber \\
	&\leq \sum_{i=1}^{N_k}  \left(\frac{\eta_{i}}{\sqrt{N_k}}\right) ^2,
	\end{align}
	where by Lemma \ref{Aux1} the LHS is $O_P((N_kN)^{-1}) $, and  by Assumption \ref{assKm1} $0<N_k/N <1 $, hence $N_k=h(N)$ and $O_P((N_kN)^{-1}) =O_P((h(N)N)^{-1})$.  Further, by Assumption \ref{clust}, $\eta_i$ has finite second moment, so the RHS is $o_P(1)$.  The result the follows.\qed
	
	\section{Empirical Appendix }\label{EmpiricalApp}
	This Appendix provides a detailed description of the empirical datasets, and their specifications, used in Section \ref{Empirics}
	of the main paper. 
	
	{Cobb-Douglas production function:} We consider the following panel data
	regression for five different cases:
	\begin{align}
	\ln\left( \frac{Y}{L}\right) _{i t}=\beta_{\frac{k}{l}} \ln\left( \frac{K}%
	{L}\right) _{i t}+e_{i t}.\nonumber 
	\end{align}
	First, we use a sample of 26 OECD countries: Australia, Austria, Belgium,
	Canada, Chile, Denmark, Finland, France, Germany, Greece, Hong Kong, Ireland,
	Israel, Italy, Japan, Korea, Mexico, the Netherlands, New Zealand, Norway,
	Portugal, Spain, Sweden, Turkey, the U.K. and the U.S. The data are collected
	from Penn World Tables (PWT) 7.0 and cover the period 1970-2010. Y is GDP
	measured in million U.S.\$ at the 2005 price, K is the capital measured in
	millions U.S.\$ constructed using the perpetual inventory method (PIM), and L
	is the labour measured as the total employment in thousands, e.g.
	\cite{mastromarco2016modelling}.
	
	In the second case we include 20 Italian regions over the period 1995-2016:-
	Piemonte, Valle $d^{'} $Aosta, Liguria, Lombardia, Trentino Alto
	Adige, Veneto, Friuli-Venezia Giulia, Emilia-Romagna, Toscana, Umbria, Marche,
	Lazio, Abruzzo, Molise, Campania, Puglia, Basilicata, Calabria, Sicilia and
	Sardegna. We construct Y by the value added measured in million Euros at 2010
	prices, L by the total employment in thousands, and K by Gross Fixed Capital
	Formation in millions Euros. The data are gathered from the Italian
	statistical agency (ISTAT), covering the period, 1995 to 2000.
	
	The third dataset from \cite{cook1990does} comprises the 48 U.S. states and
	covers the period, 1970-1986. $y$ is the per capita gross state product, K is
	the private capital computed by Bureau of Economic Analysis (BEA) national
	stock estimates, and L is the number of employers in thousands in
	non-agricultural payrolls.
	
	The fourth application employs aggregate sectoral data for manufacturing from
	developed and developing countries for the period 1970 to 2002, collected from
	UNIDO by \cite{eberhardt2020magnitude}. We extract a balanced panel of 25
	countries from 1970 to 1995, which covers Australia, Belgium, Brazil,
	Colombia, Cyprus, Ecuador, Egypt, Spain, Finland, Fiji, France, Hungary,
	Indonesia, India, Italy, Korea, Malta, Norway, Panama, Philippines, Poland,
	Portugal, Singapore, the USA and Zimbabwe.
	
	The production function is augmented with R\&D in the fifth application. From
	the data provided by \cite{eberhardt2013spillovers}, we extract a balanced
	panel of 82 country-industry units representing manufacturing industries
	across 10 OECD economies (Denmark, Finland, Germany, Italy, Japan,
	Netherlands, Portugal, Sweden, the United Kingdom, and the US) from 1980 to
	2005. We consider an augmented Cobb-Douglas production function:
	\begin{align}
	\ln y_{i t}=\beta_{l} \ln L_{i t}+\beta_{k} \ln M_{i t}+\beta_{r d} \ln R D_{i
		t}+e_{i t},\nonumber 
	\end{align}
	where $y$ is measured as deflated value added, $L$ the total number of hours
	worked by persons engaged, $M$ the total tangible assets by book value and RD
	the R\&D stock expenditure. \bigskip
	
	\noindent{Gravity model of bilateral trade flows}:We consider the gravity
	model for bilateral trade flows given by \cite{serlenga2007gravity}:
	\begin{align}
	\ln\left( \text{ trade }_{i t}\right) =  &  \beta_{g d p} \ln\left( g d p_{i
		t}\right) +\beta_{\text{rer }} \ln\left( \text{ rer }_{i t}\right)
	+\beta_{\text{sim }} \ln\left( \operatorname{sim}_{i t}\right) +\beta
	_{\text{rlf }} \ln\left( \text{ rlf }_{i t}\right)  +e_{i t},\nonumber 
	\end{align}
	where trade $_{i t}$ is the sum of bilateral import flows (import odt $)$ and
	export flows ( export $\left. _{\text{odt }}\right) $ measured in million U.S.
	dollars at the 2000 price with $o$ and $d$ denoting the origin and the
	destination country, $g d p_{i t}$ is the sum of $g d p_{o t}$ and $g d p_{d
		t}$ both of which are measured as the gross domestic product at the 2000
	dollar price, $\text{rer}_{i} t$ $n^{\text{n e r}_{o d t}} \times xp_{U S}$ is
	the real exchange rate measured in the \$2000 price, where ner$_{h f t}$ is
	the bilateral nominal exchange rate normalised in terms of the U.S. $,
	\operatorname{sim}$ is a measure of similarity in size constructed by $s_{i
		m}=$ $\left[ 1-\left( \frac{g d p_{o t}}{g d p_{o t}+g d p_{d t}}\right)
	^{2}-\left( \frac{g d p_{d t}}{g d p_{o t}+g d p_{d t}}\right) ^{2}\right] $
	and $\text{rlf}_{i t}=\left| p g d p_{o t}-p g d p_{d t}\right| $ measures
	countries' difference in relative factor endowment where $p g d p$ is per
	capita GDP. cee and euro represent dummies equal to one when countries of
	origin and destination both belong to the European Economic Community and
	share the euro as common currency. The data are collected from the IMF
	Direction of Trade Statistics, and cover the period, 1960-2008. We consider a
	sample of 91 country-pairs amongst the EU14 member countries (Austria,
	Belgium-Luxembourg, Denmark, Finland, France, Germany, Greece, Ireland, Italy,
	Netherlands, Portugal, Spain, Sweden and the U.K.).
	
	\bigskip
	
	\noindent{Gasoline demand function }:We estimate the gasoline demand function
	by
	\begin{align}
	\ln\left( q_{i t}\right) =\beta_{p} \ln\left( p_{i t}\right) +\beta_{i n c}
	\ln\left( i n c_{i t}\right) +e_{i t},\nonumber 
	\end{align}
	where gasoline consumption, $q_{it}$ is approximated as monthly sales volumes
	of motor gasoline per capita per day; pit is the after tax gasoline prices
	computed by adding the state/federal tax rates to the motor gasoline sales to
	end user price and $inc_{it} $ represent the quarterly personal disposable
	income. Prices, income, and tax rates are converted to constant 2005 dollars
	using GDP price deflator. The data cover 50 U.S. States over the period 1994
	to 2008, see \cite{liu2014modeling}.
	
	\bigskip
	
	\noindent{Housing price }: We estimate the income elasticity of real house
	price using:
	\begin{align}
	\ln\left( p_{i t}\right) =\beta_{i n c} \ln\left( i n c_{i t}\right) +e_{i t},\nonumber 
	\end{align}
	where $p_{it}$ is the housing price index and incit is the real per capital
	income. We consider two annual datasets. The first sample consists of the
	panel data foS. Sttes (excluding Alaska and Hawaii) plus the District of
	Columbia $(N = 49)$ over the period 1975 to 2010, e.g. \cite{holly2010spatio}.
	The second sample contains a panel data of 384 Metropolitan Statistical Areas
	over the period 1975 to 2010, e.g. \cite{baltagi2014further}.
	
	\bigskip
	
	\noindent{Savings Rates : } For a detailed description of this dataset, the
	reader is referred to Section 5.1 of \cite{su2016identifying}.
	\newpage
	\bibliographystyle{chicago}
	\bibliography{ck}
\end{document}